\documentclass[11pt]{article}

\usepackage{subfigure}
\usepackage{color,amsmath,amssymb,amsfonts,amsbsy}
\usepackage{bbm}
\usepackage{bm}
\usepackage[final]{graphics}
\usepackage{epsfig}
\usepackage{sidecap}
\usepackage{hyperref}
\usepackage{lipsum}

\def\empile#1\over#2{\mathrel{\mathop{\kern 0pt#1}\limits_{#2}}}

\def\beq{\begin{equation}}
\def\eeq{\end{equation}}
\def\bea{\begin{eqnarray}}
\def\eea{\end{eqnarray}}

\newcommand{\Lb}{\left(}
\newcommand{\Rb}{\right)}
\def\p{{\boldsymbol p}}

\def\d3p{\frac{d^3\p}{(2\pi)^3}E_\p}

\catcode`\@=11

\newcount\@tempcntc
\def\@citex[#1]#2{\if@filesw\immediate\write\@auxout{\string\citation{#2}}\fi
  \@tempcnta\z@\@tempcntb\m@ne\def\@citea{}\@cite{%
        \@for\@citeb:=#2\do%
    {\@ifundefined{b@\@citeb}%
        {\@citeo\@tempcntb\m@ne\@citea%
                \def\@citea{,\penalty\@m\ }{\bf ?}\@warning%
                {Citation `\@citeb' on page \thepage \space undefined}}%
        {\setbox\z@\hbox{\global\@tempcntc0\csname b@\@citeb\endcsname\relax}
     \ifnum\@tempcntc=\z@ \@citeo\@tempcntb\m@ne%
       \@citea\def\@citea{,\penalty\@m}%
       \hbox{\csname b@\@citeb\endcsname}%
     \else%
      \advance\@tempcntb\@ne%
      \ifnum\@tempcntb=\@tempcntc%
      \else\advance\@tempcntb\m@ne\@citeo%
      \@tempcnta\@tempcntc\@tempcntb\@tempcntc\fi\fi}}\@citeo}{#1}}%

\def\@citeo{\ifnum\@tempcnta>\@tempcntb\else\@citea
  \def\@citea{,\penalty\@m}%
  \ifnum\@tempcnta=\@tempcntb\the\@tempcnta\else
   {\advance\@tempcnta\@ne\ifnum\@tempcnta=\@tempcntb \else
\def\@citea{--}\fi
    \advance\@tempcnta\m@ne\the\@tempcnta\@citea\the\@tempcntb}\fi\fi}

\catcode`\@=12

\begin{document}

\title{\bf Photons from thermalizing matter \\ in heavy ion collisions}
\author{Vladimir Khachatryan$^{1}$,  Bj\"orn Schenke$^{2}$, Mickey Chiu$^{2}$, \\
Axel Drees$^{1}$, Thomas K. Hemmick$^{1}$, Norbert Novitzky$^{1}$
}

\date{}
\maketitle
\begin{enumerate}
\centering
\item Department of Physics and Astronomy, Stony Brook University, \\ Stony Brook, NY 11794, USA \\
\item Physics Department, Bldg.\,510A, Brookhaven National Laboratory, \\ Upton, NY 11973, USA 
\end{enumerate}

\begin{abstract}
We investigate the production of direct photons in heavy ion collisions within the modified bottom-up thermalization scenario, which we show to be related to the thermalizing Glasma framework. The dynamics of the parton system up to thermalization/equilibration can be described by two momentum scales, by means of which we express the photon invariant yield excluding the prompt pQCD contribution. We derive an analytic formula, which provides an estimate of photon production from thermalizing matter for a wide range of collision systems and energies. We compare the yield with that measured in the PHENIX Au+Au run04 and the combined run07+run10 data sets at $\sqrt{s_{NN}} = 200$\,GeV. We also make theory-data comparisons for Pb+Pb at 2760\,GeV and d+Au at 200\,GeV collision energies. Finally, we make predictions for the direct photon invariant yield from collisions of U+U,  Cu+Au, $^{3}$He+Au at 200\,GeV and Pb+Pb at 5020\,GeV.
\end{abstract}

\newpage
\vskip 1.0truecm
\tableofcontents

\section{Introduction}
\label{sec:Introduction}
Studies of direct photons are of great importance for understanding some of the key characteristics of evolution of the matter produced in relativistic heavy ion collisions. In the context of  heavy ion collisions  direct photons are all the photons, which do not emerge from hadronic decays. Possibly radiated from all phases of the evolution, direct photons are excellent probes for extracting information on the medium at their space-time production points because of the weak interaction of photons with the medium. 

One can obtain the direct photon yield by adding the thermal (or thermal-like) photon yield to the pQCD \cite{Vogelsang:1993} prompt photon contribution. The hard photons at high $p_{T}$ and the thermal photons at low $p_{T}$ carry information on the initial hard scattering and pre-equilibrium phase as well as on the equilibrium phase of the parton system, Quark-Gluon Plasma (QGP) and hadronic gas (HG). In particular, the thermal photons are radiated from hot and dense matter \cite{Stankus:2005} that can be in local (or out of local) equilibrium, and carry information about the space-time evolution of the medium.

The direct photon yield in p+p collisions is consistent with the NLO pQCD calculations \cite{Adare:2012}. These results serve as a crucial reference for photon yields in heavy ion collisions. Here, the yield at $p_{T} > 4$\,GeV/c is found to be consistent with NLO pQCD calculations scaled by the number of binary collisions (or by the Glauber nuclear overlap function) at different centralities. However, in the low and intermediate $p_{T}$ range it has been predicted that the photon yield is enhanced by thermal radiation from the QGP and HG \cite{Turbide:2004}. This is in agreement with experimental observations: Although the direct photon measurements are quite challenging, especially in the low transverse momentum region, the PHENIX, STAR and ALICE collaborations have observed evidence of thermal radiation from Au+Au \cite{Adare:2008qk}-\cite{STAR:2016} and Pb+Pb collisions \cite{Wilde:2013} at RHIC $\sqrt{s_{NN}} = 200$\,GeV and at LHC $\sqrt{s_{NN}} = 2760$\,GeV collision energies, respectively.

There is abundant literature on direct and thermal photon studies, for example, based on the spectral function approach \cite{Dusling:2010}-\cite{Kim:2017}, Parton-Hadron-String Dynamics transport approach \cite{Bratkovskaya:2008}-\cite{Linnyk:2015} as well as based on scenarios that use simulations in the framework of the elliptic-fireball expansion \cite{vanHees:2011}-\cite{vanHees:2015} and hydrodynamic simulations of the fireball evolution \cite{DPSYJG:2011}-\cite{SPDJG:2015}. Other results can be found in \cite{Srivastava:2001a}-\cite{Hidaka:2015}. For a recent theory overview we refer to \cite{Shen:2015}.

In this work we focus on photon production in the pre-equilibrium and equilibrium stages of the parton system. Let us briefly discuss some recent theoretical results for parametric estimates of the photon production in the early time regime of the matter produced in relativistic heavy ion collisions. Refs.\,\cite{Chiu:2013}-\cite{Schenke:2014} give an explanation of direct photon enhancement in the intermediate $p_{T}$ region based on a production mechanism of thermal photons from the thermalizing Glasma phase \cite{Blaizot:2011xf,Liao:2013}\footnote{For recent studies of the thermalization process see Refs.\,\cite{Liao:2014}-\cite{BLM:2016}.}. The Glasma \cite{Lappi:2006fp}-\cite{Gelis:2012} is theorized to exist based on gluon saturation (Color Glass Condensate) physics \cite{GLR:1983}-\cite{Gelis:2010}. In \cite{Chiu:2013,BosMc:2014}, it is shown that direct photon production from the Glasma shows geometric scaling\footnote{The geometric scaling expresses rates of particles in terms of dimensionless ratios of transverse momentum to the saturation momentum $Q_{s}$.} at different centralities and collision energies. A parametric estimate of photon production in various stages of the bottom-up thermalization scenario \cite{Baier:2001,Baier:2002} is presented in very recent studies \cite{BeReTaVe:2017}. In the bottom-up thermalization framework, a soft gluon thermal bath originates via QCD decays of A+A collision-induced primary hard gluons, the momenta of which are of the order of the saturation momentum scale. 

In this paper we propose an alternative photon production mechanism that is  based on the modified bottom-up thermalization scenario \cite{Mueller:2006}-\cite{Chiu:2015}, which is a saturation-based approach toward thermalization/equilibration of the parton system produced in relativistic heavy ion collisions. We identify two momentum scales in the modified bottom-up thermalization scenario and in this way relate it to the thermalizing Glasma \cite{Blaizot:2011xf,Liao:2013}. Consequently, we refer to the model as mBU-Glasma thermalization, where mBU stands for modified bottom-up.

The paper is organized as follows. In Sec.\,\ref{sec:mbottom-up} we discuss the photon production in the mBU-Glasma thermalization ansatz. In Sec.\,\ref{sec:Fitting} we fit/compare the derived yield with that from the PHENIX run04 Au+Au  data \cite{Adare:2008qk,Adare:2009qk} as well as from the combined run07+run10 Au+Au data \cite{Bannier:2014}, both measured at $\sqrt{s_{NN}} = 200$\,GeV. We also consider comparisons with the ALICE Pb+Pb data at $\sqrt{s_{NN}} = 2760$\,GeV \cite{Wilde:2013} and the PHENIX run08 d+Au data at $\sqrt{s_{NN}} = 200$\,GeV \cite{Yamaguchi:2012}. More specifically, for the PHENIX run04 Au+Au at $\sqrt{s_{NN}} = 200$\,GeV first we fit the theoretical yield with the data in one centrality bin (for example, $0$-$20\%$) in order to fix the normalization in the photon production formula. Then we apply the fixed normalization to the other available centrality bins. For the PHENIX run07+run10 Au+Au, run08 d+Au and for the ALICE Pb+Pb we carry out global fitting for all the available centrality bins. In Sec.\,\ref{sec:Predictions} we present predictions for the direct photon invariant yield for several symmetric A+A and asymmetric A$_{1}$+A$_{2}$ systems: namely, for U+U,  Cu+Au, $^{3}$He+Au at $\sqrt{s_{NN}} = 200$\,GeV and Pb+Pb at $\sqrt{s_{NN}} = 5020$\,GeV. We summarize our results in Sec.\,\ref{sec:Conclusions}. 

In Appendix I we show how the mBU-Glasma solutions (based on introduced ultraviolet and infrared momentum scales) are related to those of the thermalizing Glasma from \cite{Blaizot:2011xf,Liao:2013}. These relations basically mean that most of the results of this paper are at least parametrically the same for both the thermalizing Glasma and mBU-Glasma models. In Appendix II we show how these scales are utilized to estimate the thermalization time/temperature of the system produced in the most central Au+Au collisions at RHIC 200\,GeV beam energy. In Appendix III we use the ansatz developed in this work to give a qualitative estimate of  $e^{+}e^{-}$ pair (and associated photon) production in Au+Au collisions at $\sqrt{s_{NN}} = 200$\,GeV.

\section{Phenomenology of photon production in the mBU-Glasma thermalization ansatz}
\label{sec:mbottom-up}

In Ref.\,\cite{Chiu:2015} the dominant qualitative and semi-quantitative features of evolution of the parton system (produced in relativistic heavy ion collisions) toward thermalization are expressed by two momentum scales, ultraviolet $\Omega_{UV}$ and infrared $\Omega_{IR}$. As it is noted in the introduction, we refer to this framework as mBU-Glasma thermalization. In this section we use some results of \cite{Chiu:2015} in order to derive the photon rate in the mBU-Glasma thermalization ansatz. We begin with a description of the original bottom-up thermalization scenario \cite{Baier:2001,Baier:2002}, subsequently moving to the modified bottom-up thermalization and to the mBU-Glasma thermalization. The derivations of the momentum scales $\Omega_{UV}$ and $\Omega_{IR}$ are shown in Appendix I.

\subsection{The original bottom-up approach towards thermalization/equilibration}

A very important question in the physics of (ultra)relativistic heavy ion collisions is how the collision-produced system gets thermalized/equilibrated. One can investigate it in the framework of perturbative QCD. In order to do so, we invoke the bottom-up thermalization scenario \cite{Baier:2001,Baier:2002,BeReTaVe:2017}, which was recently confirmed using classical-statistical numerical lattice simulations \cite{BBSV:2014a,BBSV:2014b,BBSV:2014c,BSSV:2014}. The common argument in favor of thermalization at high collision energies is that a large number of gluons are liberated in the first moment after the collision, which then collide frequently with each other in the course of the evolution. However, the distribution of these gluons is initially very far from thermal equilibrium. In addition, the strong coupling constant decreases at high collision energies, making it more difficult to achieve thermalization. Whether the system has enough time to thermalize/equilibrate before falling apart is thus not a straightforward question requiring detailed consideration of different physical processes at early times. In the limit $Q_{s} \gg \Lambda_{QCD}$ that corresponds to very large nuclei and/or very high collision energies, it turns out that the thermalization can occur relatively fast while the whole system is still undergoing one-dimensional expansion.

The bottom-up pre-equilibrium evolution is divided into three temporal stages:
\begin{itemize} \setlength{\itemsep}{-3.0mm}
\item[(\textbf{i})] $Q_{s}^{-1} \ll \tau \ll Q_{s}^{-1} \alpha_s^{-3/2}$ \\
\item[(\textbf{ii})] $Q_{s}^{-1} \alpha_{s}^{-3/2} \ll \tau \ll Q_{s}^{-1} \alpha_{s}^{-5/2}$ \\
\item[(\textbf{iii})] $Q_{s}^{-1} \alpha_{s}^{-5/2} \ll \tau \ll Q_{s}^{-1} \alpha_{s}^{-13/5}$ .
\end{itemize} 
At $\tau < Q_{s}^{-1}$, because of the large occupation number, the collision-produced gluons (called primary hard gluons) interact so strongly that it is more appropriate to describe them as a nonlinear gluon field rather than a collection of particles. $Q_{s}^{-1}$ defines a transition boundary between the region occupied by the nonlinear gluon field and the one at which a description by the Boltzmann equation becomes applicable. In this case the field becomes almost linear, and one can start describing the gluons as particles on mass shell with a well-defined distribution $f(p_{T}/Q_{s})$. 

The beginning of stage (\textbf{i}) is highly occupied by the primary hard gluons, where the occupation number ranges from $f\sim 1/\alpha_s$ at the earliest time, $\tau\sim Q_s^{-1}$, to unity at $\tau \sim Q_s^{-1} \alpha_s^{-3/2}$. The density of the hard gluons, $n_{h}$, decreases with time because of the one-dimensional expansion. The occupation number decreases in time  as $(Q_s \tau)^{-2/3}$, which is  a consequence of broadening of the longitudinal momentum distribution by elastic scatterings among the hard gluons. These elastic scatterings modify the typical longitudinal momentum from $p_{z}\sim 1/\tau$ to $p_{z} \sim  \tau^{-1/3}$. Later on, gluons with smaller momenta, but still larger than $\Lambda_{QCD}$, are being produced. These gluons interact by elastic scatterings at small angles with exchanged momentum $k_{s} \ll Q_{s}$.

The dynamics of the stages (\textbf{ii}) and (\textbf{iii}), where the occupancy of hard gluons is less than unity, are dictated by quantum kinetic theory. In stage (\textbf{ii}) the soft gluons are being produced as a result of inelastic scatterings, via processes $hard+hard \rightarrow hard+hard+soft$. They dominate the screening by providing a larger contribution to the Debye mass relative to that from the hard gluons. The soft gluons have characteristic momenta estimated to be of the order of $\alpha_{s}^{1/2}Q_{s} > \Lambda_{QCD}$. However, the occupation number of the soft gluons is still significantly smaller than that of the hard gluons. These numbers become comparable to each other at $Q_{s}\tau \sim \alpha_s^{-5/2}$: $n_{s} \sim n_{h}$. The momentum anisotropy saturates at the value of the ratio of longitudinal to transverse pressure $P_{L}/P_{T} \sim \alpha_{s}$.

By the beginning of stage (\textbf{iii}), at $\tau > Q_{s}^{-1} \alpha_s^{-5/2}$, the soft gluons start dominating the total multiplicity:  $n_{s} > n_{h}$. They achieve thermal equilibration among themselves, forming a soft gluon thermal heat bath characterized by the temperature $T$: $n_{s} \sim T^{3}$. The remaining hard gluons collide with the soft ones, and constantly loose energy to the soft gluon thermal heat bath. A hard gluon emits a softer energy gluon, which splits into gluons with comparable momenta. The products of this branching quickly cascade further, giving all their energy to the thermal bath. Combination of expansion and infusion of energy into the thermal bath raises its temperature up to $T_{therm} = c_{T}\, c_{eq}\,\alpha_{s}^{2/5} Q_{s}$ at time $\tau_{therm} = c_{eq} \, \alpha_{s}^{-13/5} Q_s^{-1}$, where $c_{T}$ and $c_{eq}$ are constants, to be determined from experimental data \cite{Baier:2002}.

\subsection{From modified bottom-up to mBU-Glasma thermalization}

There has been a realization that collective effects in the form of magnetic plasma instabilities, known as chromo-Weibel or filamentary instabilities, necessarily play a role in the initial stage of the bottom-up thermalization scenario \cite{ArLeMo:2003} (see \cite{Mrow:2007}-\cite{Weibel:1959} for early discussions and the recent review \cite{Mrowczynski:2016etf}). The source of these instabilities is a collection of hard particles, present in the dense system produced immediately after the collision, with highly asymmetric particle momentum distributions. Numerical simulations also seemed to indicate that the plasma instabilities are effective at early times \cite{Strick:2007}-\cite{DuNa:2005}. Because of the plasma instabilities, the amount of energy transfer, flowing from initially produced hard modes to later produced soft modes, increases, potentially allowing for more rapid thermalization. In more general terms, the instabilities initially grow exponentially, generating strong transverse chromo-magnetic/electric fields at short times \cite{Strick:2007}, which can speed up local isotropization and thermalization of the initial non-equilibrium plasma by scattering the plasma particles into random directions. Nonetheless, as it was shown in \cite{ArLe:2004}, the full equilibration time in the presence of instabilities is not much shorter relative to that of the original bottom-up scenario. 

The modified bottom-up thermalization approach was developed \cite{Mueller:2006,Mueller:2006II} to include effects of anisotropic momentum distributions and instability growth. Scaling solutions, depending on one single parameter, have been proposed for following the evolution between instabilities happening in the initial phase and the system at the equilibration time. Both the collision-produced hard gluons and the soft gluons produced later during the evolution are part of the collision term in the Boltzmann equation. The scaling solutions have the following form:
\begin{eqnarray}
& & n_{s} \sim {Q_{s}^{3} \over \alpha_{s} (Q_{s}\tau)^{4/3 - \delta}}\,,\,\,\,\,\,\,\,\,\,\,\,\,\,k_{s} \sim {Q_{s} \over
(Q_{s}\tau)^{1/3 - 2\delta/5}}\,,
\nonumber\\
& & \alpha_{s}f_{s} \sim {1 \over (Q_{s}\tau)^{1/3 + \delta/5}}\,,\,\,\,\,\,\,\,\,\,\,m_{D} \sim {Q_{s} \over
(Q_{s}\tau)^{1/2 - 3\delta/10}}\,,
\label{eqn_scaling}
\end{eqnarray}
where $\delta$ is a positive real number characterizing the anisotropy of the system, $n_{s}$ is the number density of the soft gluons, $k_{s}$ is the soft gluon momentum, $f_{s}$ is the soft gluon occupation number, and $m_{D}$ is the Debye mass. 

At $\delta = 0$, the family of the $\delta$-parameter dependent scaling solutions coincides with the initial parametric form of the original bottom-up picture \cite{Baier:2001,Baier:2002}. At a time $\tau \sim Q_{s}^{-1} \alpha_{s}^{-15/2(5 - 6\delta)}$, the scaling solutions in Eq.\,(\ref{eqn_scaling}) become identical to the intermediate stage, $Q_{s}^{-1} \alpha_{s}^{-3/2} \ll \tau \ll Q_{s}^{-1} \alpha_{s}^{-5/2}$, of the original bottom-up. This is true if $0 < \delta < 1/3$. At $\delta = 1/3$, the matching happens when the soft particles are starting to thermalize/equilibrate. Solutions with $\delta > 1/3$ produce equilibration of the soft particle sector earlier than the corresponding time in the original bottom-up\footnote{That time in the original bottom-up is $\tau_{therm} \sim Q_{s}^{-1} \alpha_{s}^{-{13/5}}$.}, and in this case the matching with the bottom-up only occurs when the whole system is reaching equilibration/thermalization. In the modified bottom-up thermalization scenario the parameter $\delta$ accepts values in the range of $0 <\delta < 10/21$. The upper limit $\delta = 10/21$ sets the absolute limit of the soft gluon energy density, $N_{s}k_{s}$, being equal to the hard gluon energy density. See Appendix I for more details.

One may ask whether or not momentum scales like the thermalizing Glasma's ultraviolet $\Lambda_{UV}$ and infrared $\Lambda_{IR}$ \cite{Chiu:2013,Schenke:2014,Blaizot:2011xf}, exist in the modified bottom-up thermalization. Conceptually, this should be the case, because the main qualitative features of the solution of the Boltzmann equation can be described with an assumption that the evolution is dominated by such scales. In momentum space, the thermalizing Glasma gluon distribution is described as follows (at time $\tau > 1/Q_{s}$) \cite{Blaizot:2011xf}:
\begin{eqnarray}
\,\,\,\,\,\,\,\,\,\,& & f_{g}(p) \sim {1 \over \alpha_{s}}\,\,\,\,\,\,\,\,\,\,\,\,\,\,\,\,\,\,\,\,\,\mbox{at}\,\,p < \Lambda_{IR}\,, 
\nonumber\\
\,\,\,\,\,\,\,\,\,\,& & f_{g}(p) \sim {1 \over \alpha_{s}}{\Lambda_{IR} \over \omega_{p}}\,\,\,\,\,\,\,\,\,\,\mbox{at}\,\,\Lambda_{IR} < p < \Lambda_{UV}\,, 
\nonumber\\
\,\,\,\,\,\,\,\,\,\,& & f_{g}(p) \sim 0\,\,\,\,\,\,\,\,\,\,\,\,\,\,\,\,\,\,\,\,\,\,\,\,\,\,\,\mbox{at}\,\,p > \Lambda_{UV}\,, 
\label{eqn_Glasma_distributions_gl}
\end{eqnarray}
where $p$ is the gluon momentum, and $\omega_{p}$ its energy.

In the case of the modified bottom-up approach toward thermalization, let us assume the existence of analogous scales designated as $\Omega_{UV}$ and $\Omega_{IR}$, by which the modified bottom-up thermalization scenario can be re-expressed in a form we will call mBU-Glasma thermalization scenario:
\begin{eqnarray}
\,\,\,\,\,\,\,\,\,\,& & f_{mBU}(p) \sim {1 \over \alpha_{s}}\,\,\,\,\,\,\,\,\,\,\,\,\,\,\,\,\,\,\,\,\,\,\mbox{at}\,\,p < \Omega_{IR}\,, 
\nonumber\\
\,\,\,\,\,\,\,\,\,\,& & f_{mBU}(p) \sim {1 \over \alpha_{s}}{\Omega_{IR} \over p}\,\,\,\,\,\,\,\,\,\,\mbox{at}\,\,\Omega_{IR} < p < \Omega_{UV}\,, 
\nonumber\\
\,\,\,\,\,\,\,\,\,\,& & f_{mBU}(p) \sim 0\,\,\,\,\,\,\,\,\,\,\,\,\,\,\,\,\,\,\,\,\,\,\,\,\,\,\,\mbox{at}\,\,p > \Omega_{UV}\,.
\label{eqn_Glasma_distributions}
\end{eqnarray}
This assumption is substantiated hereinafter. Making use of the second formula of Eq.\,(\ref{eqn_Glasma_distributions_gl}), we assume that an equivalent relation (with \,$p \approx \omega_{p}$\, at \,$m_{D} \ll \omega_{p}$) exists in the mBU-Glasma thermalization picture. Then the soft gluon number density $n_{s}$ and the Debye mass $m_{D}$ can be expressed via the ultraviolet $\Omega_{UV}$ and infrared $\Omega_{IR}$ scales:
\begin{equation}
n_{s} = \int^{\Omega_{UV}} d^{3}p\,f_{mBU}(p) \sim \int^{\Omega_{UV}} p^{2}dp\,{1 \over \alpha_{s}}{\Omega_{IR} \over p} \sim {1 \over \alpha_{s}} \Omega_{UV}^{2} \Omega_{IR}\,,
\label{eqn_bottom_scales1}
\end{equation}
\begin{equation}
m_{D}^{2} = \alpha_{s}\int^{\Omega_{UV}} d^{3}p\,{f_{mBU}(p) \over p} \sim \alpha_{s}\int^{\Omega_{UV}} p^{2}dp\,{1 \over \alpha_{s}}{\Omega_{IR} \over p^{2}} \sim \Omega_{UV}\Omega_{IR}\,.
\label{eqn_bottom_scales2}
\end{equation}
For the quark occupation number we use \,$f_{q} \sim 1$,\, such that the quark number density is
\begin{equation}
n_{q} = \int^{\Omega_{UV}} d^{3}p\,f_{mBU}(p) \sim \Omega_{UV}^{3}\,.
\label{eqn_bottom_scales3}
\end{equation}
As in the case of the thermalizing Glasma, at the earliest times we have $n_{q} \sim \alpha_{s} n_{h} \ll n_{h}$ but at later times $n_{q} \sim n_{s}$, i.e., the two densities approach each other. Here, $n_{h}$ is the number density of the primary hard gluons (initial gluons).

The exact time dependence of the scales $\Omega_{UV}$ and $\Omega_{IR}$ in the mBU-Glasma thermalization scenario will be derived in Appendix I. Here we merely state the most basic relations, starting from Eqs.\,(\ref{eqn_bottom_scales1}) and (\ref{eqn_bottom_scales2}):
\begin{equation}
n_{s} \sim {1 \over \alpha_{s}}\Omega_{UV}^{2} \Omega_{IR}\,,
\label{eqn_bottomgluon}
\end{equation}
\begin{equation}
m_{D}^{2} \sim \Omega_{UV}\Omega_{IR}\,,
\label{eqn_bottomDebye}
\end{equation}
The scales have to fulfill the following requirements:
\begin{equation}
\Omega_{UV}(\tau_{0}) = \Omega_{IR}(\tau_{0}) \sim Q_{s}\,,
\label{eqn_bottom_initial}
\end{equation}
and
\begin{equation}
\Omega_{UV}(\tau_{therm}) \sim T_{therm} \approx T_{in,QGP}\,,\,\,\,\,\,\Omega_{IR}(\tau_{therm}) \sim \alpha_{s}\, \Omega_{UV}(\tau_{therm})\,.
\label{eqn_bottom_boundary}
\end{equation}

Thereby, by utilizing Eqs.\,(\ref{eqn_bottomgluon}), (\ref{eqn_bottomDebye}), (\ref{eqn_bottom_initial}) and (\ref{eqn_bottom_boundary}) at the initial time $\tau_{0} \sim Q_{s}^{-1}$, we shall have the ``initial conditions":
\begin{equation}
n_{h} \sim { Q_{s}^{3} \over \alpha_{s}(Q_{s}\tau_{0})}\,,
\label{eqn_bottomgluon_in}
\end{equation}
\begin{equation}
m_{D}^{2} \sim {Q_{s}^{2} \over (Q_{s}\tau_{0})}\,,
\label{eqn_bottomDebye_in}
\end{equation}
which exist in the original bottom-up thermalization as well. In addition to these ``initial conditions", at the thermalization/equilibration time there exist also the following conditions, which we name as ``thermalization conditions":
\begin{equation}
n_{s} \sim T_{therm}^{3}\,,
\label{eqn_bottomgluon_fin}
\end{equation}
\begin{equation}
m_{D}^{2} \sim \alpha_{s} T_{therm}^{2}\,.
\label{eqn_bottomDebye_fin}
\end{equation}

\subsection{The photon production at $\mathbf{p_{T} \gtrsim 1\,GeV/c}$}

By using the modified bottom-up scaling solutions (see Eq.\,(\ref{eqn_scaling})) for the Debye mass $m_{D}$ and soft gluon number density $n_{s}$ \cite{Mueller:2006} as well as the expressions for $m_{D}$ from Eq.\,(\ref{eqn_bottom_scales2}) and $n_{s}$ from Eq.\,(\ref{eqn_bottom_scales1}) expressed by the scales $\Omega_{UV}$ and $\Omega_{IR}$, one can obtain the following relations for $Q_{s}\tau$:
\begin{equation}
m_{D}^{2} \sim {Q_{s}^{2} \over (Q_{s}\tau)^{1- 3\delta/5}}\,\,\,\,\,\,\,\,\,\,\,\,\mbox{and}\,\,\,\,\,\,\,\,\,\,\,\,m_{D}^{2} \sim \Omega_{UV}\Omega_{IR}\,,
\label{eqn_MD}
\end{equation}
giving
\begin{equation}
Q_{s}\tau \sim \Lb {Q_{s}^{2} \over \Omega_{UV}\Omega_{IR}} \Rb^{1 \over 1- 3\delta/5}\,,
\label{eqn_Qstau1}
\end{equation}
and 
\begin{equation}
n_{s} \sim {Q_{s}^{3} \over \alpha_{s} (Q_{s}\tau)^{4/3 - \delta}}\,\,\,\,\,\,\,\,\,\,\,\,\mbox{and}\,\,\,\,\,\,\,\,\,\,\,\,n_{s} \sim {1 \over \alpha_{s}} \Omega_{UV}^{2} \Omega_{IR}\,,
\label{eqn_Ns}
\end{equation}
giving
\begin{equation}
Q_{s}\tau \sim \Lb {Q_{s}^{3} \over \Omega_{UV}^{2}\Omega_{IR}} \Rb^{1 \over 4/3 - \delta}\,.
\label{eqn_Qstau2}
\end{equation}
From Eq.\,(\ref{eqn_Qstau1}) and Eq.\,(\ref{eqn_Qstau2}) we get the following result:
\begin{equation}
\Omega_{IR} \sim \Omega_{UV}^{3\delta - 10 \over 6\delta - 5}Q_{s}^{3\delta + 5 \over 6\delta - 5}\,.
\label{eqn_OmegaIR}
\end{equation}
Then using Eq.\,(\ref{eqn_Ns}) we find
\begin{equation}
\Omega_{UV} \sim Q_{s} {\Lb 1 \over Q_{s}\tau \Rb}^{5 - 6\delta \over 15}\,.
\label{eqn_Omega}
\end{equation}
We would arrive at the same parametric form if we looked at both formulas of $m_{D}$ from Eq.\,(\ref{eqn_MD}) along with using the result from Eq.\,(\ref{eqn_OmegaIR}). Finally Eq.\,(\ref{eqn_Omega}) can be rewritten as an expression for $\tau$:
\begin{equation}
\tau \sim {1 \over Q_{s}}\Lb {Q_{s} \over \Omega_{UV}} \Rb^{15 \over 5 - 6\delta}\,,
\label{eqn_tau}
\end{equation}
and
\begin{equation}
d\tau \sim \Lb {-15 \over 5 - 6\delta} \Rb {1 \over Q_{s}}\Lb {Q_{s} \over \Omega_{UV}} \Rb^{15 \over 5 - 6\delta}{d\Omega_{UV} \over \Omega_{UV}}\,.
\label{eqn_dtau}
\end{equation}

We can now estimate the rate of the photon production in the mBU-Glasma scenario in a similar way as it has been accomplished for the thermalizing Glasma \cite{Chiu:2013}\footnote{In the thermalizing Glasma during the overpopulated stage all the way toward thermalization, the system behaves as a strongly interacting fluid, even though the coupling constant is small. The system may evolve for a long time with a fixed anisotropy between average longitudinal and transverse momenta. In the mBU-Glasma thermalization picture the coupling constant is also small, and the fixed anisotropy between the longitudinal and transverse momenta is actually related to that of the thermalizing Glasma, by which in both scenarios the system at thermalization has parametrically the same gluon density, Debye mass, thermalization time, energy density and entropy density (see Appendix I). Both scenarios have scaling solutions to the Boltzmann transport equation. The only key difference between the two is that the thermalizing Glasma includes a transient component of the system, which is a Bose-Einstein condensate of gluons \cite{Chiu:2013,Blaizot:2011xf,Liao:2013}.}, starting with the rate of photon production at finite temperature derived for thermal emission from a QGP in a fixed box \cite{Kapusta:1991,Srivastava:2001b}:
\begin{equation}
{d^{7}N \over {d^{4}x dy d^{2}p_{T}}} = {{\alpha \alpha_s} \over {2\pi^2}}\,T^2 e^{-E/T} h(E/T)\,,
\label{eqn_Photon_rate}
\end{equation}
where $h$ is a slowly varying function (dependent on $E/T$) of order one. The factor $\alpha_s$ arises from interaction of quarks with the medium during the photon production process. In the mBU-Glasma, this is compensated for by the high gluon density $\sim 1/\alpha_s$. Note that the quark number density, up to an overall constant, is identical to that of the QGP with the substitution $T \rightarrow \Omega_{UV}$ (as in the thermalizing Glasma). Our assumption is that for emission of the mBU-Glasma photons, emanated during the evolution of the matter toward and at thermalization, we can use the following formula\footnote{A heuristic derivation of a similar formula is given in section $A$ of the Appendix in \cite{Chiu:2013}.}:
\begin{equation}
{d^{7}N \over {d^{4}x dy d^{2}p_{T}}} \sim \alpha\,\Omega_{IR} \Omega_{UV}\,u(E/\Omega_{UV})\,,
\label{eqn_Photon_rate1}
\end{equation}
where $u(E/\Omega_{UV})$ is a function that cuts off when the photon energy is of the order of the ultraviolet scale $\Omega_{UV}$. It  incorporates an exponential function $e^{-E/\Omega_{UV}}$ and a logarithmic function dependent on $E/\Omega_{UV}$. At thermalization when $T_{therm} \sim \Omega_{UV} \sim \Omega_{IR}/\alpha_s$ (see Eq.\.(\ref{eqn_bottom_boundary})) it reduces to that of the thermal emission in Eq.\,(\ref{eqn_Photon_rate}). 


In order to compute the photon yield, we first integrate over longitudinal coordinates by assuming that the early time expansion is purely longitudinal, and that in the integration the space-time rapidity is strongly correlated with the momentum-space rapidity.
We make an assumption that Eq.\,(\ref{eqn_Photon_rate1}) is $\eta$-independent, which means that the yield is boost-invariant.

Thus, using the four-volume element expressed by
\begin{equation}
d^{4}x = dt\,d^{2}r_{T}\,dz \approx d\tau\,d^{2}r_{T}\,\tau\,d\eta\,,
\label{eqn_element}
\end{equation}
and integrating both sides of Eq.\,(\ref{eqn_Photon_rate1}) with $\tau d\tau d\eta$, we obtain the following result at midrapidity:
\begin{equation}
{d^{5}N \over {d^{2}r_{T} dyd^{2}p_{T}}} \sim \alpha \int \tau d\tau\,\Omega_{IR}\Omega_{UV}\,u(p_T/\Omega_{UV})\,,
\label{eqn_Photon_rate2}
\end{equation}
Inserting the expressions for $\Omega_{IR}$  from Eq.\,(\ref{eqn_OmegaIR}), $\tau$ from Eq.\,(\ref{eqn_tau}), and $d\tau$ from Eq.\,(\ref{eqn_dtau}) into the r.h.s of Eq.\,(\ref{eqn_Photon_rate2}) yields
\bea
{d^{5}N \over {d^{2}r_{T} dyd^{2}p_{T}}} & \sim & \alpha \int \Lb {1 \over Q_{s}}\Lb {Q_{s} \over \Omega_{UV}} \Rb^{15 \over 5 - 6\delta} \Rb \times
\nonumber \\
& & \times \Lb \Lb {-15 \over 5 - 6\delta} \Rb {1 \over Q_{s}}\Lb {Q_{s} \over \Omega_{UV}} \Rb^{15 \over 5 - 6\delta}{d\Omega_{UV} \over \Omega_{UV}} \Rb  \Lb \Omega_{UV}^{3\delta - 10 \over 6\delta - 5}Q_{s}^{3\delta + 5 \over 6\delta - 5} \Rb \Omega_{UV}\,u\!\!\Lb {p_T \over \Omega_{UV}} \Rb \sim
\nonumber \\
& \sim & \alpha \Lb {-15 \over 5 - 6\delta} \Rb Q_{s}^{{15 + 9\delta \over 5 - 6\delta}} \int \Omega_{UV}^{-{20 + 3\delta \over 5 - 6\delta}}\,u\!\!\Lb {p_T \over \Omega_{UV}} \Rb d\Omega_{UV}\,,
\label{eqn_Photon_rate3}
\eea
which reduces to

\begin{equation}
 {d^{5}N \over {d^{2}r_{T} dy d^{2}p_{T}}} \sim \alpha{\Lb 3 + 2\Delta \over 3 \Rb} \left(Q_{s} \over p_{T} \right)^{\Delta}\,,
\label{eqn_Photon_rate4}
\end{equation}
where $\Delta=(15 + 9\delta)/(5 - 6\delta)$. When explicitly evaluating this integral, we have to make sure that the dominant part of the contribution does not overlap with the end points of integration. In the thermal case the dominant range of integration is at $p_{T} \sim 6T$ \cite{Chiu:2013}. We use the same argumentation for $\Omega_{UV}$, assuming also that the smallest possible value for $\Omega_{UV}$ is of the order of the QCD transition (deconfinement) temperature $\sim 170$\,MeV. Besides, one can take $Q_{s} \sim 1$\,GeV \cite{Baier:2002}, e.g., at RHIC energies. Consequently, we can rely on Eq.\,(\ref{eqn_Photon_rate4}) for estimating photon production in the range of 1\,GeV $\lesssim p_{T} \lesssim$ 6\,GeV. An analogous result has been derived for photon production from the thermalizing Glasma \cite{Chiu:2013}.

Integrating over $d^{2}r_{T}$ in Eq.\,(\ref{eqn_Photon_rate4}), and identifying the overlap cross section as being proportional to the number of participants, $N_{part}$, we ultimately obtain
\begin{equation}
{{d^{3}N_{\gamma}} \over {dy d^{2}p_{T}}} \sim \alpha R_{0}^{2} N_{part}^{2/3}{\Lb 3 + 2\Delta \over 3 \Rb} \left(Q_{s} \over p_T \right)^{\Delta}\,.
\label{eqn_Photon_rate44}
\end{equation}
Here the constant $R_{0}$ has dimensions of length that can be of the order of $0.1\,\rm{fm} \div 1\,\rm{fm}$, nevertheless, cannot be determined precisely because of the approximations used. The factor of $N_{part}^{2/3}$ arises because the number of participants in a collision is proportional to the nuclear volume: $N_{part} \sim R_{N}^{3}$, and so $R_{N}^{2} = R_{0}^{2}N_{part}^{2/3}$. The exponent of the ratio \,$Q_{s}/p_{T}$\, ranges as
\begin{equation}
3 \leqslant \Delta \leqslant 9\,,
\label{eqn_Delta}
\end{equation}
with the two limits $\Delta = 3$ and $\Delta = 9$ corresponding to $\delta=0$ and $\delta = 10/21$, respectively. 

Note that once the exponent of $Q_{s}/p_{T}$ in Eq.\,(\ref{eqn_Photon_rate44}) is determined from fitting with photon data, then using the known relation $Q_{s}^2 \sim N_{part}^{1/3}$ we shall have the photon invariant yield scaling as $N_{part}^{2/3 + \Delta/6}$, which depending on the value of $\Delta$ can be a very rapid dependence on the number of participants. 


\section{Direct photon theory-data comparisons}
\label{sec:Fitting}
In this section we fit/compare the mBU-Glasma thermal photon yield with experimental data on direct photons measured at various centralities at $\sqrt{s_{NN}} = 200$\,GeV, starting with the PHENIX run04 Au + Au data \cite{Adare:2008qk,Adare:2009qk} (from Fig.\,\ref{Photons_pt} of Appendix III). Then we continue with global fitting of the theoretical yield with the PHENIX combined run07+run10  Au + Au data \cite{Bannier:2014} as well as with the ALICE Pb+Pb data at $\sqrt{s_{NN}} = 2760$\,GeV \cite{Wilde:2013} and the PHENIX d+Au data at $\sqrt{s_{NN}} = 200$\,GeV \cite{Yamaguchi:2012}.

\subsection{Fitting/comparison with the PHENIX run04 Au+Au data}
\label{sec:Fitting1}

The saturation scale in Eq.\,(\ref{eqn_Photon_rate44}) can be parametrized as follows \cite{McPrasz:2010,Prasz:2011}:
\begin{equation}
Q_{s}^{2}[p_{T}, N_{part}, \sqrt{s}] = Q_{0}^{2}[N_{part}]\Lb \frac{10^{-3}\!\cdot\!\sqrt{s}}{p_{T}} \Rb^{\lambda}\,,
\label{eqn_momentum}
\end{equation}
where $Q_{0}$ depends on centrality, and the saturation parameter $\lambda$ characterizes the growth of $Q_{s}$ with decreasing Bjorken $\mathrm{x}$, being in the range of $\lambda \sim 0.2 \div 0.35$. 

Inserting Eq.\,(\ref{eqn_momentum}) into Eq.\,(\ref{eqn_Photon_rate44}) leads to
\bea
{d^{3}N_{\gamma} \over dy\,d^{2}p_{T}} & \equiv & Y_{\rm mBU}(p_T, N_{part}) =
\nonumber \\
 & = & C_{\gamma} N_{part}^{2/3}{\Lb 3 + 2\Delta \over 3 \Rb}\Lb \sqrt{Q_{0}^{2}[N_{part}]\Lb 10^{-3}\!\cdot\!\sqrt{s} \Rb^{\lambda}} \Rb^{\Delta} /  p_{T}^{\Delta(1+ \lambda/2)}\,.
\label{eqn_E}
\eea
where we introduce the proportionality constant $C_{\gamma}  = {\rm const}\!\cdot\!\alpha R_{0}^{2}$, which is determined by fitting in one centrality bin (for example, $0$-$20\%$), and then applying it to the other available centrality bins. 

One also needs to include the photon production in the absence of a medium, which can be properly achieved by using a $T_{AA}$-scaled\footnote{$T_{AA}$ is the Glauber nuclear overlap function in A+A collisions. See Appendix III.} p + p fit at various centralities of Au + Au collisions. That is why we use the following (modified) power-law function from \cite{Adare:2008qk,Adare:2009qk} for parametrizing this ``background'' prompt photon contribution, $Y_{\rm prompt}$, at $\sqrt{s_{NN}} = 200$\,GeV:

\begin{equation}
Y_{\rm prompt}(p_{T}, N_{part}) = T_{AA} \times \frac{A_{pp}}{(1 + p_{T}^{2}/p_{0})^{n}}\,,
\label{eqn_B}
\end{equation}
where $T_{AA}$ depends on centrality. 
This part of the contribution has been studied in \cite{Adare:2008qk,Adare:2009qk} and the parameters determined to be $A_{pp} = 0.0133264\,{\rm mb\,GeV^{-2}}$, $p_{0} = 1.5251\, \rm GeV^{2}$ and $n = 3.24692$. The general formula for the total photon yield is the sum of the two contributions: $Y_{\rm mBU} + Y_{\rm prompt}$. 

The data we wish to fit/compare are the direct photon data measured in Au + Au collisions \cite{Adare:2008qk,Adare:2009qk} (see Fig.\,\ref{Photons_pt}) in the centrality classes $0$-$20\%$, $20$-$40\%$ and $0$-$92\%$ (minimum bias) as a function of $p_{T}$ at $\sqrt{s_{NN}} = 200$\,GeV. The procedure is discussed in \cite{Chiu:2013}, and described in the following:

\begin{itemize}
\item[a)] For the key parameters $\lambda$ and $\Delta$ in Eq.\,(\ref{eqn_E}), we test a wide range within $0.2 \leqslant \lambda \leqslant 0.4$ and $3 \leqslant \Delta \leqslant 9$. Performing $\chi^{2}$/d.o.f. (chi square per degrees of freedom) fitting, using the data from all three centrality classes, we determine the best fit values for $\lambda$ and $\Delta$. For given values of $\lambda$ and $\Delta$ in Eq.\,(\ref{eqn_E}), we also simultaneously fix the coefficient $C_{\gamma}$ from fitting with the $0$-$20\%$ centrality case, and then make predictions for the other centrality classes. This strategy will provide a critical test of the geometric scaling properties of the direct photon data.
For the fit we focus on the ``enhancement region" 1\,GeV/c\,$< p_{T} <$\,4.5\,GeV/c (see Fig.\,\ref{Photons_pt}), which includes ten data points (six points in $0$-$20\%$, and four points in $20$-$40\%$ centrality class).
\item[b)] The values of $N_{part}$ and $T_{AA}$ are obtained from the PHENIX Glauber model \cite{Reygers:2005}:
\begin{itemize}
\centering
\item[] $N_{part} = 279.9$ and $T_{AA} = 18.550$\,mb$^{-1}$ in $0$-$20\%$;
\item[] $\,N_{part} = 140.4$ and $T_{AA} = 7.065$\,mb$^{-1}$ in $20$-$40\%$;
\item[] $\!N_{part} = 109.1$ and $T_{AA} = 6.140$\,mb$^{-1}$ in $0$-$92\%$.
\end{itemize}
\item[c)] We determine the values of $Q_{0}$ for each centrality class, starting from the relation $Q_{0}^{2} \propto N_{part}^{1/3}$ and the reference value $Q_{0}^{2} = 1$\,GeV$^{2}$ at $|\vec{b}| = 0$ and $\sqrt{s} = 130$\,GeV that has been used in \cite{Baier:2002}. Note that at $|\vec{b}| = 0$ we use $N_{part} = 378.4$ \cite{Kharzeev:2001}. Above three $N_{part}$ values lead to: $Q_{0}^{2}$$(0$-$20\%) \simeq 0.90\,$GeV$^{2}$, $Q_{0}^{2}$$(20$-$40\%) \simeq 0.72\,$GeV$^{2}$ and  $Q_{0}^{2}$$(0$-$92\%) \simeq 0.66\,$GeV$^{2}$.
\end{itemize}

\begin{figure}[h!]
\vspace{0.0cm}
\begin{center}
\scalebox{1.0}[1.0] {
\hspace{-0.4cm}
  \includegraphics[width=5.0in,height=3.5in]{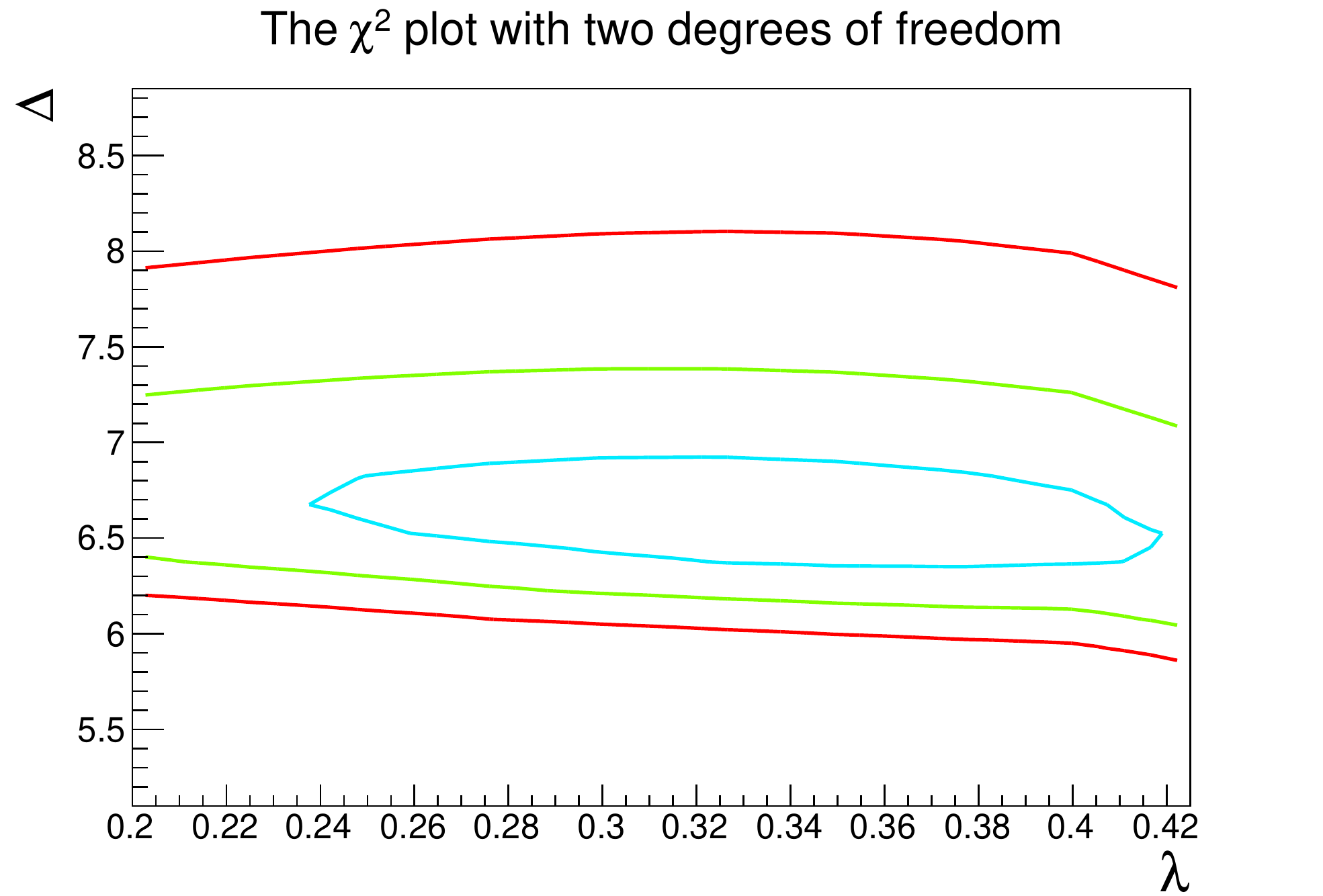} }
\end{center}
\caption{(Color online). The result of the $\chi^{2}$/d.o.f. analysis in the $\Delta$-$\lambda$ parameter space for the discussed PHENIX Au + Au direct photon data in the aforementioned three centrality classes from fitting with the mBU-Glasma thermalization model: where the blue, green, and red contours indicate $1$-, $2$-, and $3$-$\sigma$ errors, respectively.}
\label{Chi2}
\vspace{0.0cm}
\end{figure}

Fig.\,\ref{Chi2} shows the result of the $\chi^{2}$/d.o.f. analysis in the \,$\Delta$-$\lambda$\, parameter space\footnote{This figure is similar to the one in \cite{Chiu:2013}.} for the PHENIX run04 data in the three centrality classes, by plotting three contours corresponding to $1$-$\sigma$(blue), $2$-$\sigma$(green), and $3$-$\sigma$(red) errors. The formula which gives these contours has the following form:

\begin{equation}
\frac{\chi^{2}}{\mbox{d.o.f.}} = \frac{\sum_{k=1}^{10}{\left[ ({\rm data_{k}} - {\rm fit_{k}})/{\rm err_{k}} \right]^{2}}}{\mbox{d.o.f.}}\,,
\label{eqn_Chi2}
\end{equation}
where the data and error values are taken from \cite{Adare:2009qk}, and the theoretical results $\rm fit_{k}$ are calculated from $Y_{\rm mBU} + Y_{\rm prompt}$. As regards the degrees of freedom, by having the total number of the free parameters $\lambda$ and $\Delta$ equal to two, the d.o.f. are determined to be
\begin{equation}
\mbox{d.o.f.\,\,(8)} = \mbox{number\,\,of\,\,data\,\,points\,\,(10)} \,-\, \mbox{number\,\,of\,\,parameters\,\,(2)}\,.
\label{eqn_dof}
\end{equation}
Consequently, based on this analysis, one can identify the best fit with $\lambda_{fit} \simeq 0.33 \pm 0.09$ and $\Delta_{fit} = 6.63 \pm 0.28$.

\begin{figure}[ht]
\vspace{0.0cm}
\begin{center}
\scalebox{1.0}[1.0] {
\hspace{-0.4cm}
  \includegraphics[scale=0.95]{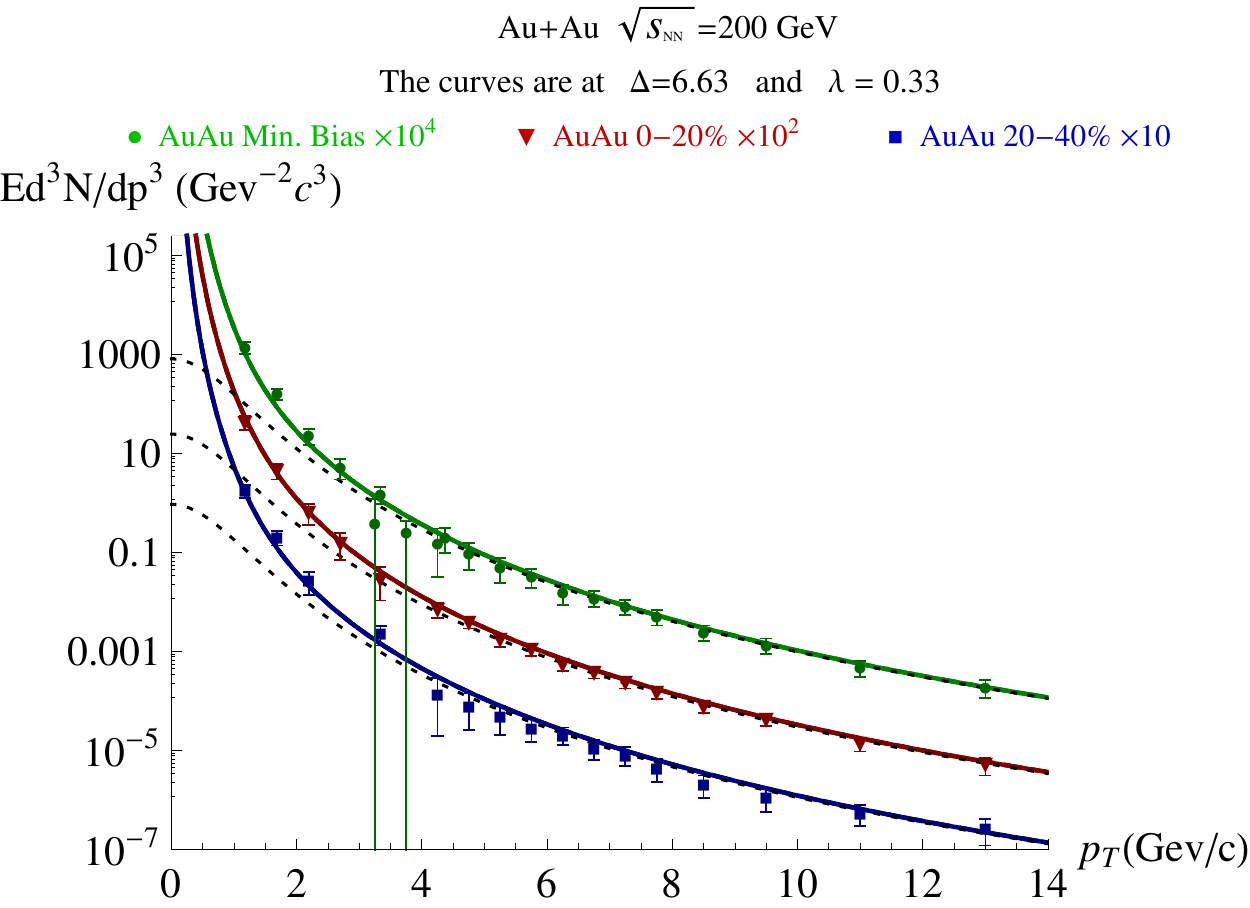} }
\end{center}
\caption{(Color online). The fit and comparison with direct photon data in three centrality bins: $0$-$20\%$, $20$-$40\%$ and $0$-$92\%$.  The data are from \cite{Adare:2008qk,Adare:2009qk}. In each centrality the dashed curve represents the $T_{AA}$-scaled p + p yield from Eq.\,(\ref{eqn_B}), and the solid curve represents the full yield including also the mBU-Glasma contribution from Eq.\,(\ref{eqn_E}).}
\label{fit}
\vspace{0.2cm}
\end{figure}

Having identified the optimal parameters, we present the fit of our model to the data in Fig.\,\ref{fit}. It is done using the central values of the two parameters at $1$-$\sigma$, which will be $\overline{\lambda_{fit}} = 0.33$ and $\overline{\Delta_{fit}} = 6.63$ (correspondingly $\overline{\delta_{fit}} = 0.372$ from $\Delta=(15 + 9\delta)/(5 - 6\delta)$ of Eq.\,(\ref{eqn_Photon_rate4})). In Fig.\,\ref{fit} we use the notations $\overline{\Delta_{fit}} \equiv \Delta$ and $\overline{\lambda_{fit}} \equiv \lambda$. We also compare the PHENIX data to the contribution from the $T_{AA}$-scaled p + p yield described by Eq.\,(\ref{eqn_B}).

\subsection{Global fitting/comparison with the PHENIX combined run07+run10 Au+Au data, and with the ALICE Pb+Pb and PHENIX d+Au data}
\label{sec:Fitting2}

Before turning to the data comparison we begin this section by discussing additional considerations for asymmetric systems. In Ref.\,\cite{BosMc:2014} it is assumed that the saturation momentum for asymmetric d+Au collisions is:
\begin{equation}
Q_{s}^{2} = \sqrt{Q_{s,d}^{2}\,Q_{s,Au}^{2}}\,,
\label{eqn_asym}
\end{equation}
The squared saturation scale for colliding nuclei changes with $\sim A^{1/3}$, nevertheless, in the case of asymmetric nuclei this scale is reduced since only the size of the nucleon along the boost direction is relevant. The scaling factor between symmetric Au+Au (at 200\,GeV) and asymmetric d+Au (at some energy) collisions can be estimated if one considers the saturation scales of the individual partners of the d+Au system. Making use of the known relation from saturation physics
\begin{equation}
Q_{0}^{2}[N_{part}] \sim N_{part}^{1/3}\,,
\label{eqn_Q0_Npart}
\end{equation}
one can write down the excess photon yield for an $A_{1}A_{2}$ asymmetric system at some collision energy:
\begin{eqnarray}
& & \!\!\!\!\!\!\!\!\!\!\!\!
\frac{d^{3}N_{\gamma}}{dy d^{2}p_{T}}|_{A_{1}A_{2}} \sim \frac{d^{3}N_{\gamma}}{dy d^{2}p_{T}}|_{AuAu} \times
\nonumber \\
& \times &
\Lb \frac{\Lb 4 N_{part,A_{1}}\!\cdot\!N_{part,A_{2}} \Rb^{\Delta/12} \Lb 2 N_{part,A_{1}} \Rb^{2/3}}
{N_{part,AuAu}^{(\Delta/6) + (2/3)}}\frac{\sqrt{s}^{\,(\lambda\Delta)/2}}{200^{\,(\lambda\Delta)/2}} \Rb\,
\label{eqn_asym_factor2}
\end{eqnarray}
where for a colliding symmetric system, A$_{1}$ = A$_{2}$ = A, we have $N_{part,A_{1}} = N_{part,A_{2}} = N_{part}/2$. 

The functional form of the factor appearing in  Eq.\,(\ref{eqn_asym_factor2}) can be understood as follows. The geometry of the collision in the transverse plane and hence the cross section, is dictated by the smaller nucleus A$_1$ (e.g., a deuteron). The expression $(2N_{part,A_1})^{2/3}$ thus represents the area of the interaction region, while the other expression $(4N_{part,A_1}\!\cdot\!N_{part,A_2})^{\Delta/12}/N_{part,AuAu}^{(\Delta/6)+(2/3)}$ takes care of the change in $Q_s$. In the limit A$_1$ = A$_2$ and at 200\,GeV collision energy, the whole factor in the parentheses in Eq.\,(\ref{eqn_asym_factor2}) becomes one.

For A$_{1}$ = d and A$_{2}$ = Au in minimum bias collisions at 200\,GeV, Eq.\,(\ref{eqn_asym_factor2}) reduces to a form, which is somewhat different (the numerator in the parentheses) than the following formula  obtained in \cite{BosMc:2014}:
\begin{equation}
\frac{d^{3}N_{\gamma}}{dy d^{2}p_{T}}|_{dAu} = \frac{d^{3}N_{\gamma}}{dy d^{2}p_{T}}|_{AuAu} \times \Lb \frac{3.2^{2/3}\,1.6^{\Delta^{\prime}/12}\,197^{\Delta^{\prime}/12}}{N_{part,AuAu}^{(\Delta^{\prime}/6) + (2/3)}} \Rb\,,
\label{eqn_asym_factor1}
\end{equation}
where the average number of participants from the deuteron is $\langle N_{part}[d]\rangle = 1.62 \pm 0.01$, which gives the total d+Au overlap area, $\pi R_{d}^{2}$, to be proportional to $3.24^{2/3}$.

Ultimately, based on  Eq.\,(\ref{eqn_asym_factor2}) we can utilize a generalized form of Eq.\,(\ref{eqn_E}) for A$_{1}$+A$_{2}$ collisions (for A$_{1}$$\,\leq\,$A$_{2}$), where the mBU-Glasma thermal excess photon invariant yield is given by this phenomenological formula:
\begin{eqnarray}
\frac{d^{3}N_{\gamma}}{dy d^{2}p_{T}}|_{A_{1}A_{2}} & \equiv & Y_{\rm mBU}(p_T, N_{part})|_{A_{1}A_{2}} =
\nonumber \\
& = & F_{\gamma} \Lb {3 + 2\Delta \over 3} \Rb \Lb 2 N_{part}[A_{1}] \Rb^{2/3} \times
\nonumber \\
& \times & 
\Lb 4 N_{part}[A_{1}]\,N_{part}[A_{2}] \Rb^{\Delta/12}\,
{\Lb 10^{-3}\!\cdot\!\sqrt{s} \Rb^{\lambda_{eff}[p_{T}] \cdot \Delta/2}\over p_{T}^{\,\Delta\,(1 + \lambda_{eff}[p_{T}]/2) }}\,,
\label{eqn_scaling_law}
\end{eqnarray}
with the constant $F_{\gamma}$ (different from $C_{\gamma}$) having a dimension of GeV$^{-2+\Delta}$. In this final form for the yield we assume that the saturation parameter $\lambda$ can be substituted by the $p_{T}$-dependent ``running" $\lambda_{eff}[p_{T}]$:
\begin{equation}
\lambda_{eff}[p_{T}] = 0.13 + 0.1 \Lb \frac{(2p_{T})^{2}}{10} \Rb^{0.35}\,.
\label{eqn_lambda_pT}
\end{equation}
This reasoning and Eq.\,(\ref{eqn_lambda_pT}) come from Ref.\,\cite{Prasz:2011}, where it is shown that the quality of geometric scaling in hadronic collisions improves if the exponent $\lambda$ becomes $p_{T}$-dependent. 

\begin{table}[h]
\caption{Some symmetric and asymmetric collision systems for which there are measured direct photon data. $N_{part}$ is calculated from the TGlauberMC model \cite{TGlauberMC:2008}. The values of $N_{part}$ in different centrality classes in Au+Au at $\sqrt{s_{NN}} = 200$\,GeV are only slightly different from those calculated in the PHENIX Glauber model.}
\begin{center}
\begin{tabular}{|c|c|c|c|}
\hline
$\sqrt{s_{NN}}$ & System & $ N_{part}$ & Experiment \\
\hline\hline\hline
200\,GeV & Au+Au (0-20\%) & 280.200 & PHENIX (RHIC) \cite{Bannier:2014} \\
\hline
200\,GeV & Au+Au (20-40\%) & 140.600 & PHENIX (RHIC) \cite{Bannier:2014} \\
\hline
200\,GeV & Au+Au (40-60\%) & 61.180 & PHENIX (RHIC) \cite{Bannier:2014} \\
\hline
200\,GeV & Au+Au (60-92\%) & 14.886 & PHENIX (RHIC) \cite{Bannier:2014} \\
\hline\hline

2760\,GeV & Pb+Pb (0-20\%) & 154.188 & ALICE (LHC) \cite{Wilde:2013} \\
\hline
2760\,GeV & Pb+Pb (20-40\%) & 78.825 & ALICE (LHC) \cite{Wilde:2013} \\
\hline
2760\,GeV & Pb+Pb (40-80\%) & 22.950 & ALICE (LHC) \cite{Wilde:2013} \\
\hline\hline

100.7+100.0\,GeV & d+Au (0-100\%) & 8.647 & PHENIX (RHIC) \cite{Yamaguchi:2012} \\
         &                              &  $(N_{part}[d] = 1.658)$  &                            \\
         &                              & $(N_{part}[Au] = 6.989)$ &                            \\
\hline
\end{tabular}
\label{tab:Table1}
\end{center}
\end{table}

\begin{table}[h!]
\caption{Some other symmetric and asymmetric collision systems, for which there may be measured direct photon data in the coming future. $T_{A_{1}A_{2}}$ and $N_{part}$ are calculated from the TGlauberMC model \cite{TGlauberMC:2008}. The designation in the table is that $N_{part} \equiv N_{part}[A_{1}+A_{2}]$.}
\begin{center}
\begin{tabular}{|c|c|c|c|c|}
\hline
System & $T_{A_{1}A_{2}}$\,(mb$^{-1}$) & $N_{part}$ & $N_{part}[A_{1}]$ & $N_{part}[A_{2}]$ \\
\hline\hline\hline
U+U (0-20\%) & 23.628 & 334.30 & 167.150 & 167.150 \\
\hline
U+U (20-40\%) & 8.958 & 166.90 & 83.450 & 83.450 \\
\hline
U+U (40-60\%) & 2.784 & 71.50 & 35.750 & 35.750 \\
\hline
U+U (60-90\%) & 0.501 & 18.73 & 9.365 & 9.365 \\
\hline\hline

Pb+Pb (0-20\%) & 18.891 & 310.30 & 155.15 & 155.15 \\
\hline
Pb+Pb (20-40\%) & 6.828 & 158.90 & 79.45 & 79.45 \\
\hline
Pb+Pb (40-60\%) & 1.996 & 70.02 & 35.01 & 35.01 \\
\hline
Pb+Pb (60-80\%) & 0.410 & 22.78 & 11.39 & 11.39 \\
\hline\hline

Cu+Au (0-20\%) & 7.885 & 153.590 & 55.000 & 98.590 \\
\hline
Cu+Au (20-40\%) & 3.150 & 79.250 & 32.700 & 46.550 \\
\hline
Cu+Au (40-60\%) & 1.037 & 35.270 & 15.260 & 20.010 \\
\hline
Cu+Au (60-93\%) & 0.208 & 9.672 & 4.335 & 5.337 \\
\hline\hline

${}^{3}$He+Au (0-20\%) & 0.553 & 21.203 & 2.993 & 18.210 \\
\hline
${}^{3}$He+Au (20-40\%) & 0.382 & 15.379 & 2.909 & 12.470 \\
\hline
${}^{3}$He+Au (40-60\%) & 0.210 & 9.456 & 2.516 & 6.940 \\
\hline
${}^{3}$He+Au (60-88\%) & 0.085 & 4.673 & 1.668 & 3.005 \\
\hline
\end{tabular}
\label{tab:Table2}
\end{center}
\end{table}

The results of the global fitting with the PHENIX combined run07+run10 Au+Au data set \cite{Bannier:2014}, along with the ALICE Pb+Pb and PHENIX d+Au data sets from \cite{Wilde:2013} and \cite{Yamaguchi:2012}, are shown in Fig.\,\ref{fig:fig_AuAu2_200GeV} and Fig.\,\ref{fig:fig_PbPb2_2760GeV}. Table\,\ref{tab:Table1} shows the centrality classes of the collision systems under consideration. $N_{part}$ of each centrality class is calculated from the updated version of the TGlauberMC model \cite{TGlauberMC:2008}\footnote{TGlauberMC is a ROOT-based implementation of the PHOBOS Glauber Monte Carlo.}. The Glauber nuclear overlap function used to scale the p+p yield\footnote{We use slightly different values of the parameters $A_{pp}$, $p_{0}$ and $n$ determined in \cite{Bannier:2014}: $A_{pp} = (8.3 \pm 7.5)\!\cdot\!10^{-3}$\,mb\,GeV$^{-2}$, $p_{0} = 2.26 \pm 0.78$\,GeV$^{2}$, and $n = 3.45 \pm 0.08$. The systematic uncertainties of these quantities are highly correlated but for fitting we use only the central values.} in Fig.\,\ref{fig:fig_AuAu2_200GeV} is also calculated using the TGlauberMC model:  $T_{AuAu}$ = 18.651\,mb$^{-1}$ in $0$-$20\%$, $T_{AuAu}$ = 7.073\,mb$^{-1}$ in $20$-$40\%$, $T_{AuAu}$ = 2.209\,mb$^{-1}$ in $40$-$60\%$ and $T_{AuAu}$ = 0.360\,mb$^{-1}$ in $60$-$92\%$. The pQCD yield in Fig.\,\ref{fig:fig_PbPb2_2760GeV}, obtained from extrapolation down to $p_{T} = 1$\,GeV/c from Ref.\,\cite{Paquet:2017}, is scaled by $N_{coll}$ from \cite{Wilde:2013}: $N_{coll}$ = 1210.8 in $0$-$20\%$, $N_{coll}$ = 438.4 in $20$-$40\%$ and $N_{coll}$ = 77.2 in $40$-$80\%$.

For each centrality class in Fig.\,\ref{fig:fig_AuAu2_200GeV} and Fig.\,\ref{fig:fig_PbPb2_2760GeV} the dotted curve represents the $T_{AA}(N_{coll})$-scaled prompt photon yield, and the solid curve the total yield including also the mBU-Glasma contribution (dash-dotted line) from Eq.\,(\ref{eqn_scaling_law}). The number in the top left corner of each sub-figure of both Fig.\,\ref{fig:fig_AuAu2_200GeV} and Fig.\,\ref{fig:fig_PbPb2_2760GeV} is the year of the corresponding data publication.

From the global fitting the constant $F_{\gamma}$ (introduced in Eq.\,(\ref{eqn_scaling_law})) and the parameter $\delta$ are determined to be $F_{\gamma} = (5.19 \cdot 10^{-5}\,\pm\,3.99 \cdot 10^{-6})$\,GeV$^{-2+\Delta}$, $\delta = 0.287 \pm 0.009$ and $\chi^{2}/\mbox{d.o.f.} \approx 0.75$. The relation between the mBU-Glasma $\delta$ and the thermalizing Glasma $\delta^{\prime}$ is given by Eq.\,(\ref{eqn_delta_deltaprime}) and Eq.\,(\ref{eqn_deltaprime_delta}) in Appendix I. We note that $\Delta = (15 + 9\delta)/(5 - 6\delta) = 5.36$ is smaller from that found in the previous fit and used in Fig.\,\ref{fit}. 

In Appendix II we discuss how to estimate the thermalization time and thermalization temperature as a function of $\delta$ at RHIC $\sqrt{s_{NN}} = 200$\,GeV.

\section{Predictions for other systems}
\label{sec:Predictions} 

\begin{figure}[h!]
\vspace{-1.0mm}
\begin{center}
   \subfigure[]{\includegraphics[width=0.475\textwidth,height=0.45\textwidth]{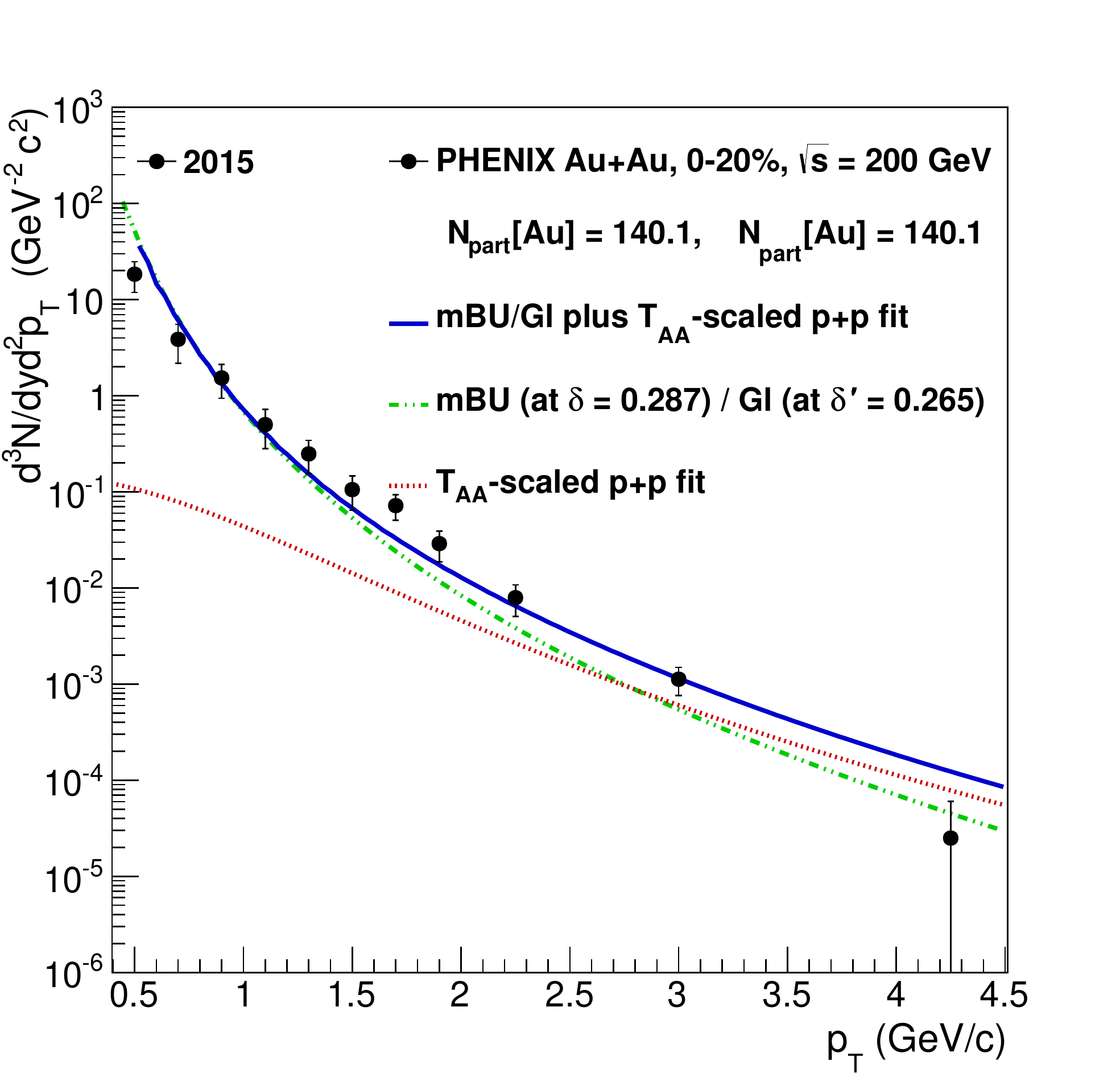} \label{fig:fig_AuAu2_200GeV_020}}
\hspace{-0.025\textwidth}
\vspace{-0.025\textwidth}
   \subfigure[]{\includegraphics[width=0.475\textwidth,height=0.45\textwidth]{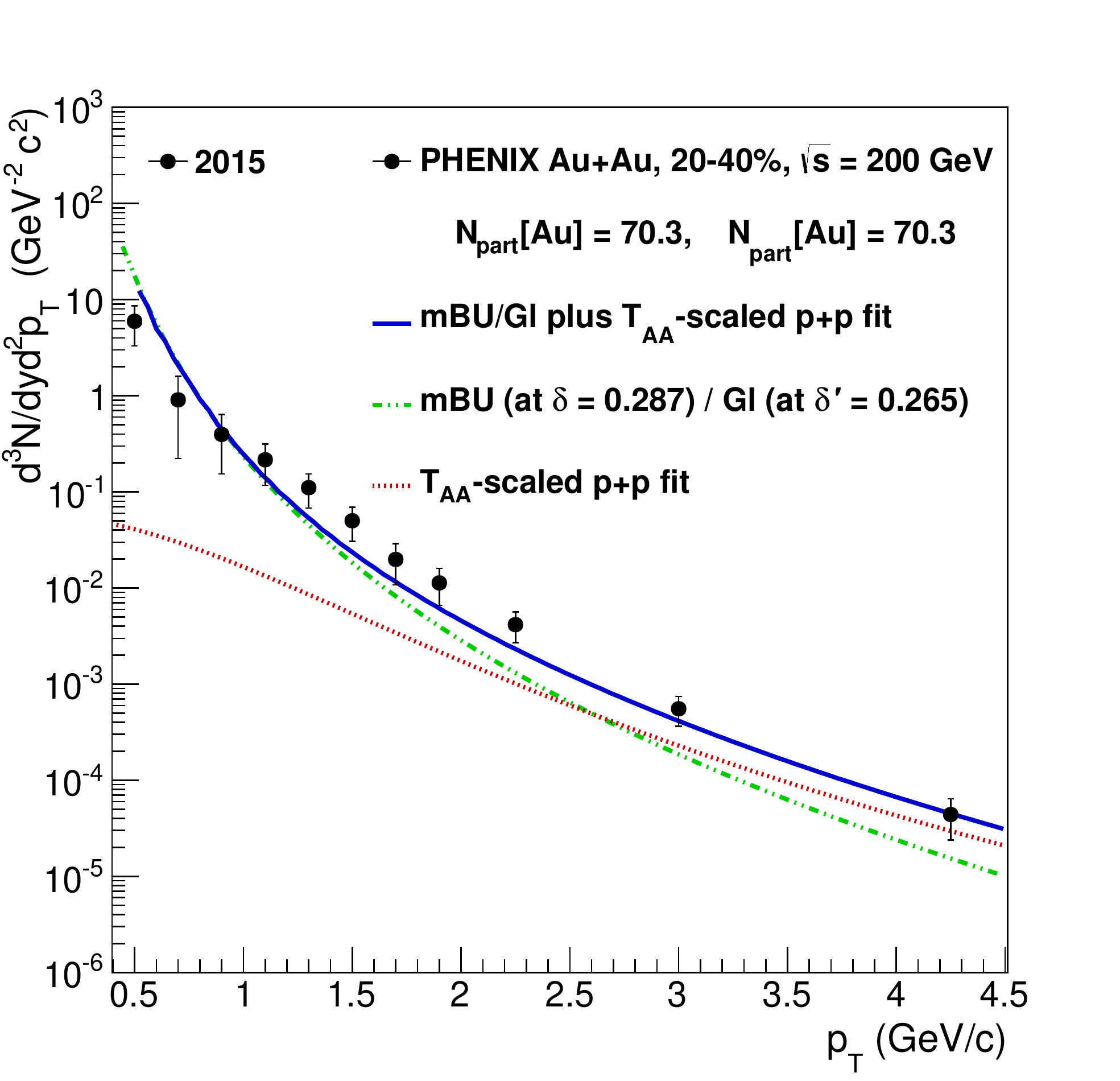} \label{fig:fig_AuAu2_200GeV_2040}}
\\
   \subfigure[]{\includegraphics[width=0.475\textwidth,height=0.45\textwidth]{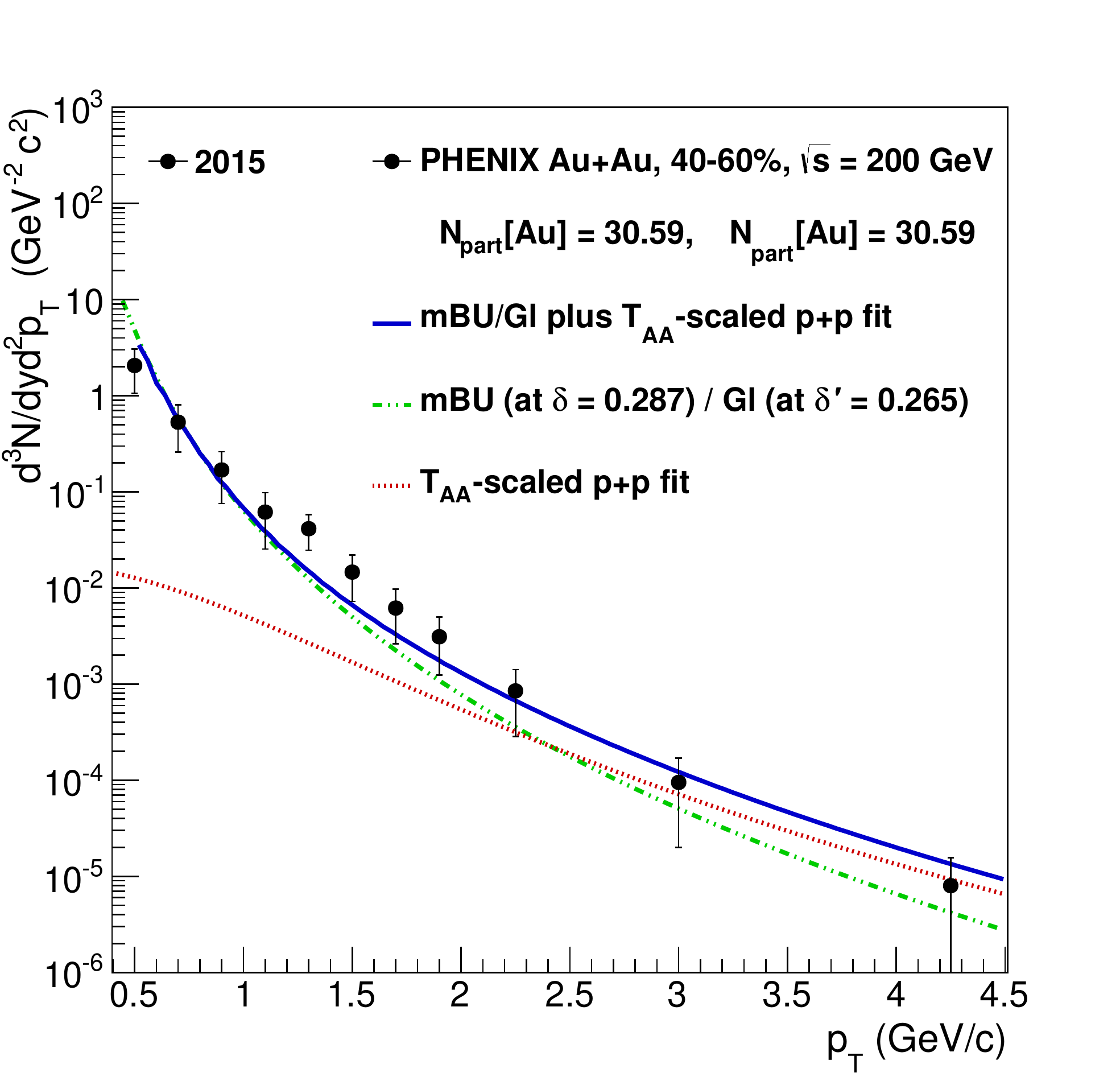} \label{fig:fig_AuAu2_200GeV_4060}}
\hspace{-0.025\textwidth}
\vspace{-0.025\textwidth}
   \subfigure[]{\includegraphics[width=0.475\textwidth,height=0.45\textwidth]{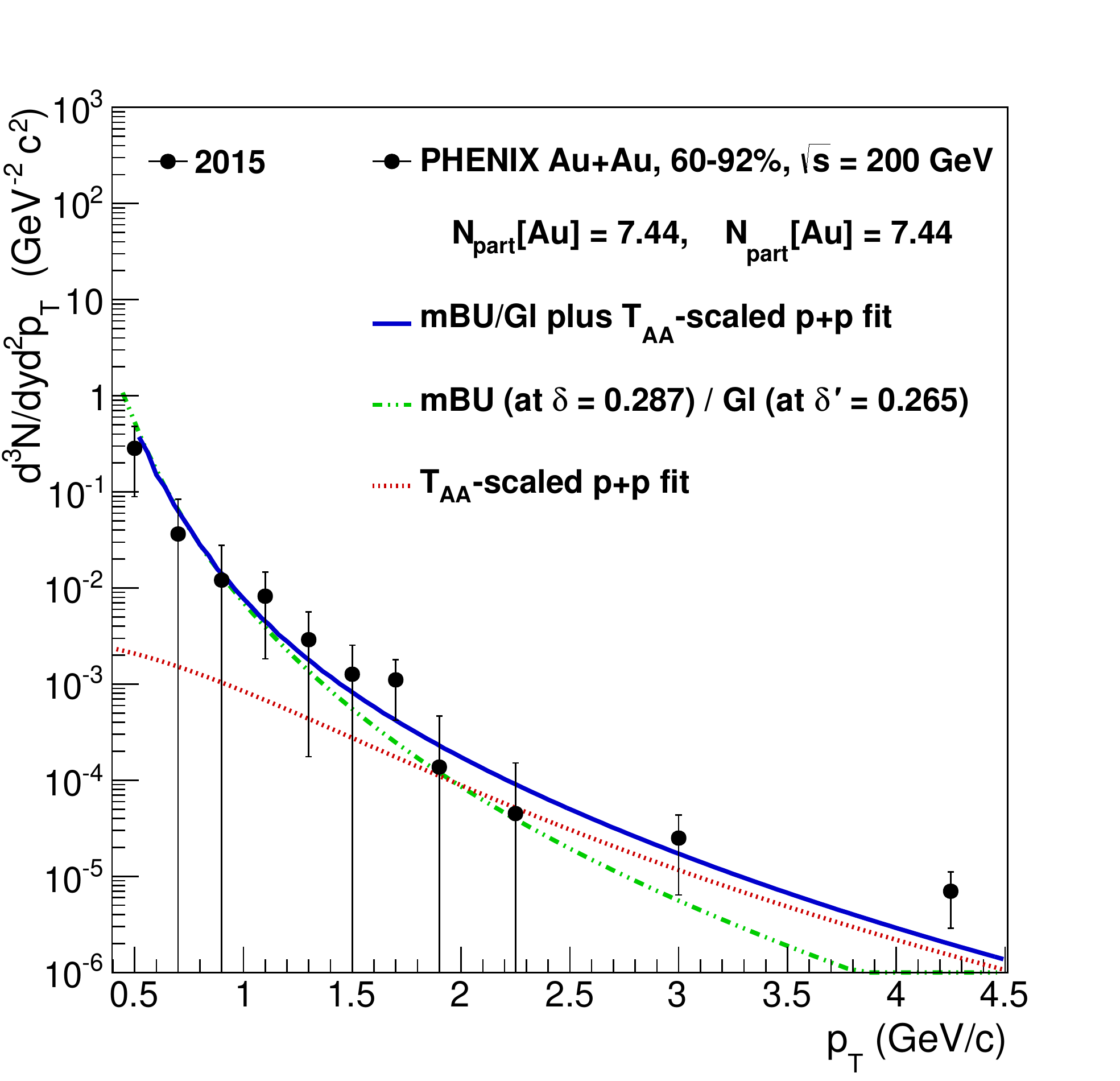} \label{fig:fig_AuAu2_200GeV_6092}}
\end{center}
\vspace{-1.0mm}
\caption{(Color online) The direct photon data comparisons for Au+Au at $\sqrt{s_{NN}} = 200$\,GeV (see Table\,\ref{tab:Table1}) obtained from the global fitting of the direct photon data from \cite{Bannier:2014}, \cite{Wilde:2013}, \cite{Yamaguchi:2012} with the mBU-Glasma model in the four centrality bins shown in Fig.\,\ref{fig:fig_AuAu2_200GeV_020}, Fig.\,\ref{fig:fig_AuAu2_200GeV_2040}, Fig.\,\ref{fig:fig_AuAu2_200GeV_4060} and Fig.\,\ref{fig:fig_AuAu2_200GeV_6092}. In each centrality selection the dotted curve represents the $T_{AA}$-scaled p+p yield from Eq.\,(\ref{eqn_B}) (with the parameters given in Sec.\,\ref{sec:Fitting2}), and the solid curve represents the full yield including also the mBU-Glasma contribution (dash-dotted line) from Eq.\,(\ref{eqn_scaling_law}).}
\label{fig:fig_AuAu2_200GeV}
\end{figure}

\begin{figure}[h!]
\vspace{-1.0mm}
\begin{center}
   \subfigure[]{\includegraphics[width=0.475\textwidth,height=0.45\textwidth]{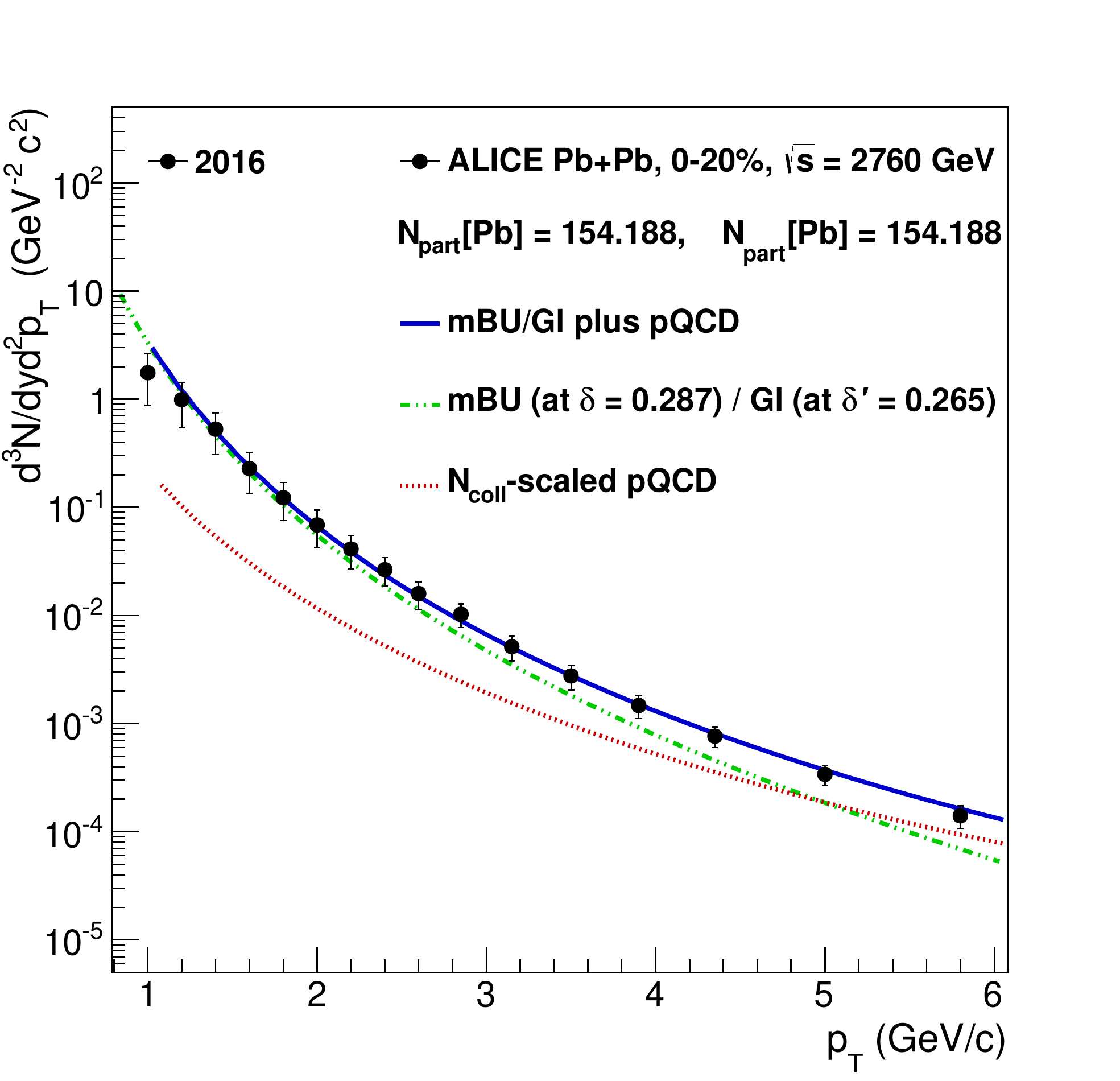} \label{fig:fig_PbPb2_2760GeV_020}}
\hspace{-0.025\textwidth}
\vspace{-0.025\textwidth}
   \subfigure[]{\includegraphics[width=0.475\textwidth,height=0.45\textwidth]{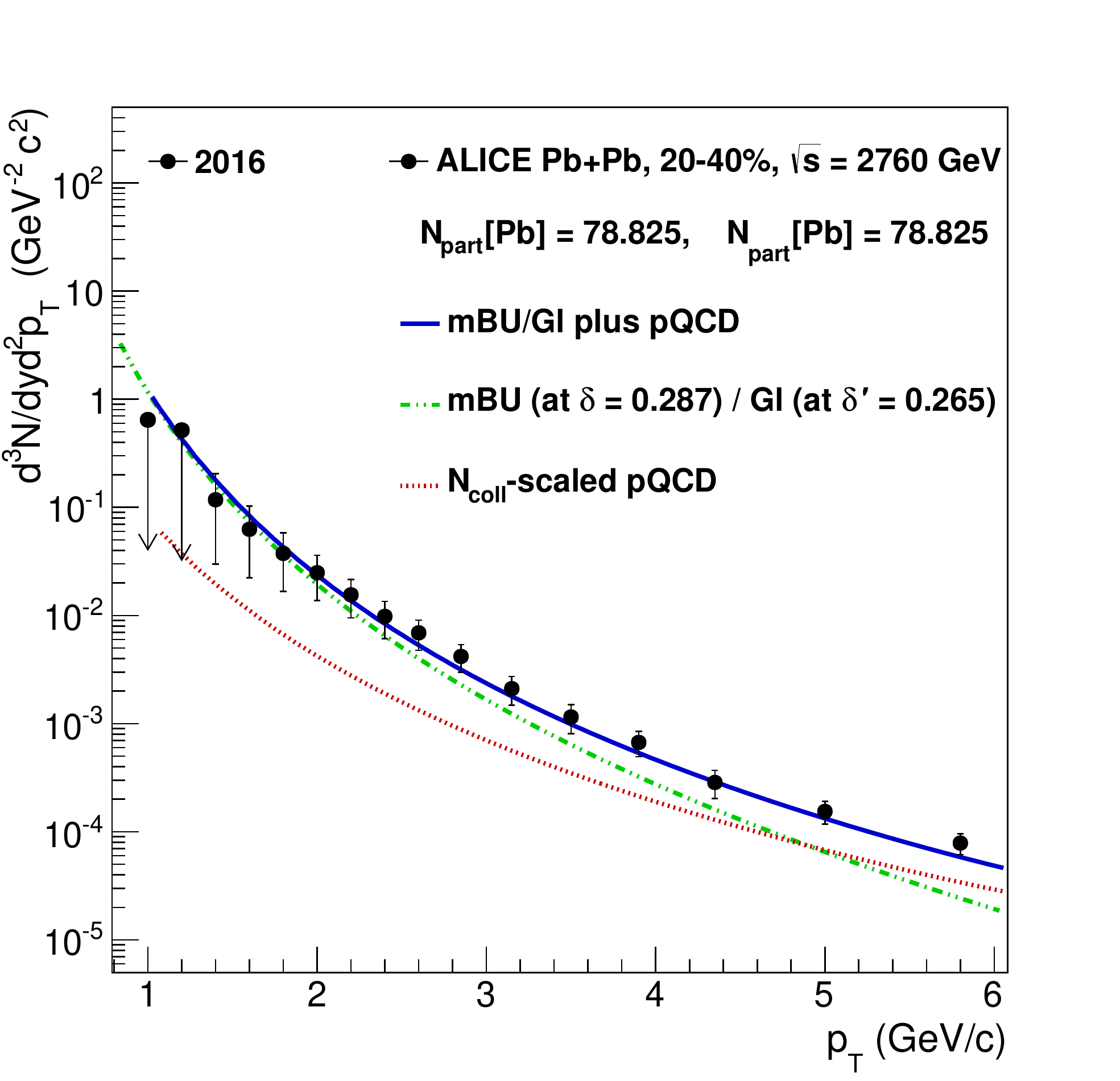} \label{fig:fig_PbPb2_2760GeV_2040}}
\\
   \subfigure[]{\includegraphics[width=0.475\textwidth,height=0.45\textwidth]{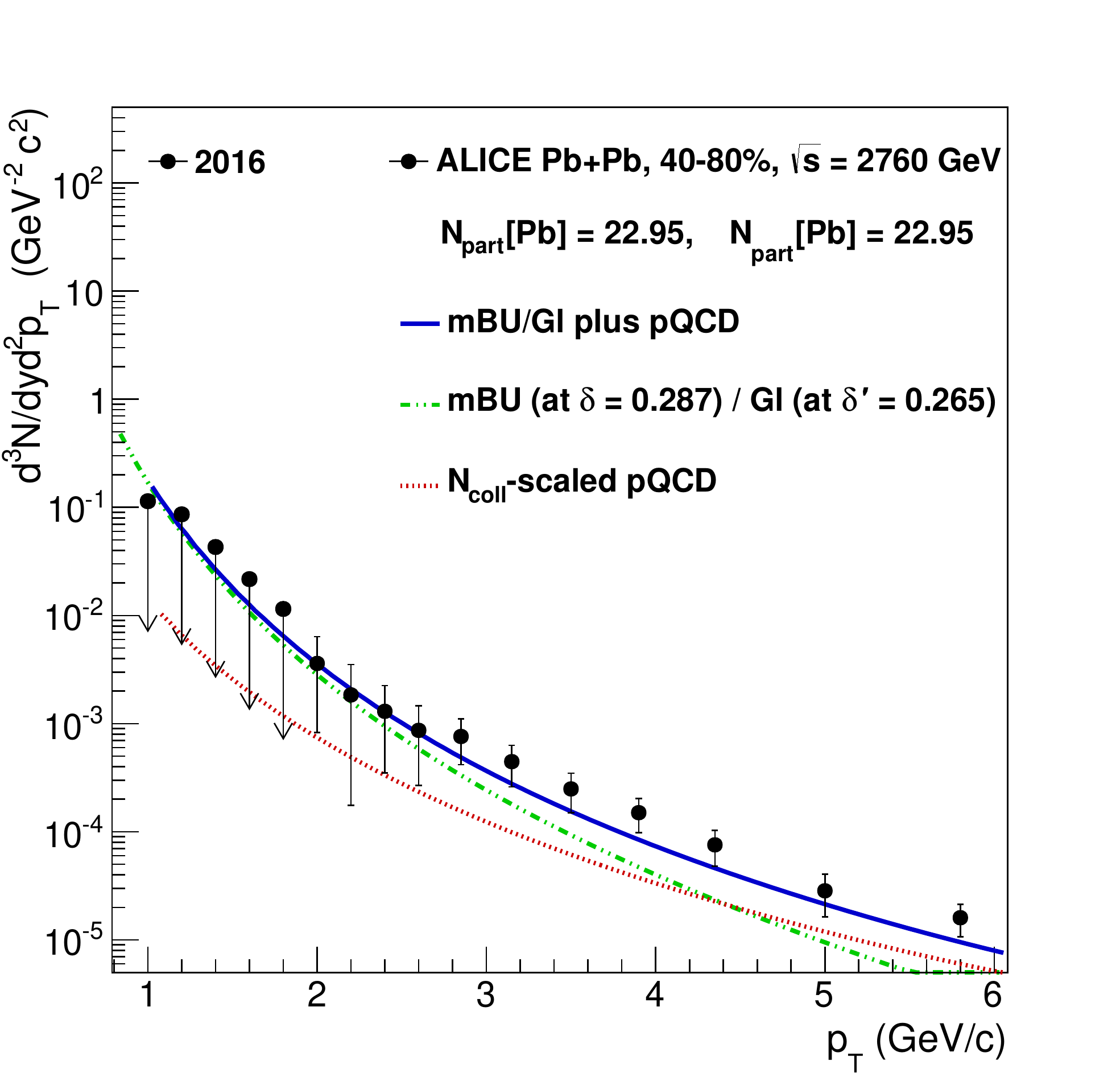} \label{fig:fig_PbPb2_2760GeV_4080}}
\hspace{-0.025\textwidth}
\vspace{-0.025\textwidth}
   \subfigure[]{\includegraphics[width=0.475\textwidth,height=0.45\textwidth]{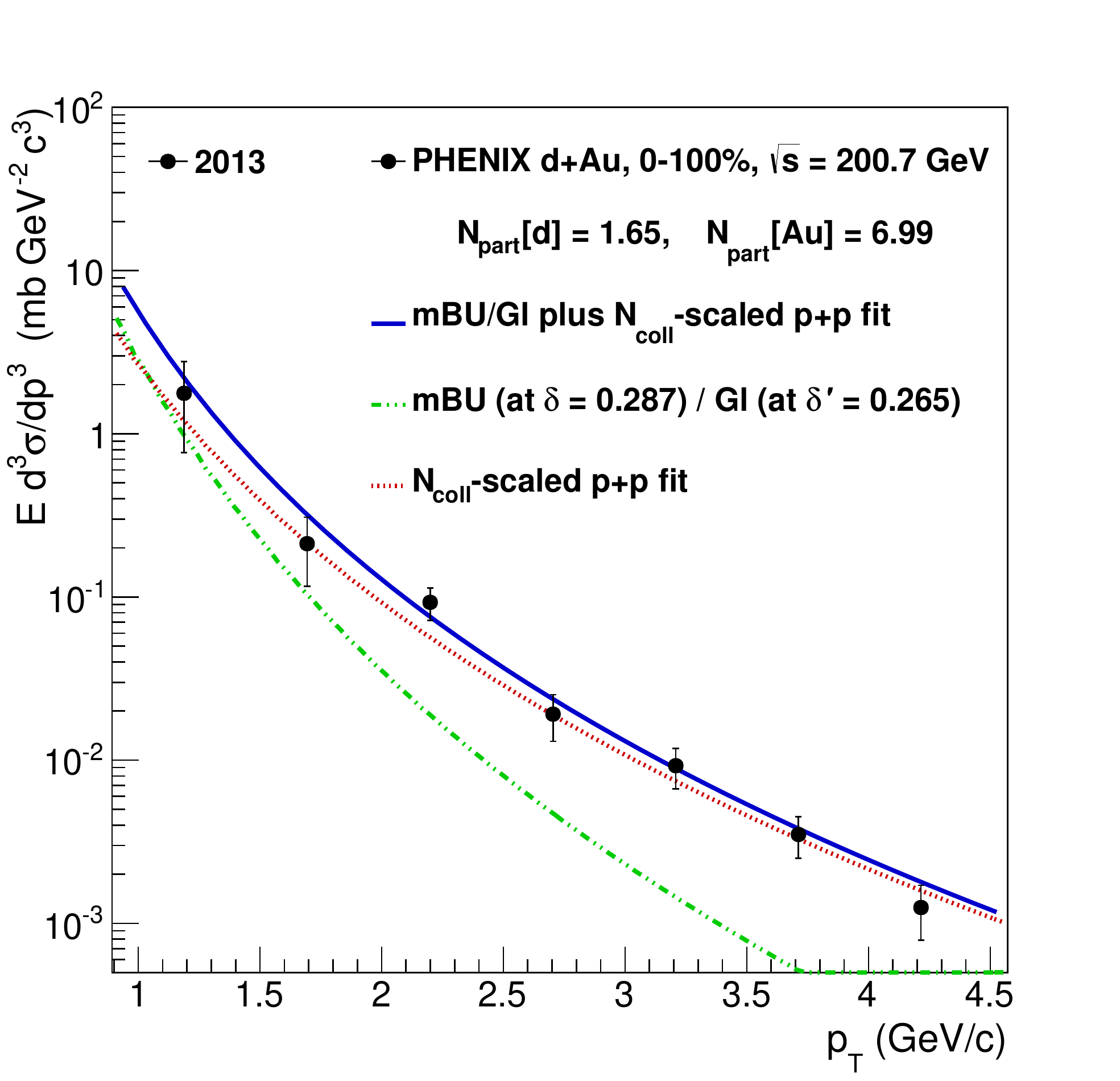} \label{fig:fig_dAu2_200GeV}}
\end{center}
\vspace{-1.0mm}
\caption{(Color online) The direct photon data comparisons for Pb+Pb at $\sqrt{s_{NN}} = 2760$\,GeV and d+Au at $\sqrt{s_{NN}} = 200.7$\,GeV (see Table\,\ref{tab:Table1}) obtained from the global fitting of the direct photon data from \cite{Bannier:2014}, \cite{Wilde:2013}, \cite{Yamaguchi:2012} with the mBU-Glasma model shown in Fig.\,\ref{fig:fig_PbPb2_2760GeV_020} (for Pb+Pb 0-20\%), Fig.\,\ref{fig:fig_PbPb2_2760GeV_2040} (for Pb+Pb 20-40\%), Fig.\,\ref{fig:fig_PbPb2_2760GeV_4080} (for Pb+Pb 40-80\%) and Fig.\,\ref{fig:fig_dAu2_200GeV} (for d+Au 0-100\%). The pQCD yield is from \cite{Paquet:2017}. The nomenclature is the same as in Fig.\,\ref{fig:fig_AuAu2_200GeV}.}
\label{fig:fig_PbPb2_2760GeV}
\end{figure}

\begin{figure}[h!]
\vspace{-1.0mm}
\begin{center}
   \subfigure[]{\includegraphics[width=0.475\textwidth,height=0.45\textwidth]{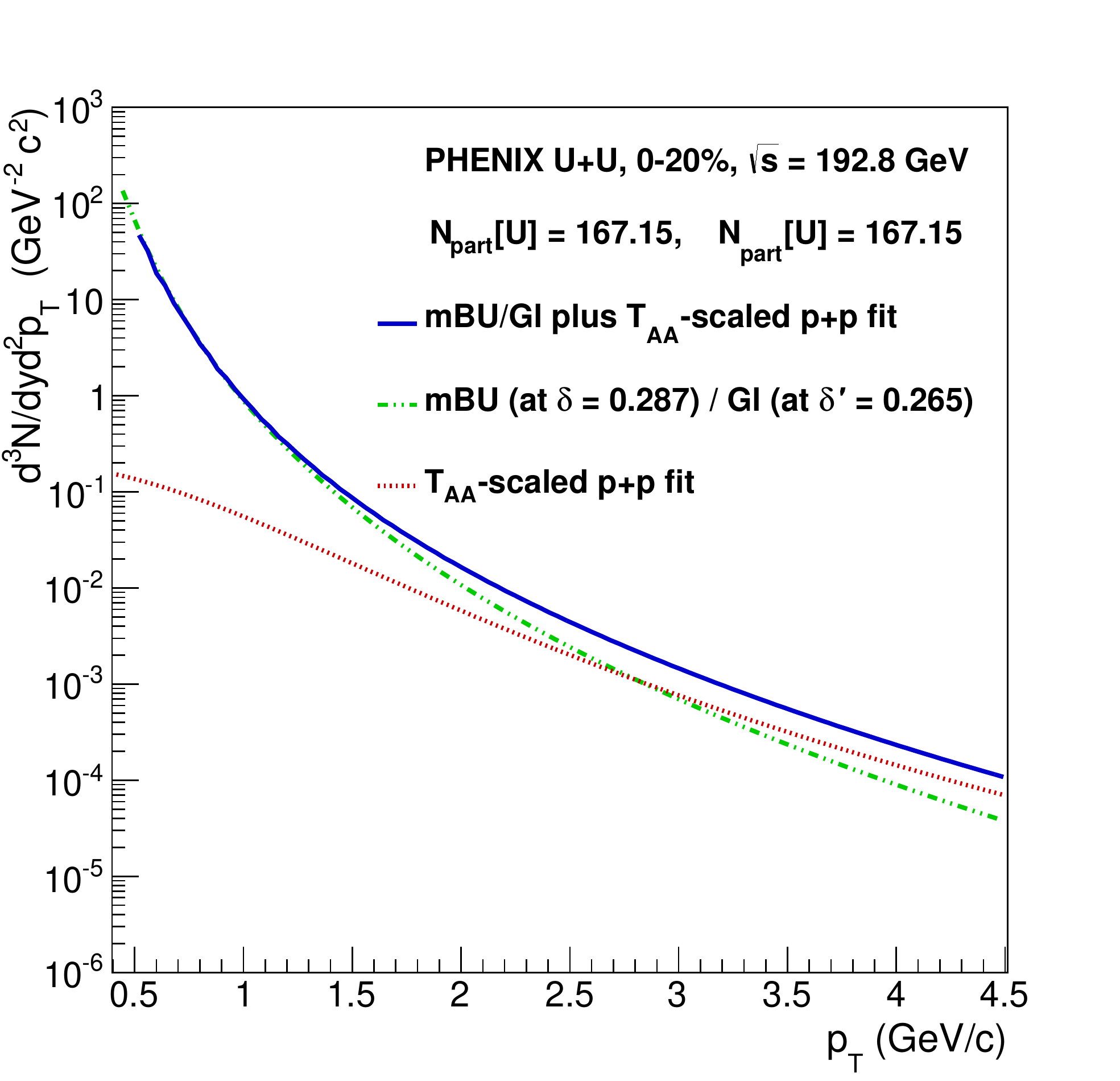} \label{fig:fig_UU_200GeV_020}}
\hspace{-0.025\textwidth}
\vspace{-0.025\textwidth}
   \subfigure[]{\includegraphics[width=0.475\textwidth,height=0.45\textwidth]{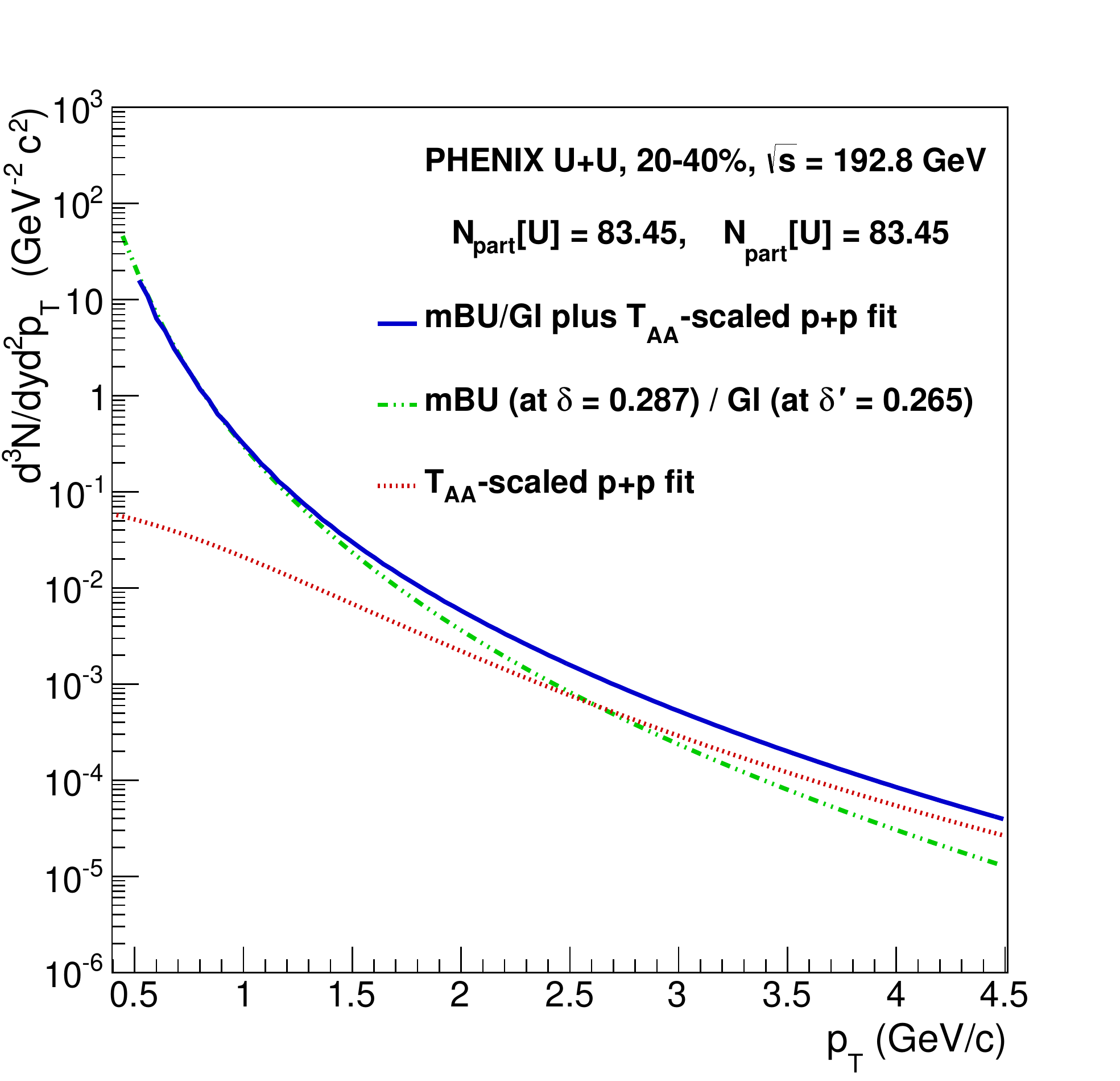} \label{fig:fig_UU_200GeV_2040}}
\\
   \subfigure[]{\includegraphics[width=0.475\textwidth,height=0.45\textwidth]{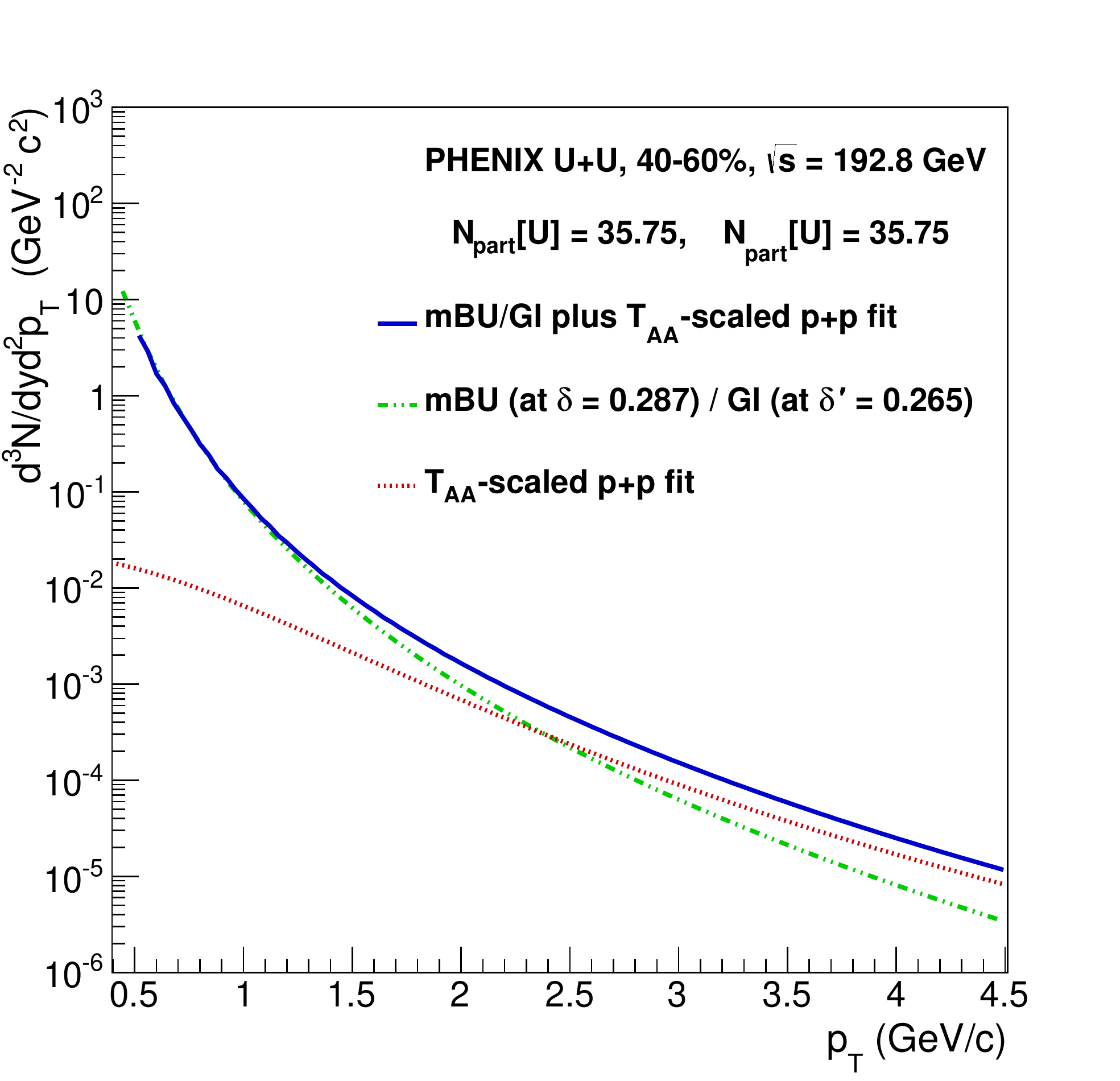} \label{fig:fig_UU_200GeV_4060}}
\hspace{-0.025\textwidth}
\vspace{-0.025\textwidth}
   \subfigure[]{\includegraphics[width=0.475\textwidth,height=0.45\textwidth]{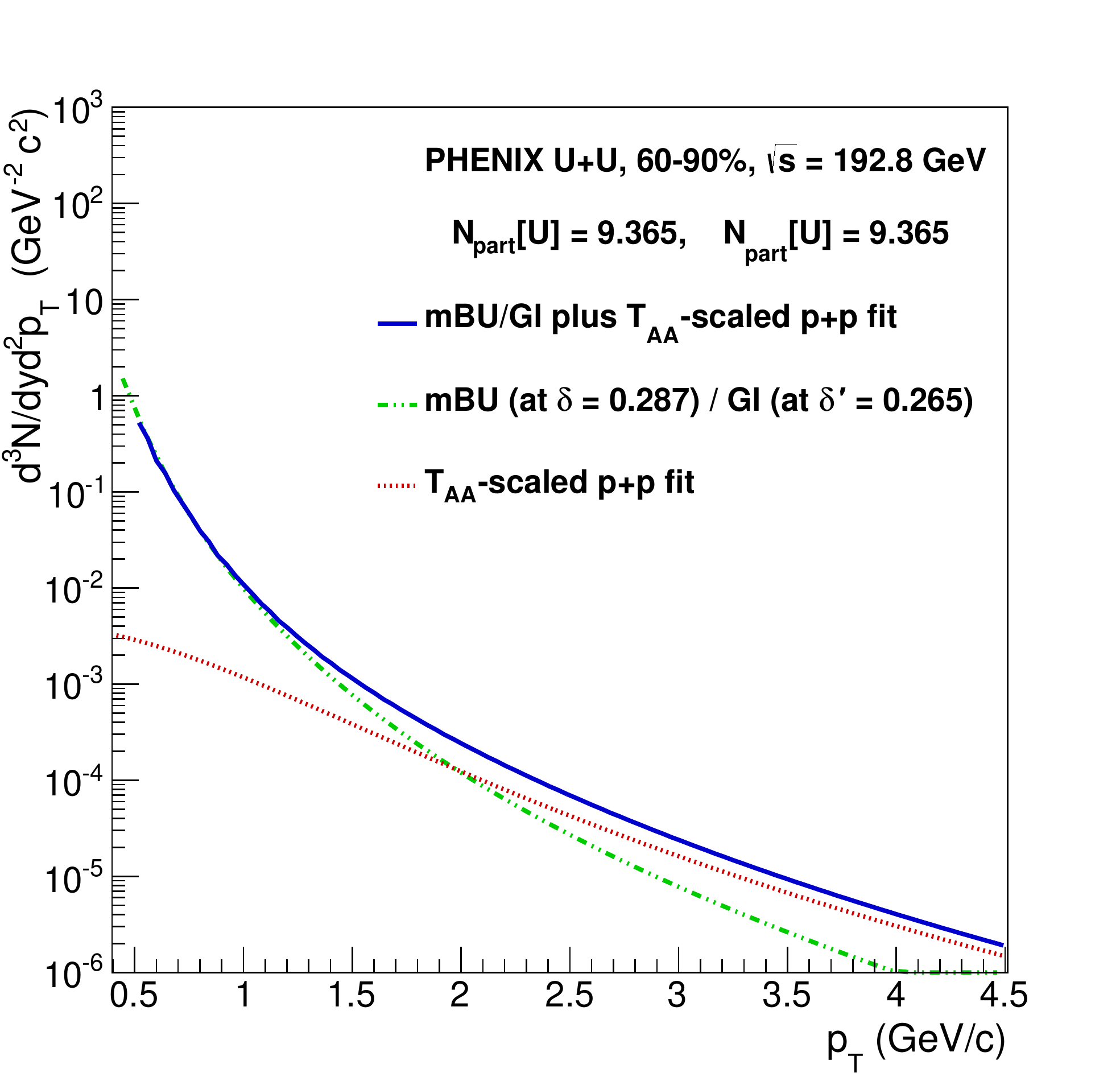} \label{fig:fig_UU_200GeV_6090}}
\end{center}
\vspace{-1.0mm}
\caption{(Color online) The prediction for the direct photon invariant yield for U+U at $\sqrt{s_{NN}} = 192.8$\,GeV (see Table\,\ref{tab:Table2}) obtained from the global fitting of the direct photon data from \cite{Bannier:2014}, \cite{Wilde:2013}, \cite{Yamaguchi:2012} with the mBU-Glasma model in the four centrality bins shown in Fig.\,\ref{fig:fig_UU_200GeV_020}, Fig.\,\ref{fig:fig_UU_200GeV_2040}, Fig.\,\ref{fig:fig_UU_200GeV_4060} and Fig.\,\ref{fig:fig_UU_200GeV_6090}. The nomenclature is the same as in Fig.\,\ref{fig:fig_AuAu2_200GeV}.}
\label{fig:fig_UU_200GeV}
\end{figure}

\begin{figure}[h!]
\vspace{-1.0mm}
\begin{center}
   \subfigure[]{\includegraphics[width=0.475\textwidth,height=0.475\textwidth]{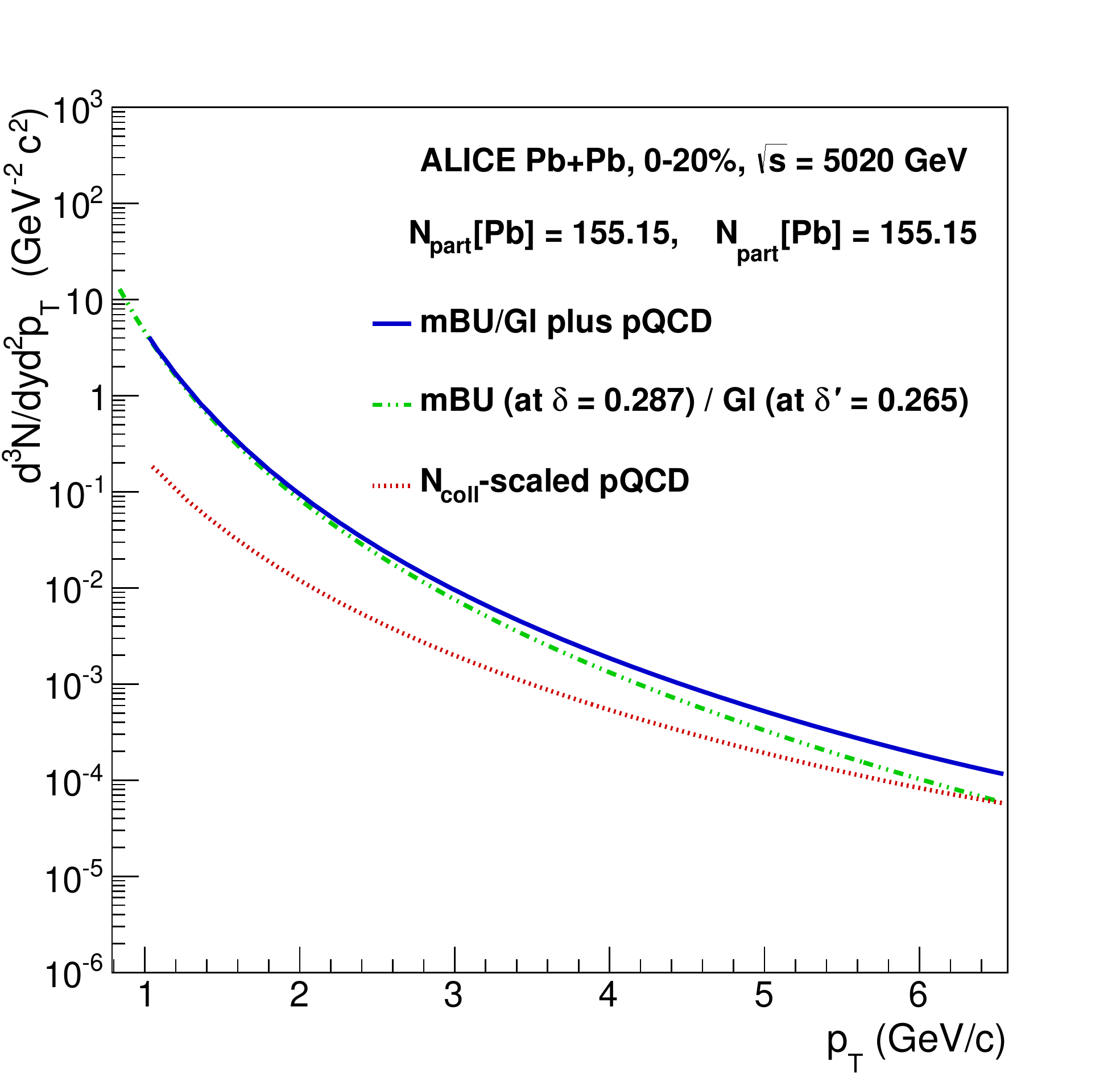} \label{fig:fig_PbPb_5020GeV_020}}
\hspace{-0.025\textwidth}
\vspace{-0.025\textwidth}
   \subfigure[]{\includegraphics[width=0.475\textwidth,height=0.475\textwidth]{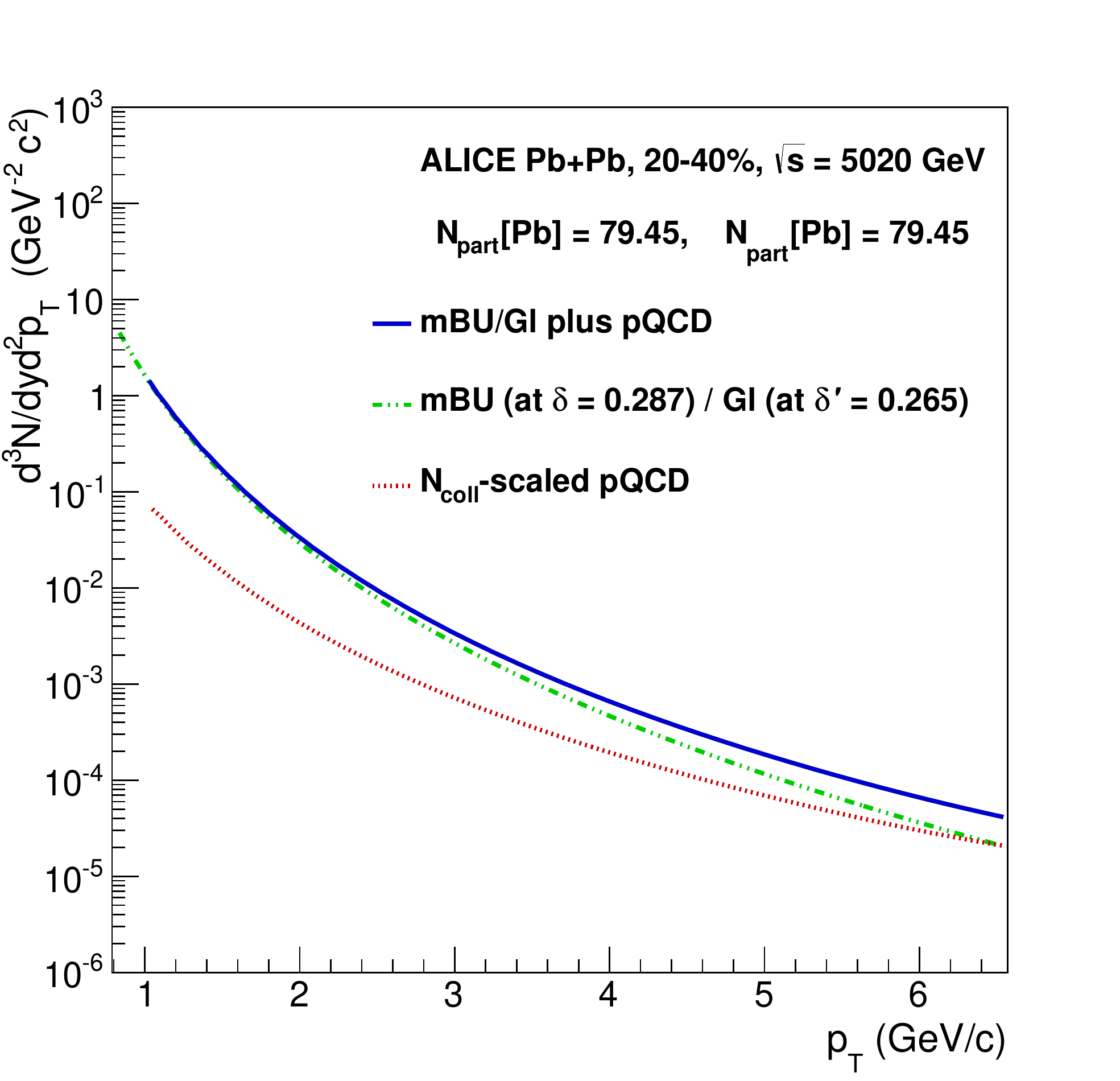} \label{fig:fig_PbPb_5020GeV_2040}}
\\
   \subfigure[]{\includegraphics[width=0.475\textwidth,height=0.475\textwidth]{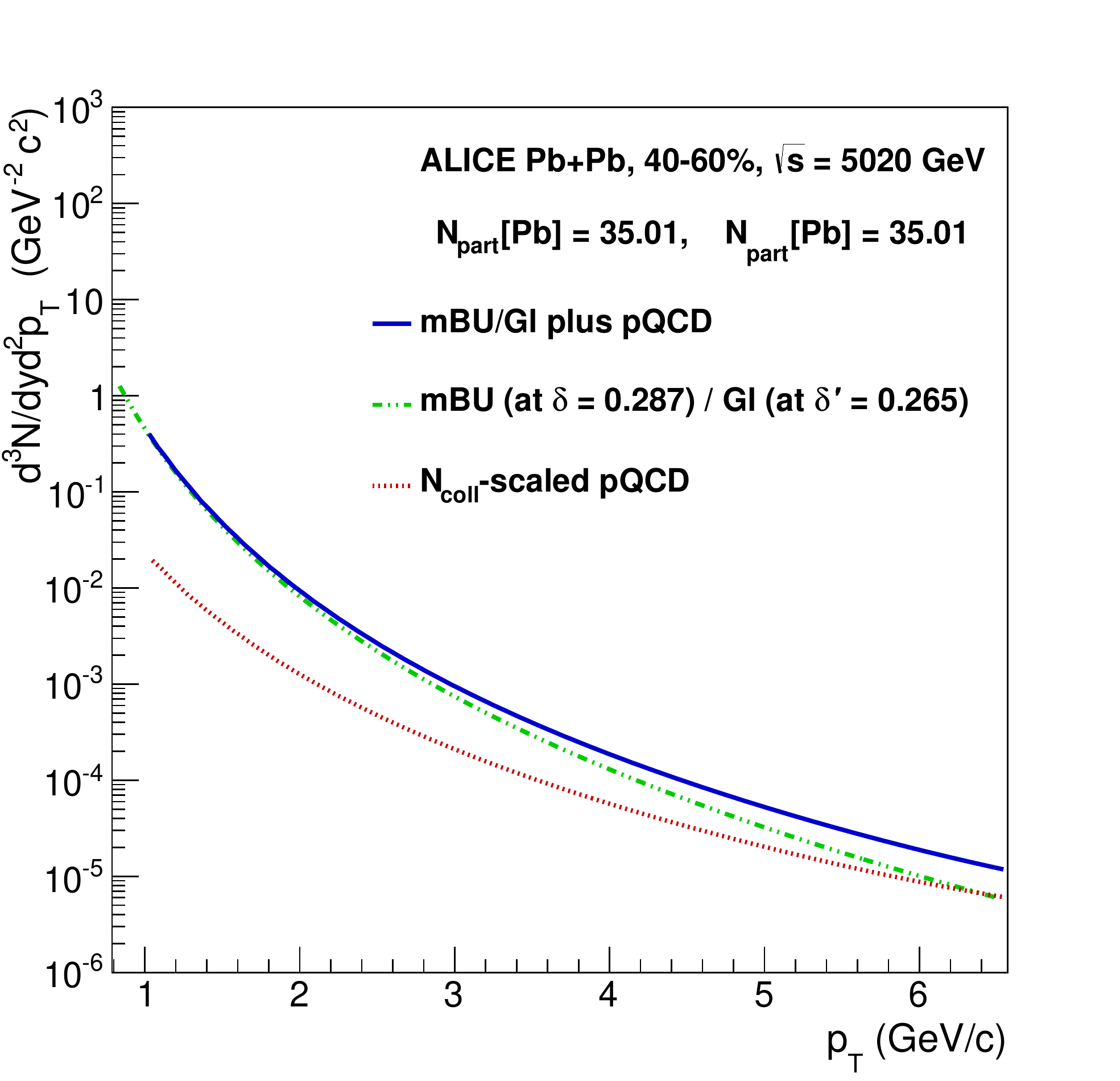} \label{fig:fig_PbPb_5020GeV_4060}}
\hspace{-0.025\textwidth}
\vspace{-0.025\textwidth}
   \subfigure[]{\includegraphics[width=0.475\textwidth,height=0.475\textwidth]{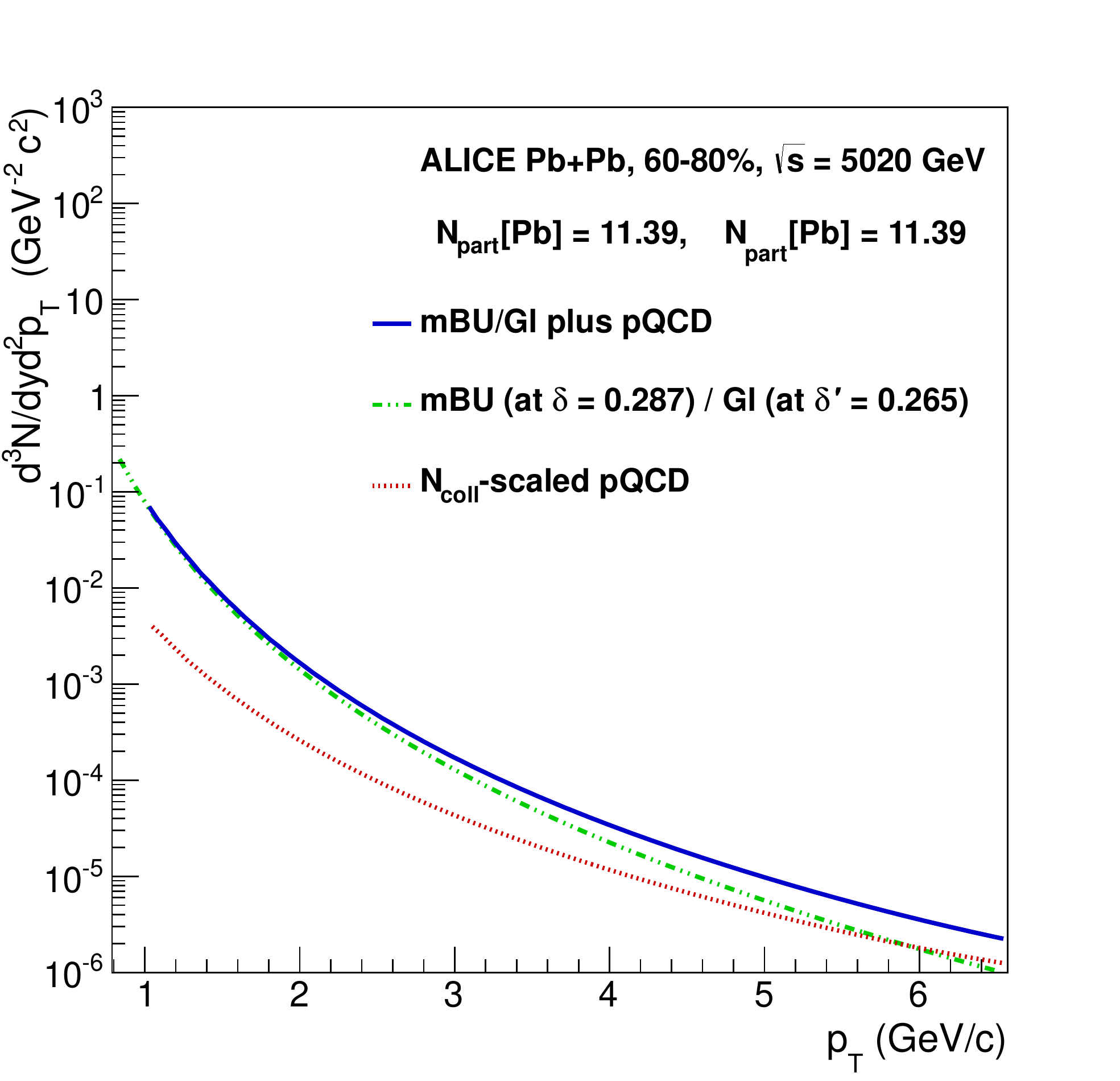} \label{fig:fig_PbPb_5020GeV_6080}}
\end{center}
\vspace{-1.0mm}
\caption{(Color online) The prediction for the direct photon invariant yield for Pb+Pb at $\sqrt{s_{NN}} = 5020$\,GeV (see Table\,\ref{tab:Table2}) obtained from the global fitting of the direct photon data from \cite{Bannier:2014}, \cite{Wilde:2013}, \cite{Yamaguchi:2012} with the mBU-Glasma model in the four centrality bins shown in Fig.\,\ref{fig:fig_PbPb_5020GeV_020}, Fig.\,\ref{fig:fig_PbPb_5020GeV_2040}, Fig.\,\ref{fig:fig_PbPb_5020GeV_4060} and Fig.\,\ref{fig:fig_PbPb_5020GeV_6080}. The pQCD yield is from \cite{Paquet:2017}. The nomenclature is the same as in Fig.\,\ref{fig:fig_AuAu2_200GeV}.}
\label{fig:fig_PbPb_5020GeV}
\end{figure}

\begin{figure}[h!]
\vspace{-1.0mm}
\begin{center}
   \subfigure[]{\includegraphics[width=0.475\textwidth,height=0.475\textwidth]{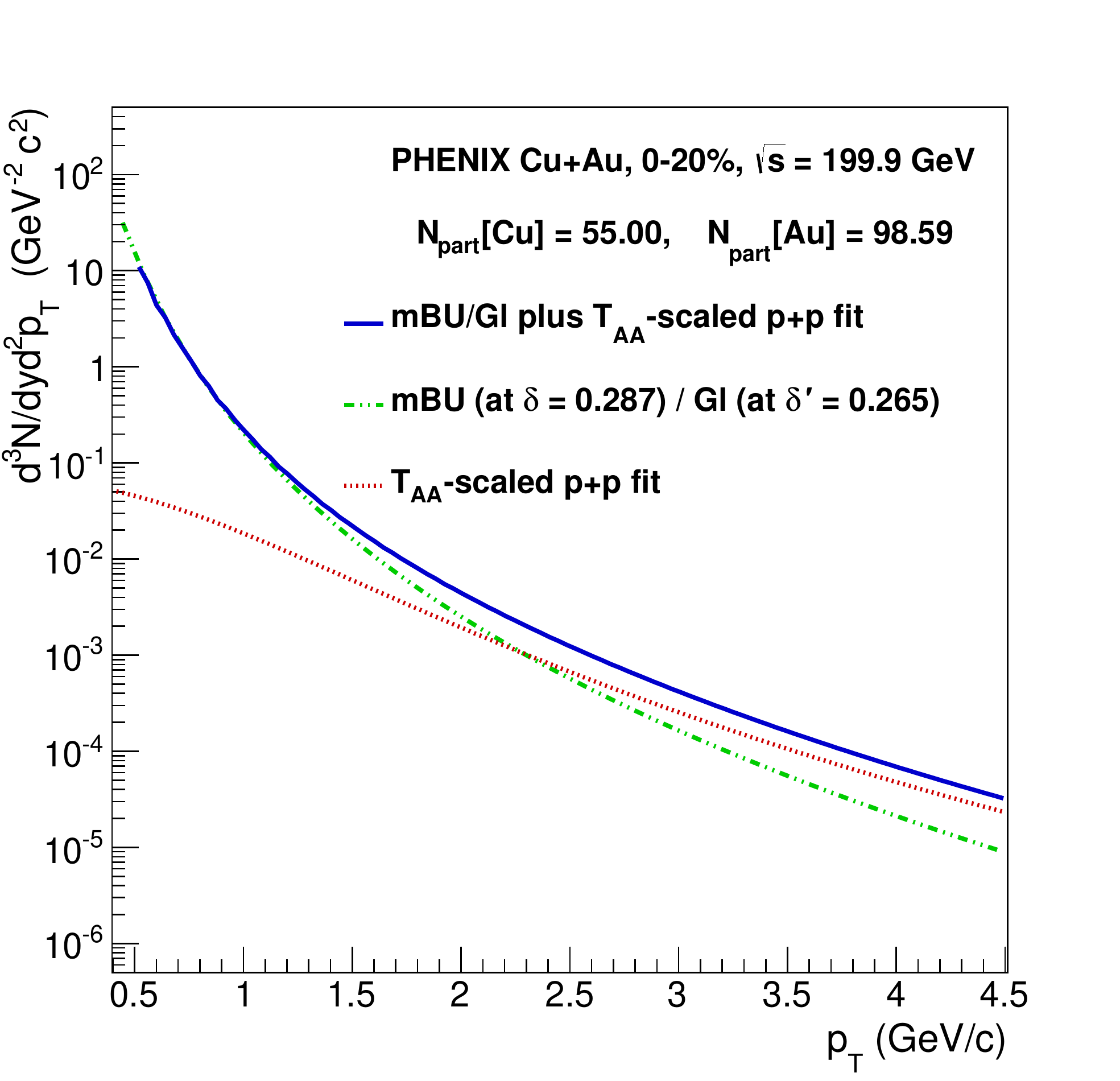} \label{fig:fig_CuAu_200GeV_020}}
\hspace{-0.025\textwidth}
\vspace{-0.025\textwidth}
   \subfigure[]{\includegraphics[width=0.475\textwidth,height=0.475\textwidth]{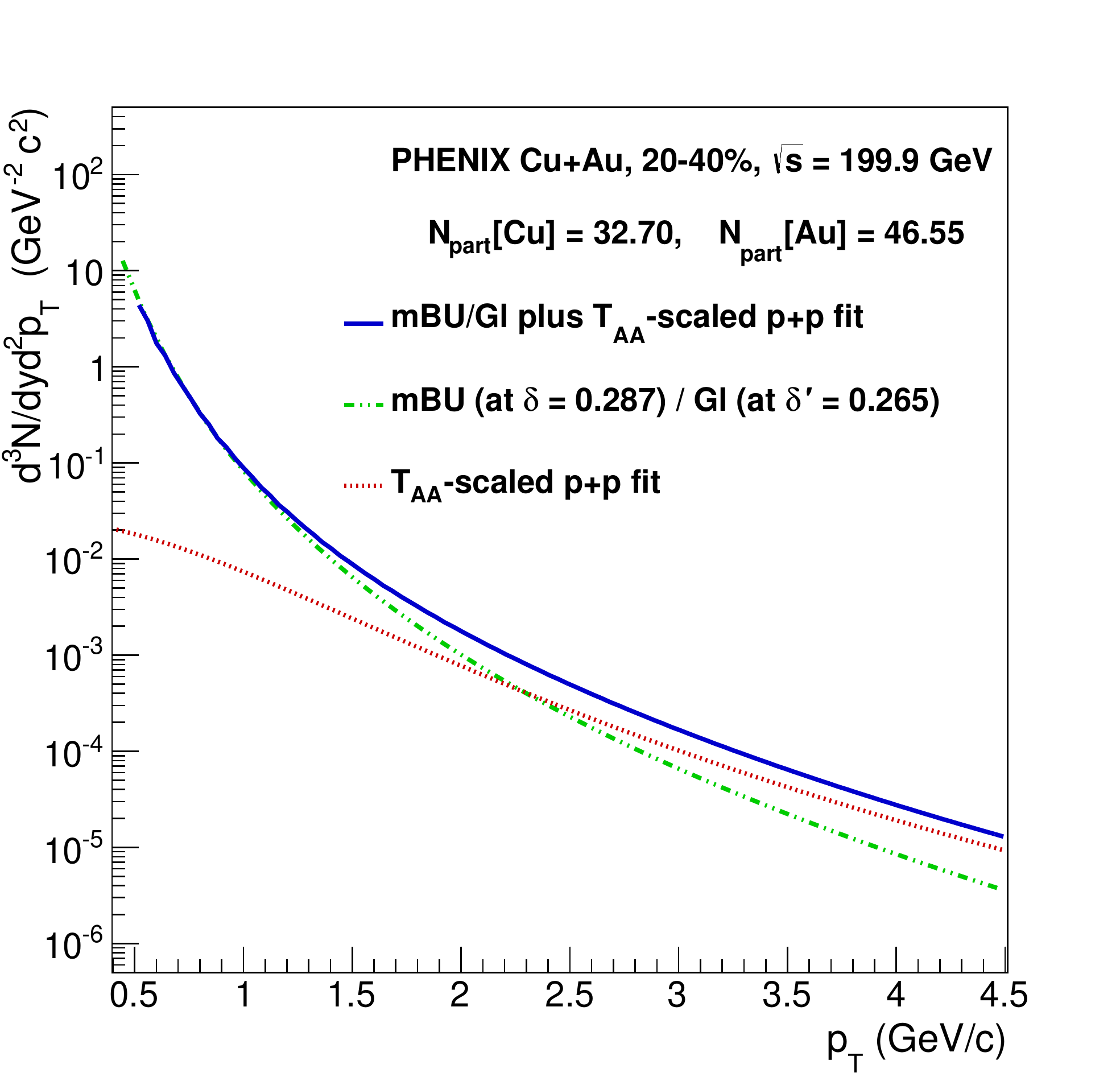} \label{fig:fig_CuAu_200GeV_2040}}
\\
   \subfigure[]{\includegraphics[width=0.475\textwidth,height=0.475\textwidth]{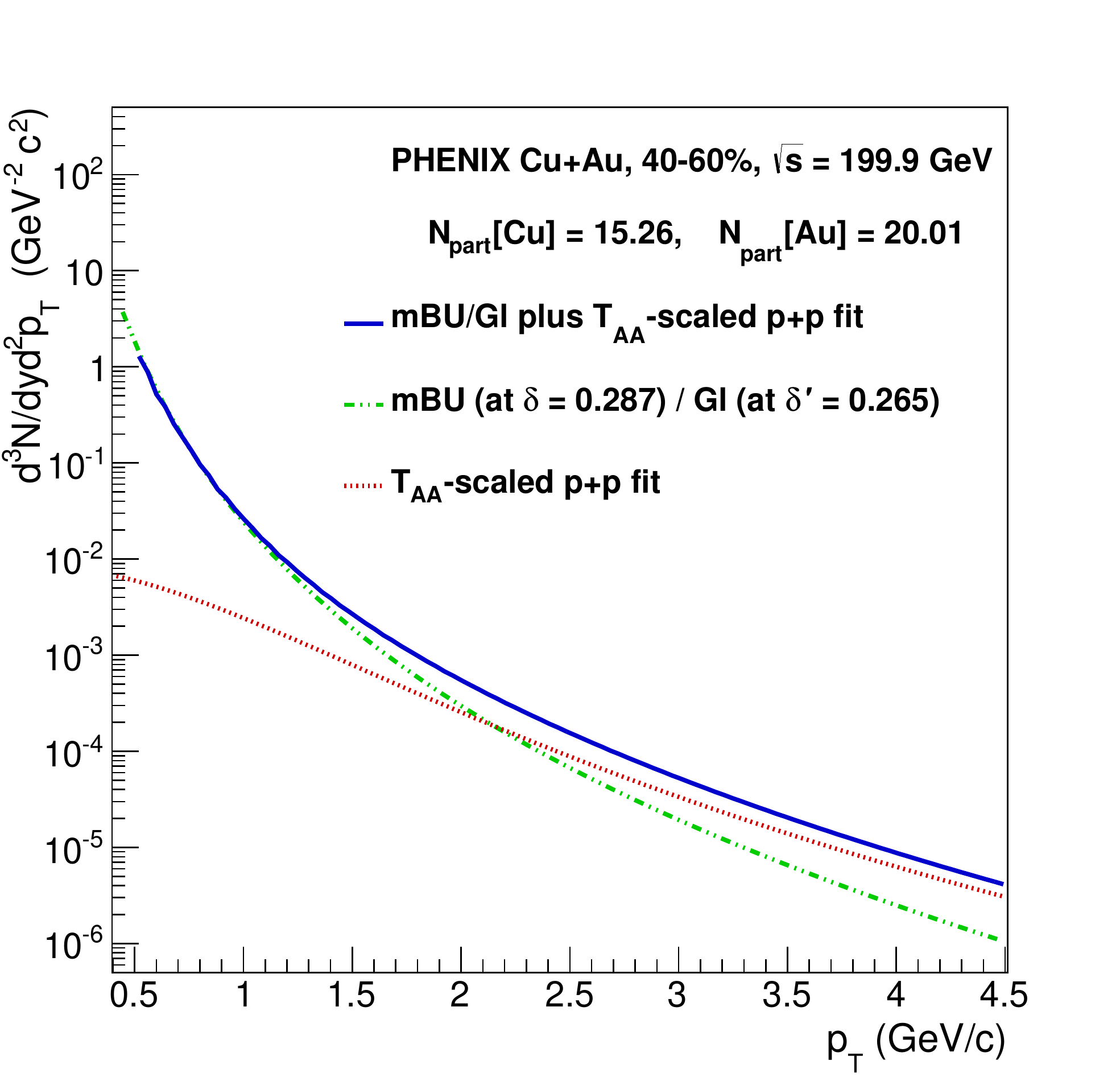} \label{fig:fig_CuAu_200GeV_4060}}
\hspace{-0.025\textwidth}
\vspace{-0.025\textwidth}
   \subfigure[]{\includegraphics[width=0.475\textwidth,height=0.475\textwidth]{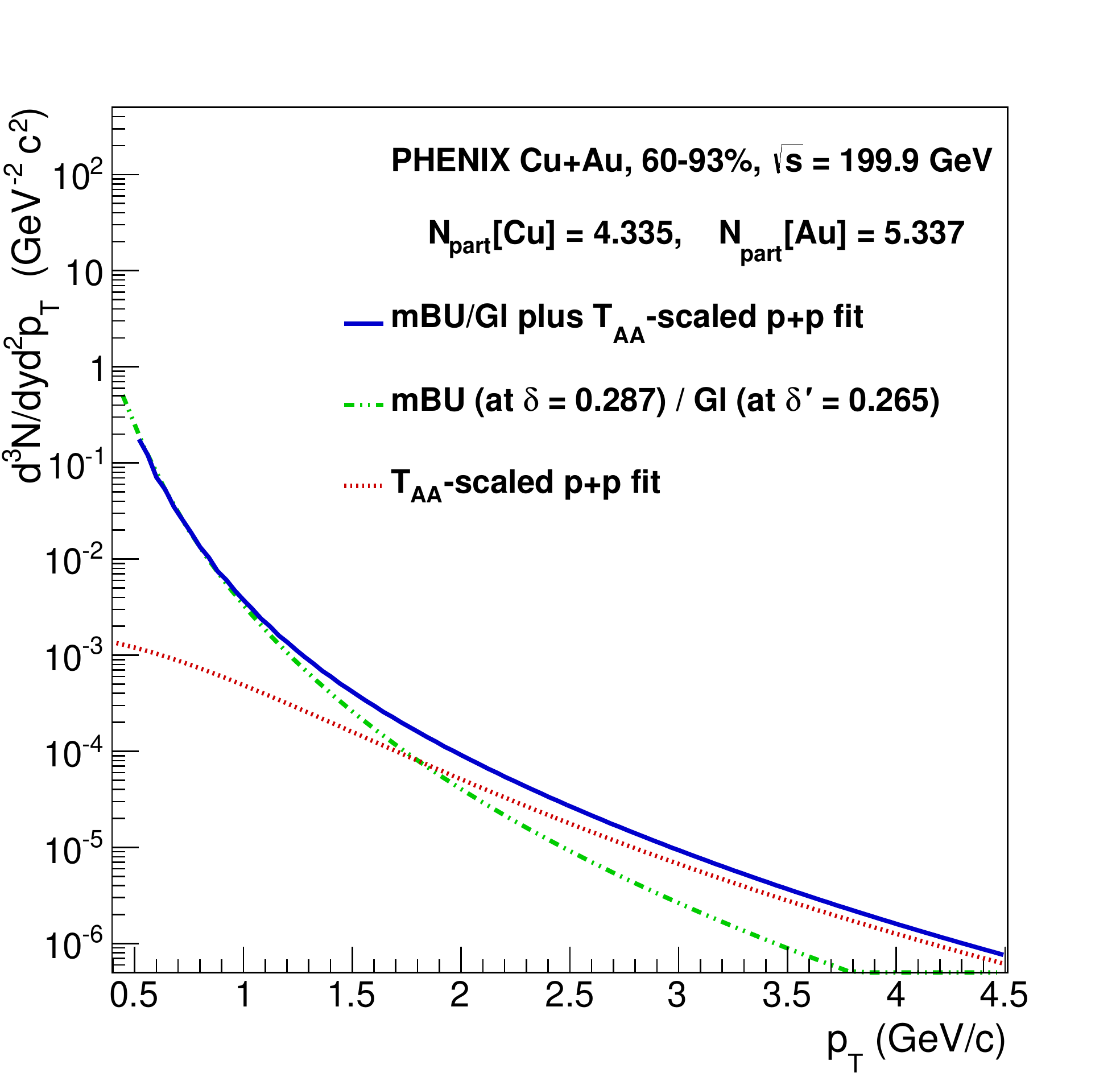} \label{fig:fig_CuAu_200GeV_6093}}
\end{center}
\vspace{-1.0mm}
\caption{(Color online) The prediction for the direct photon invariant yield for Cu+Au at $\sqrt{s_{NN}} = 199.9$\,GeV (see Table\,\ref{tab:Table2}) obtained from the global fitting of the direct photon data from \cite{Bannier:2014}, \cite{Wilde:2013}, \cite{Yamaguchi:2012} with the mBU-Glasma model in the four centrality bins shown in Fig.\,\ref{fig:fig_CuAu_200GeV_020}, Fig.\,\ref{fig:fig_CuAu_200GeV_2040}, Fig.\,\ref{fig:fig_CuAu_200GeV_4060} and Fig.\,\ref{fig:fig_CuAu_200GeV_6093}. The nomenclature is the same as in Fig.\,\ref{fig:fig_AuAu2_200GeV}.}
\label{fig:fig_CuAu_200GeV}
\end{figure}

\begin{figure}[h!]
\vspace{-1.0mm}
\begin{center}
   \subfigure[]{\includegraphics[width=0.475\textwidth,height=0.475\textwidth]{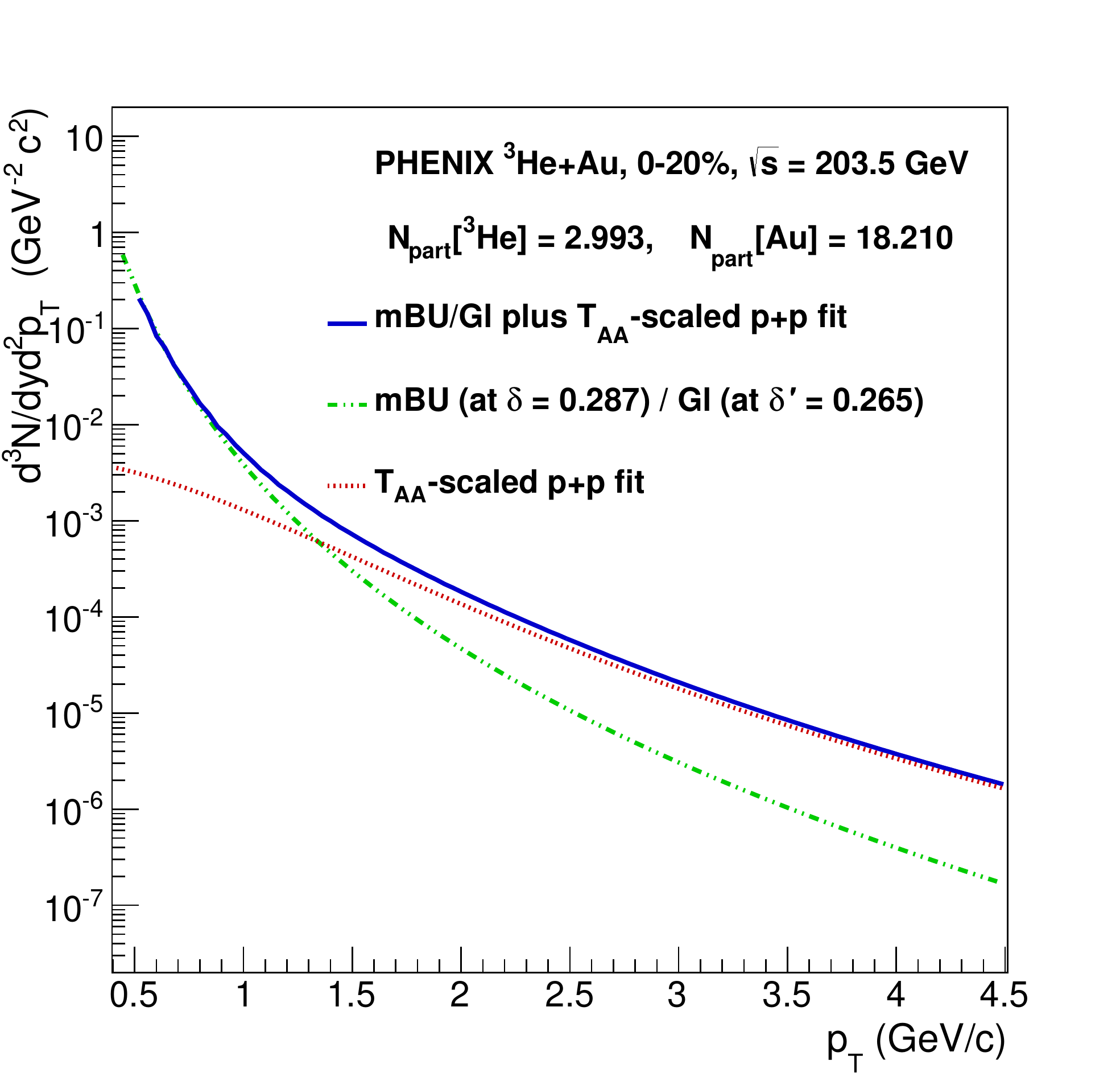} \label{fig:fig_HeAu_200GeV_020}}
\hspace{-0.025\textwidth}
\vspace{-0.025\textwidth}
   \subfigure[]{\includegraphics[width=0.475\textwidth,height=0.475\textwidth]{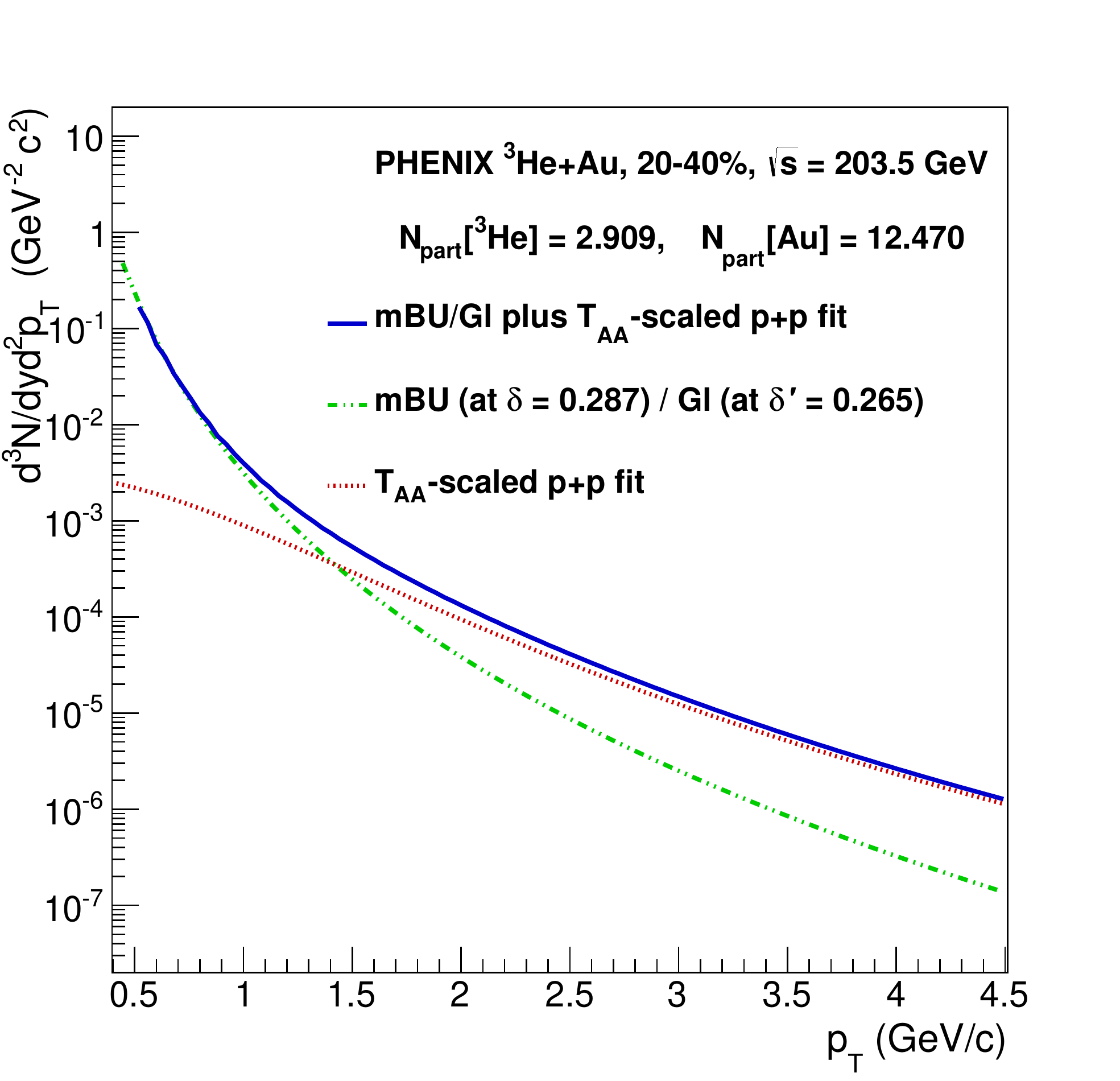} \label{fig:fig_HeAu_200GeV_2040}}
\\
   \subfigure[]{\includegraphics[width=0.475\textwidth,height=0.475\textwidth]{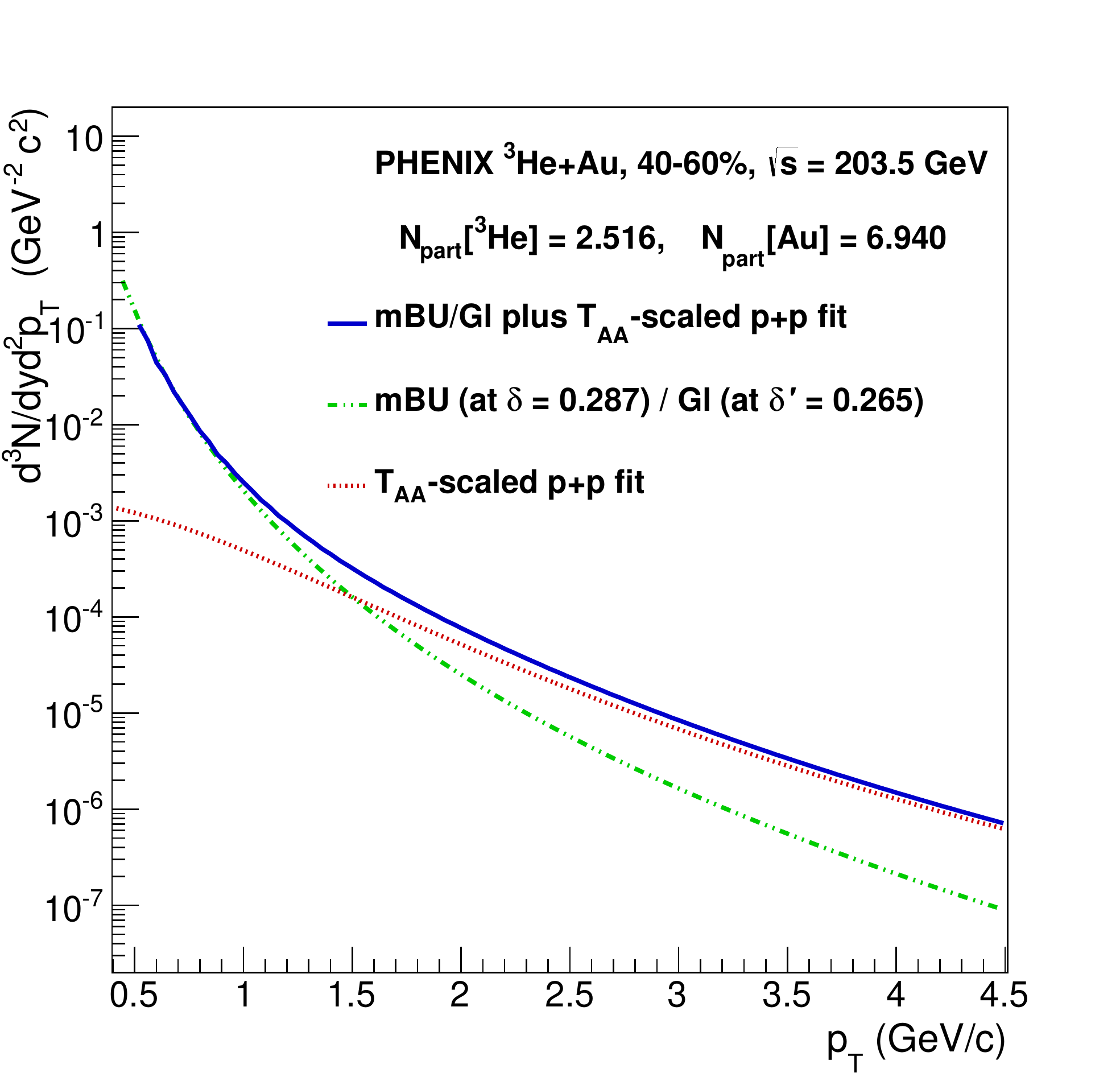} \label{fig:fig_HeAu_200GeV_4060}}
\hspace{-0.025\textwidth}
\vspace{-0.025\textwidth}
   \subfigure[]{\includegraphics[width=0.475\textwidth,height=0.475\textwidth]{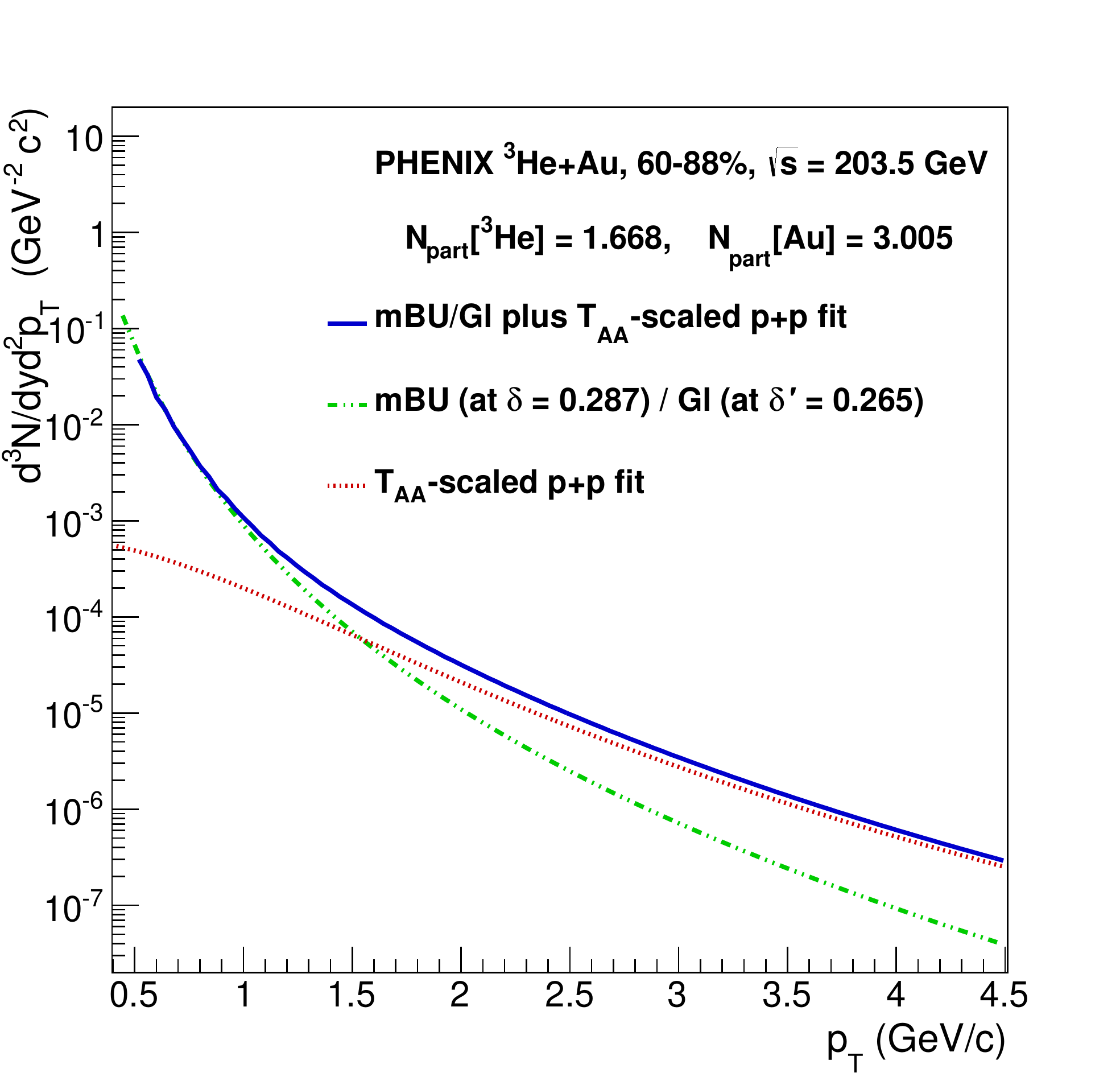} \label{fig:fig_HeAu_200GeV_6088}}
\end{center}
\vspace{-1.0mm}
\caption{(Color online) The prediction for the direct photon invariant yield for ${}^{3}$He+Au at $\sqrt{s_{NN}} = 203.5$\,GeV (see Table\,\ref{tab:Table2}) obtained from the global fitting of the direct photon data from \cite{Bannier:2014}, \cite{Wilde:2013}, \cite{Yamaguchi:2012} with the mBU-Glasma model in the four centrality bins shown in Fig.\,\ref{fig:fig_HeAu_200GeV_020}, Fig.\,\ref{fig:fig_HeAu_200GeV_2040}, Fig.\,\ref{fig:fig_HeAu_200GeV_4060} and Fig.\,\ref{fig:fig_HeAu_200GeV_6088}. The nomenclature is the same as in Fig.\,\ref{fig:fig_AuAu2_200GeV}.}
\label{fig:fig_HeAu_200GeV}
\end{figure}

In this section we show predictions for the direct photon invariant yield in collision systems listed in Table\,\ref{tab:Table2}. Here we use the same $F_{\gamma}$ and $\delta$ obtained form the global fitting in the previous section. Table\,\ref{tab:Table2} has the total $N_{part}$ values of both colliding nuclei as well as the $N_{part}$ values of each nucleus in selected centrality bins. $T_{A_{1}A_{2}}$ and $N_{part}$ are calculated using the TGlauberMC model as before. The predictions for the direct photon invariant yield ($p_{T}$ spectra) are shown for the following collision systems:
\begin{itemize}
\centering
\item[] \hskip -1truecm U+U at $\sqrt{s_{NN}}$ = 192.8\,GeV, Fig.\,\ref{fig:fig_UU_200GeV}; \,\,\,\,\,\,\,\,Cu+Au at $\sqrt{s_{NN}}$ = 99.9+100.0\,GeV, Fig.\,\ref{fig:fig_CuAu_200GeV};
\item[] \hskip -1truecm Pb+Pb at $\sqrt{s_{NN}}$ = 5020\,GeV, Fig.\,\ref{fig:fig_PbPb_5020GeV}; \,\,\,\,\,\,\,\,${}^{3}$He+Au at $\sqrt{s_{NN}}$ = 103.5+100.0\,GeV, Fig.\,\ref{fig:fig_HeAu_200GeV}.
\end{itemize}
The nomenclature and color coding of Fig.\,\ref{fig:fig_AuAu2_200GeV} is applied for all these figures as well.

\section{Conclusions and outlook}
\label{sec:Conclusions}

We have studied the photon production from the thermalizing partonic system produced in (ultra-) relativistic heavy ion collisions in the modified bottom-up framework of \cite{Mueller:2006,Mueller:2006II}. By introducing two momentum scales, we have demonstrated that the modified bottom-up scaling solutions of \cite{Mueller:2006,Mueller:2006II} are equivalent to the thermalizing Glasma solutions of \cite{Blaizot:2011xf}, excluding the Bose-Einstein condensate. Based on this formulation, the mBU-Glasma framework, we have derived the main result of the paper, an analytical formula (Eq.\,(\ref{eqn_scaling_law})) for the invariant yield of the produced excess photons in the low- and intermediate-$p_{T}$ regions for a system undergoing thermalization.

The yield is a function of $p_{T}$, $N_{part}$ and $\sqrt{s_{NN}}$. We constrain the free parameters using a global fit to the PHENIX combined run07+run10 direct photon Au+Au data in four centrality bins at $\sqrt{s_{NN}} = 200$\,GeV with the photon yield estimated from the mBU-Glasma model in combination with the prompt photon yield, as shown in Fig.\,\ref{fig:fig_AuAu2_200GeV}. Also included in the global fit is the ALICE Pb+Pb data in three centrality bins at $\sqrt{s_{NN}} = 2760$\,GeV and the PHENIX run08 $\sqrt{s_{NN}} = 200$\,GeV d+Au data at minimum bias. The comparison of the fit to this experimental data is shown in Fig.\,\ref{fig:fig_PbPb2_2760GeV}. 

With the parameters constrained, we have a powerful analytic formula for the photon yield, which can be used to make predictions for a wide variety of collision systems. Here we presented predictions for the direct photon invariant yield for four centrality classes for the collisions of U+U (Fig.\,\ref{fig:fig_UU_200GeV}), Cu+Au (Fig.\,\ref{fig:fig_CuAu_200GeV}), $^{3}$He+Au (Fig.\,\ref{fig:fig_HeAu_200GeV}) at $\sqrt{s_{NN}} \approx 200$\,GeV and of Pb+Pb (Fig.\,\ref{fig:fig_PbPb_5020GeV}) at $\sqrt{s_{NN}} = 5020$\,GeV.

We have further demonstrated the close connection between the modified bottom up thermalization scenario and the thermalizing Glasma, which have thus far been only discussed separately in the literature. Besides the Bose-Einstein condensate contributon, sometimes included in the latter, both models contain the same physics.

In Fig.\,\ref{fit} we showed that in our mBU-Glasma framework we get a very similar result to that from \cite{Chiu:2013}, where the corresponding figure was obtained using only the thermalizing Glasma ansatz. Thus, we have shown that the two frameworks are not only related with each other in terms of the equations shown in Appendix I, but also that they produce quantitatively similar results, as demonstrated in Figs.\,\ref{Chi2} and \ref{fit}. Besides, we have improved the accuracy of the fit presented in Fig.\,\ref{Chi2} over that in \cite{Chiu:2013}.

The approximations that we have used can be improved upon in the future. First, in the integration in Eq.\,(\ref{eqn_Photon_rate2}) one can use an explicit function for $u(p_{T}/\Omega_{UV})$ similar to that obtained based on a thermal field theory calculation shown in Sec.\,4 of \cite{Schenke:2014}. Along with this the integration limits need to be determined. This will lead to an increased predictive power of the model since the overall normalization will not be merely a free parameter. Also, the assumption that Eq.\,(\ref{eqn_Photon_rate1}) is $\eta$-independent should be relaxed. Another possible improvement is the derivation of the precise $N_{part}$ dependence in asymmetric systems.

These changes will lead to a more complicated expression than that shown in Eq.\,(\ref{eqn_scaling_law}), which may have to be solved numerically. It is expected that the fits and data comparisons (including the predictions) will improve and become more reliable after performing more detailed calculations. For example, the sharp rise of the yield at very low-$p_{T}$, seen in Fig.\,\ref{fit} and Fig.\,\ref{fig:fig_AuAu2_200GeV}, may be reduced.
It would certainly be very important to theoretically calculate (if possible at all) the values of the parameter $\Delta$ (or $\delta$) and constant $F_{\gamma}$ of Eq.\,(\ref{eqn_scaling_law}) as well as the values of the parameter $\mu$ and constant $D_{\gamma}$ of Eq.\,(\ref{eqn_lowpT1}) that we discuss in Appendix III. 

Notwithstanding these efforts to find the partonic photon rates in the mBU-Glasma thermalization scenario, one needs to also include the contribution from the hadron gas \cite{Turbide:2004,Dusling:2010,Lee:2014,Kim:2017,vanHees:2015}. One should also perform full 3+1D hydro simulations including the photon contributions from all partonic and hadronic channels. It will be more challenging to investigate such a combined evolutionary picture, however, for diverse collision systems at various center-of-mass energies, such framework will give the most realistic description of the data (including the very low-$p_{T}$ region) for the direct photon invariant yield/cross section and will allow the computation of the direct photon elliptic flow.

\section*{Acknowledgments}
We greatly appreciate valuable and fruitful discussions with Al Mueller on the subject matter of the paper. We are also grateful to Gabor David, Jinfeng Liao and Larry McLerran for very helpful comments. The research of Vladimir Khachatryan, Axel Drees, Thomas K. Hemmick and Norbert Novitzky is supported under DOE Contract No. DE-FG02-96-ER40988. The research of Bj\"orn Schenke is supported under DOE Contract No. DE-SC0012704. The research of Mickey Chiu is supported under DOE Contract No. DE-AC02-98-CH10886.

\clearpage
\section*{Appendix I: The derivation of the soft gluon number density, Debye mass, energy density and entropy density, and correspondence to previous results in the literature}
\label{sec:Appendix I}

We proceed with $n_{s}$ and $m_{D}$ from Eq.\,(\ref{eqn_scaling}), along with the soft gluon thermal bath temperature $T(\tau)$ from \cite{Mueller:2006,Mueller:2006II}:
\begin{equation}
n_{s} \sim {Q_{s}^{3} \over \alpha_{s} (Q_{s}\tau)^{4/3 - \delta}}\,,
\label{eqn_bottomNs}
\end{equation}
\begin{equation}
m_{D}^{2} \sim {Q_{s}^{2} \over (Q_{s}\tau)^{1 - 3\delta/5}}\,,
\label{eqn_bottomMD}
\end{equation}
\begin{equation}
T^{2} \sim Q_{s}^{2}\,\alpha_{s}^{2(35 - 78\delta)/(39\delta - 10)}(Q_{s}\tau)^{2(15 - 36\delta)/(39\delta - 10)}\,.
\label{eqn_bottom_temp}
\end{equation}
As it was already noted, independent of what value $\delta$ takes, as long as $\delta > 1/3$, the scaling solutions in Eq.\,(\ref{eqn_scaling}) (and the temperature in Eq.\,(\ref{eqn_bottom_temp})) match onto the final stage of the original bottom-up only at the final time \,$Q_{s}\tau \sim \alpha_{s}^{-13/5}$.\, In this case the temperature in Eq.\,(\ref{eqn_bottom_temp}) reduces to $T \sim Q_{s} \alpha_{s}^{2/5}$, which is independent of $\delta$. 

\begin{figure}[h]
\begin{center}
\vspace{-2.5mm}
\includegraphics[width=10cm]{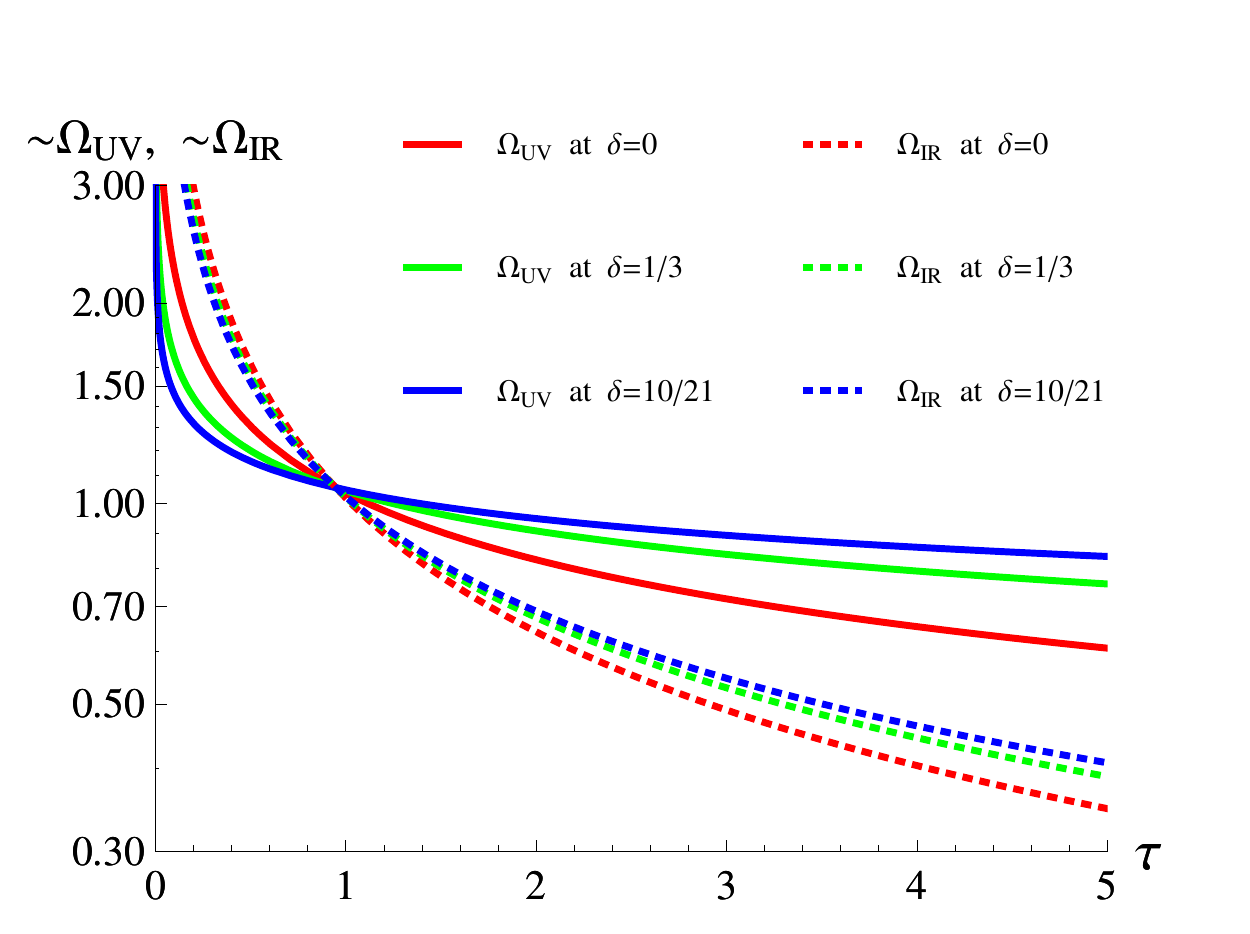}
\caption{The two scales $\Omega_{UV}$ and $\Omega_{IR}$ as functions of time $\tau$, at three values of $\delta$. The number $10/21$ is the absolute limit the parameter $\delta$ may accept in the modified bottom-up thermalization.}
\label{OmegaOmegas}
\end{center} 
\vspace{-0.52cm}
\end{figure}

Analogously, there should exist a solution for $n_{s}$ and $m_{D}$ that is $\delta$-independent. Using the two expressions for $m_D$, Eq.\,(\ref{eqn_bottomMD}) and Eq.\,(\ref{eqn_bottomDebye})
\begin{equation}
m_{D}^{2} \sim{Q_{s}^{2} \over (Q_{s}\tau)^{1- 3\delta/5}}\,\,\,\,\,\,\,\,\,\,\,\,\mbox{and}\,\,\,\,\,\,\,\,\,\,\,\,m_{D}^{2} \sim \Omega_{UV} \Omega_{IR}\,,
\label{eqn_eqns1}
\end{equation}
and solving for $\Omega_{IR}$ we find
\begin{equation}
\Omega_{IR} \sim {1 \over \Omega_{UV}} {Q_{s}^2 \over (Q_{s}\tau)^{1- 3\delta/5}}\,.
\label{eqn_Omegas}
\end{equation}
Making use of  Eqs.\,(\ref{eqn_bottomNs}) and (\ref{eqn_bottomgluon}) 
\begin{equation}
n_{s} \sim {Q_{s}^{3} \over \alpha_{s} (Q_{s}\tau)^{4/3 - \delta}}\,\,\,\,\,\,\,\,\,\,\,\,\mbox{and}\,\,\,\,\,\,\,\,\,\,\,\,n_{s} \sim {1 \over \alpha_{s}} \Omega_{UV}^{2} \Omega_{IR}\,,
\label{eqn_eqns2}
\end{equation}
along with (\ref{eqn_Omegas}) and solving for $\Omega_{UV}$, we find the scale $\Omega_{UV}$ as a function of the proper time $\tau$:
\begin{equation}
\Omega_{UV} \sim Q_{s} {\Lb 1 \over Q_{s}\tau \Rb}^{(5 - 6\delta)/15}\,.
\label{eqn_Omegatau}
\end{equation}
Using $\Omega_{UV}$ from Eq.\,(\ref{eqn_Omegatau}) in the denominator in the r.h.s. of Eq.\,(\ref{eqn_Omegas}), we obtain the scale $\Omega_{IR}$ as a function of $\tau$.
\begin{equation}
\Omega_{IR} \sim Q_{s} {\Lb 1 \over Q_{s}\tau \Rb}^{(10 - 3\delta)/15}\,.
\label{eqn_Omegastau}
\end{equation}
In Fig.\,\ref{OmegaOmegas}, these time-dependent scales are parametrically depicted at three selected values of the parameter $\delta$.

Given the time dependence of the scales, we can parametrically find the thermalization time of the mBU-Glasma thermalization scenario. Using the second condition of Eq.\,(\ref{eqn_bottom_boundary}):
\begin{equation}
Q_{s} {\Lb 1 \over Q_{s}\tau_{therm} \Rb}^{(10 - 3\delta)/15} \sim \alpha_{s} Q_{s} {\Lb 1 \over Q_{s}\tau_{therm} \Rb}^{(5 - 6\delta)/15}\,,
\end{equation}
leads to
\begin{equation}
Q_{s}\tau_{therm} \sim \alpha_{s}^{-{15/(5 + 3\delta)}}\,.
\label{eqn_bottom_therm}
\end{equation}

Utilizing $Q_{s}\tau_{therm}$ from Eq.\,(\ref{eqn_bottom_therm}), as a next step we divide Eq.\,(\ref{eqn_bottomNs}) by Eq.\,(\ref{eqn_bottom_temp}) (with $T^{3}$) and Eq.\,(\ref{eqn_bottomMD}) by Eq.\,(\ref{eqn_bottom_temp}). This results in the following expressions:
\beq
n_{s} \sim \alpha_{s}^{-\left[ (1) + (3(35 - 78\delta)/(39\delta - 10)) \right]} \left( \alpha_{s}^{-15/(5 + 3\delta)} \right)^{-\left[ (3(15 - 36\delta)/(39\delta - 10)) + ((4 - 3\delta)/3) \right]} T_{therm}^{3}\,,
\label{eqn_bottomNsright}
\eeq
(which reduces to \,$n_{s} \sim T_{therm}^{3}$,\, that is to say the same as Eq.\,(\ref{eqn_bottomgluon_fin})),
\beq
m_{D}^{2} \sim \alpha_{s}^{ -2(35 - 78\delta)/(39\delta - 10)} \left( \alpha_{s}^{-15/(5 + 3\delta)} \right)^{ - \left[ (1) - (3\delta/5) + (2(15 - 36\delta)/(39\delta - 10)) \right]} T_{therm}^{2}\,,
\label{eqn_bottomMDright}
\eeq
(which reduces to \,$m_{D}^{2} \sim \alpha_{s} T_{therm}^{2}$,\, the same as Eq.\,(\ref{eqn_bottomDebye_fin})). Alternatively, an analogous outcome can be derived if we use the $n_{s}$ and $m_{D}$ from Eqs.\,(\ref{eqn_bottomgluon}) and (\ref{eqn_bottomDebye}), instead of $n_{s}$ and $m_{D}$ from the scaling solutions of Eq.\,(\ref{eqn_scaling}).
\bea
\!\!\!\!\!\!\!\!\!\!
\frac{n_{s}}{T^{3}}|_{therm} & \sim & \Lb {1 \over \alpha_{s}} \Omega_{UV}^{2} \Omega_{IR} \Rb / \Lb Q_{s}^{3}\,\alpha_{s}^{3(35 - 78\delta)/(39\delta - 10)}\,(Q_{s}\tau)^{3(15 - 36\delta)/(39\delta - 10)} \Rb|_{therm} \Rightarrow
\nonumber\\ 
n_{s} & \sim & T_{therm}^{3}\,,
\label{eqn_NsT}
\eea
and
\bea
\!\!\!\!\!\!\!\!\!\!
\frac{m_{D}^{2}}{T^{2}}|_{therm} & \sim & \Lb \Omega_{UV}\Omega_{IR} \Rb / \Lb Q_{s}^{2}\,\alpha_{s}^{2(35 - 78\delta)/(39\delta - 10)}\,(Q_{s}\tau)^{2(15 - 36\delta)/(39\delta - 10)} \Rb|_{therm} \Rightarrow
\nonumber\\ 
m_{D}^{2} & \sim & \alpha_{s} T_{therm}^{2}\,,
\label{eqn_MDT}
\eea
Obviously the results in Eqs.\,(\ref{eqn_bottomNsright}) and (\ref{eqn_bottomMDright}) are independent of any value the $\delta$ accepts. Note that the thermalization temperature $T_{therm}$ of the mBU-Glasma parton matter is assumed to be parametrically of the same order as the initial temperature of the Quark-Gluon Plasma $T_{in,QGP}$, as in \cite{Chiu:2013}.

An equilibrated system of soft gluons at thermalization temperature $T_{therm}$ must satisfy the condition $\epsilon_{s} \sim T_{therm}^{4}$ for the energy density. Needless to say that we should also be able to prove the validity of this relation in our ansatz. From \cite{Mueller:2006,Mueller:2006II} we find the energy density of the soft sector expressed via the soft gluon number density and momentum, $\epsilon_{s} \sim n_{s}k_{s}$. Then from the scaling solutions of Eq.\,(\ref{eqn_scaling}) we have the following expression:
\begin{equation}
\epsilon_{s} \sim {Q_{s}^{4} \over \alpha_{s} (Q_{s}\tau)^{(25 - 21\delta)/15}}\,,
\label{eqn_en_density1}
\end{equation}
It is clear also that the same result comes from $\epsilon_{s} \sim (1/\alpha_{s})\Omega_{IR}\Omega_{UV}^{3}$. By dividing Eq.\,(\ref{eqn_en_density1}) by Eq.\,(\ref{eqn_bottom_temp}) (with $T^{4}$) along with using $Q_{s}\tau_{therm}$ from Eq.\,(\ref{eqn_bottom_therm}), the energy density becomes
\beq
\epsilon_{s} \sim \alpha_{s}^{-\left[ (1) + (4(35 - 78\delta)/(39\delta - 10)) \right]} \left( \alpha_{s}^{-15/(5 + 3\delta)} \right)^{-\left[ ((25 - 21\delta)/15) + (4(15 - 36\delta)/(39\delta - 10)) \right]} T_{therm}^{4}\,,
\label{eqn_en_density2}
\eeq
which reduces to \,$\epsilon_{s} \sim T_{therm}^{4}$.\, 

Besides, the entropy density in the thermal soft gluon bath at thermalization is derived as
\bea
\frac{s}{T^{3}}|_{therm} & \sim & \Omega_{UV}^{3} / \Lb Q_{s}^{3}\,\alpha_{s}^{3(35 - 78\delta)/(39\delta - 10)}(Q_{s}\tau)^{3(15 - 36\delta)/(39\delta - 10)} \Rb|_{therm} \Rightarrow
\nonumber\\ 
s & \sim & T_{therm}^{3}\,.
\label{eqn_entropy}
\eea

We now turn to the comparison to results from the Glasma evolution in [42]. First we assume that the scale $\Omega_{UV}$ from Eq.\,(\ref{eqn_Omegatau}) and the Glasma ultraviolet scale $\Lambda_{UV} \sim Q_{s} \Lb 1/Q_{s}\tau \Rb^{(1 + 2\delta^{\prime})/7}$ from \cite{Blaizot:2011xf} are parametrically the same:
\begin{equation}
\Omega_{UV} \sim \Lambda_{UV}\,\,\,\,\,\,\,\,\,\,\Rightarrow\,\,\,\,\,\,\,\,\,\,Q_{s} {\Lb 1 \over Q_{s}\tau \Rb}^{(5 - 6\delta)/15} \sim Q_{s}{\Lb 1 \over Q_{s}\tau \Rb}^{(1 + 2\delta^{\prime})/7}\,.
\label{eqn_OmegaLambda}
\end{equation}
Mathematically this assumption is valid if 
\begin{equation}
\delta = \frac{10 - 15\delta^{\prime}}{21}\,,
\label{eqn_delta_deltaprime}
\end{equation}
or vice versa
\begin{equation}
\delta^{\prime} = \frac{10 - 21\delta}{15}\,.
\label{eqn_deltaprime_delta}
\end{equation}
Inserting the same relation into the r.h.s. of Eq.\,(\ref{eqn_Omegastau}) we find
\begin{equation}
\Omega_{IR} \sim Q_{s} {\Lb 1 \over Q_{s}\tau \Rb}^{(10 - 3\delta)/15}\,\,\,\,\,\rightarrow\,\,\,\,\,Q_{s} {\Lb 1 \over Q_{s}\tau \Rb}^{(4 + \delta^{\prime})/7}\,.
\label{eqn_OmegasLambda}
\end{equation}
We find that the r.h.s of Eq.\,(\ref{eqn_OmegasLambda}) is indeed the infrared scale $\Lambda_{IR}$ from \cite{Blaizot:2011xf}. Based on this finding we will now make explicit the correspondence of our expressions for the gluon density, Debye mass, thermalization time and energy density with those in \cite{Blaizot:2011xf}:
\begin{equation}
n_{g}\Lb \mbox{from\,Ref.}\,\cite{Blaizot:2011xf} \Rb \sim {Q_{s}^{3} \over \alpha_{s} (Q_{s}\tau)^{(6 + 5\delta^{\prime})/7}}\,\,\,\,\,\,\,\,\,\,\,\,\,\,\,\,\,\,\,\,m_{D}^{2} \Lb \mbox{from\,Ref.}\,\cite{Blaizot:2011xf} \Rb \sim {Q_{s}^{2} \over (Q_{s}\tau)^{(5 + 3\delta^{\prime})/7}}\,,
\label{eqn_Glasma_NgMD}
\end{equation}
\begin{equation}
Q_{s}\tau_{therm}\Lb \mbox{from\,Ref.}\,\cite{Blaizot:2011xf} \Rb \sim \alpha_{s}^{-{7/(3 - \delta^{\prime})}}\,.
\label{eqn_Glasma_tautherm}
\end{equation}
The energy density in gluon modes  is given by 
\begin{equation}
\epsilon_{g}\Lb \mbox{from\,Ref.}\,\cite{Blaizot:2011xf} \Rb \sim \frac{1}{\alpha_{s}}\Lambda_{IR}\Lambda_{UV}^{3}\,,
\label{eqn_Glasma_en_density1}
\end{equation}
which, in the case of the longitudinal expansion under assumption of the parameter $\delta^{\prime}$ being independent of time, is represented as a formula for evolution of the energy density:
\begin{equation}
\epsilon_{g}(\tau)\Lb \mbox{from\,Ref.}\,\cite{Blaizot:2011xf} \Rb \sim \epsilon_{g}(\tau_{0}) \Lb \frac{1}{Q_{s}\tau} \Rb^{1 + \delta^{\prime}}\,.
\label{eqn_Glasma_en_density2}
\end{equation}
Thus, by making use of Eq.\,(\ref{eqn_deltaprime_delta}) it is straigtforward to see the one-to-one correspondence in what follows:
\begin{equation}
n_{g}\,\,\mbox{of\,\,Eq.\,(\ref{eqn_Glasma_NgMD})}\,\,\,\,\,\rightarrow\,\,\,\,\,n_{s}\,\,\mbox{of\,\,Eq.\,(\ref{eqn_scaling})}\,;
\label{eqn_NgNs}
\end{equation}
\begin{equation}
m_{D}\,\,\mbox{of\,\,Eq.\,(\ref{eqn_Glasma_NgMD})}\,\,\,\,\,\rightarrow\,\,\,\,\,m_{D}\,\,\mbox{of\,\,Eq.\,(\ref{eqn_scaling})}\,;
\label{eqn_MDMD}
\end{equation}
\begin{equation}
\tau_{therm}\,\,\mbox{of\,\,Eq.\,(\ref{eqn_Glasma_tautherm})}\,\,\,\,\,\rightarrow\,\,\,\,\,\tau_{therm}\,\,\mbox{of\,\,Eq.\,(\ref{eqn_bottom_therm})}\,;
\label{eqn_thermtherm}
\end{equation}
\begin{equation}
\epsilon_{g}\,\,\mbox{of\,\,Eq.\,(\ref{eqn_Glasma_en_density1})}\,\,\,\,\,\rightarrow\,\,\,\,\,\epsilon_{s}\,\,\mbox{of\,\,Eq.\,(\ref{eqn_en_density1})}\,.
\label{eqn_epsilong_epsilons}
\end{equation}
These transformations are valid in the range of $10/21 > \delta > 5/21$, which corresponds to the range of \,$0 < \delta^{\prime} < 1/3$,\, This can be seen as follows:
\begin{quote}
at \,$\delta^{\prime} = 0$\, \,$\rightarrow$\, \,$\delta = 10/21$,\,\,\,\,\,\,\,
at \,$\delta^{\prime} = 1/3$\, \,$\rightarrow$\, \,$\delta = 5/21$,\,\,\,\,\,\,\,
at \,$\delta^{\prime} = 2/3$\, \,$\rightarrow$\, \,$\delta = 0$.
\end{quote}
On the other hand, if one formally needs to recover the static case by setting $\delta^{\prime} = -1$, corresponding to constant energy density $\epsilon_{g}(\tau) \rightarrow \epsilon_{g}(\tau_{0})$, then it will also be the case for the mBU-Glasma, $\epsilon_{s}(\tau) \rightarrow \epsilon_{s}(\tau_{0}) \sim Q_{s}^{4}/\alpha_{s}$, because in that case $\delta =25/21$.

At the end of this section we mention one of the results from \cite{Mueller:2006II}. If $\delta > 1/3$, the family of the scaling solutions in Eq.\,(\ref{eqn_scaling}) changes its character at a time $\tau^{\ast}$ given by
\begin{equation}
Q_{s}\tau^{\ast} \sim (1/\alpha_{s})^{15/(5 + 3\delta)}\,,
\label{eqn_bottom_tau1}
\end{equation}
in which case $f_{s} \sim 1$. In this instance the scaling solutions go into evolution much like the final phase of the original bottom-up picture, where the soft gluons are thermalized and the hard gluons feed energy into the soft thermalized system causing the temperature to rise with time, until the whole system is thermalized. Notice that Eq.\,(\ref{eqn_bottom_therm}), which we obtain after introducing the scales $\Omega_{UV}$ and $\Omega_{IR}$ into the mBU-Glasma ansatz, turns out to be parametrically the same as Eq.\,(\ref{eqn_bottom_tau1}), though in our approach the corresponding time is already derived as the thermalization time of the system based on the thermalization condition of Eq.\,(\ref{eqn_bottom_boundary}). In this sense, our discussed mBU-Glasma evolution reconciles the modified bottom-up scaling solutions of \cite{Mueller:2006,Mueller:2006II} with all the thermalizing Glasma solutions of \cite{Blaizot:2011xf} as it was noted in Sec.\,\ref{sec:Conclusions}. This reconciliation attempt led to Eq.\,(\ref{eqn_Photon_rate44}) (or Eq.\,(\ref{eqn_scaling_law})), which is parametrically the same as Eq.\,(3.6) of Ref.\,\cite{Chiu:2013}. Furthermore, the parameter $\eta$ that can be found in Eq.\,(3.6) is actually the parameter $\Delta$ of Eq.\,(\ref{eqn_Photon_rate44}). It can be demonstrated by means of Eq.\,(\ref{eqn_deltaprime_delta}), namely
\begin{equation}
\eta = \frac{9 - 3\delta^{\prime}}{1 + 2\delta^{\prime}}\,\,\,\,\,\,\,\,\,\,\,\,\rightarrow\,\,\,\,\,\,\,\,\,\,\,\,
\frac{9 - 3((10 - 21\delta)/15)}{1 + 2((10 - 21\delta)/15)}\,\,\,\,\,\,\,\,\,\,\,\,\rightarrow\,\,\,\,\,\,\,\,\,\,\,\,
\Delta = \frac{15 + 9\delta}{5 - 6\delta}\,.
\label{eqn_eta_Delta}
\end{equation}
%

\section*{Appendix II: The thermalization time and temperature of the mBU-Glasma evolution in the most central Au+Au collisions at RHIC $\mathbf{\sqrt{s_{NN}} = 200\,GeV}$}
\label{sec:Appendix II}

One can estimate the values of the thermalization time (from Eq.\,(\ref{eqn_bottom_therm})) and thermalization temperature (from Eq.\,(\ref{eqn_bottom_temp})) at RHIC $\sqrt{s_{NN}} = 200$\,GeV collision energy, by following the procedure that has been employed in \cite{Baier:2002}. First of all for the thermalization time we have
\begin{equation}
\tau_{therm} = C_{therm}\,\alpha_{s}^{-15/(5 + 3\delta)} Q_{s}^{-1}\,,
\label{eqn_therm_time}
\end{equation}
where the $C_{therm}$ is the thermalization constant. For the thermalization temperature we have
\begin{equation}
T_{therm} = C_{T}\,\alpha_{s}^{(35 - 78\delta)/(39\delta - 10)}(Q_{s}\tau_{therm})^{(15 - 36\delta)/(39\delta - 10)}Q_{s}\,,
\label{eqn_therm_temp1}
\end{equation}
which by using Eq.\,(\ref{eqn_therm_time}) reduces to
\begin{equation}
T_{therm} \simeq 0.16543\,C\,C_{therm}^{(15 - 36\delta)/(39\delta - 10)}\,\alpha_{s}^{(5 - 6\delta)/(5 + 3\delta)} Q_{s}\,,
\label{eqn_therm_temp2}
\end{equation}
where we use the numerical constant $C_{T}$ expressed by the ``gluon liberation'' coefficient $C$ \cite{Baier:2002}:
\begin{equation}
C_{T} \simeq {15 \over 8\pi^{5}}\,N_{c}^{3}\,C \simeq 0.16543\,C\,.
\label{eqn_numerical}
\end{equation}
The coefficient $C$ links the number of gluons in the nucleus wave function to the number of gluons, which are freed during a collision. The number of gluons increases with time because the primary hard gluons degrade, and the soft ones are formed starting to dominate in the system. Such an increase of the number of gluons must be equal to or larger than two, and can be found as the following ratio:
\bea
R & = & {\left[ n_{s}(\tau)(Q_{s}\tau) \right]|_{\tau_{therm}} \over \!\!\!\!\!\left[ n_{h}(\tau)(Q_{s}\tau) \right]|_{\tau_{0}}} \geq 2\,\,\,\,\,\Rightarrow
\nonumber\\
R & \simeq & 0.13061\,C^{2}\,C_{therm}^{(35 - 69\delta)/(39\delta - 10)}\,\alpha_{s}^{(5 - 15\delta)/(5 + 3\delta)} \geq 2\,\,\,\,\,\Rightarrow
\nonumber\\
2 & \leq & 0.13061 \left( {4\pi \over 9} \right)^{(5 - 15\delta)/(5 +3\delta)} C^{2}\,C_{therm}^{(35 - 69\delta)/(39\delta - 10)} \left( \ln{\!\!\left( {Q_{s}^{2} \over \Lambda_{QCD}^{2}} \right)} \right)^{(15\delta - 5)/(5 + 3\delta)}
\label{eqn_ratio}
\eea
where we made use of Eqs.\,(\ref{eqn_therm_time}) and (\ref{eqn_therm_temp2}) along with the formulas shown in what follows:
\begin{equation}
n_{s}(\tau_{therm}) = 2(N_{c}^{2} - 1){\zeta(3) \over \pi^{2}}\,T_{therm}^{3}\,,
\label{eqn_nsoft}
\end{equation}
\begin{equation}
n_{h}(\tau_{0}) = C\,{(N_{c}^{2} - 1)Q_{s}^{3} \over 4\pi^{2}N_{c}\alpha_{s}\,(Q_{s}\tau_{0})}\,,
\label{eqn_nhard}
\end{equation}
\begin{equation}
\alpha_{s}(Q_{s}^{2}) \simeq \frac{4\pi}{\left( 11- {2 \over 3}N_{f} \right) \ln{\!\!\left( {Q_{s}^{2} \over \Lambda_{QCD}^{2}} \right)}}\,.
\label{eqn_alpha}
\end{equation}
$n_{s}$ is the number density of the soft gluons, and $n_{h}$ is the number density of the primary hard gluons of the original bottom-up thermalization \cite{Baier:2001,Baier:2002}. For $N_{c} = 3$ we take $N_{f} = 3$. Also, by using the well-known relation at midrapidity
\begin{equation}
Q_{s}^{2}(A,\sqrt{s}) = Q_{0}^{2}(A,\sqrt{s_{0}}) \left( \frac{\sqrt{s}}{\sqrt{s_{0}}} \right)^{\lambda/(1 + \lambda/2)}\,,
\label{eqn_sat}
\end{equation}
with $\lambda = 0.288$ \cite{Kharzeev:2004}, we shall have the saturation momentum $Q_{s}^{2} \simeq 1.115\,\mbox{GeV}^{2}$ at $\sqrt{s_{NN}} = 200$\,GeV for the most central Au+Au collisions, which is obtained from $Q_{0}^{2} = 1\,\mbox{GeV}^{2}$ at $\sqrt{s_{NN}} = 130$\,GeV that has been used in \cite{Baier:2002} as a reference. Thus, the ratio $R$ in Eq.\,(\ref{eqn_ratio}) can be two or larger than two if the overall constant $C^{2}\,C_{therm}^{(35 - 69\delta)/(39\delta - 10)}$ is taken adequately.

We need to have also one more formula with $C$ and $C_{therm}$ from which these parameters can be determined. It is derived by comparing the theoretically calculable charged hadron multiplicity at midrapidity at $\sqrt{s_{NN}} = 200$\,GeV with the multiplicity from the corresponding RHIC Au+Au data. As a reference value we use the result by the PHOBOS collaboration \cite{PHOBOS:2002}, the same as used in \cite{Baier:2002}, namely
\begin{equation}
\bigg\langle {2 \over N_{part}}{dN_{ch} \over d\eta} \bigg\rangle |_{exp} = 3.78 \pm 0.25\,(syst)\,.
\end{equation}
The charged hadron multiplicity can be calculated as follows:
\begin{equation}
\bigg\langle {2 \over N_{part}}{dN_{ch} \over d\eta} \bigg\rangle \simeq {R\,C \over 3}\,\ln{\!\!\Lb {Q_{s}^{2} \over \Lambda_{QCD}^{2}} \Rb}\,,
\label{eqn_multip1}
\end{equation}
which, using $R$ from Eq.\,(\ref{eqn_ratio}), becomes
\bea
& & \!\!\!\!\!\!\!\!\!\!\!
\bigg\langle {2 \over N_{part}}{dN_{ch} \over d\eta} \bigg\rangle \simeq
\nonumber\\
& \simeq & 0.04354 \left( {4\pi \over 9} \right)^{(5 - 15\delta)/(5 +3\delta)} C^{3}\,C_{therm}^{(35 - 69\delta)/(39\delta - 10)} \left( \ln{\!\!\left( {Q_{s}^{2} \over \Lambda_{QCD}^{2}} \right)} \right)^{18\delta/(5 + 3\delta)}\,. 
\label{eqn_multip2}
\eea
This charged hadron multiplicity will be equal to the experimental value 3.78 if the overall constant $C^{3}\,C_{therm}^{(35 -69\delta)/(39\delta - 10)}$ is taken adequately.

\begin{figure}[h!]
\begin{center}
\vspace{0.0cm}
\includegraphics[width=8.25cm]{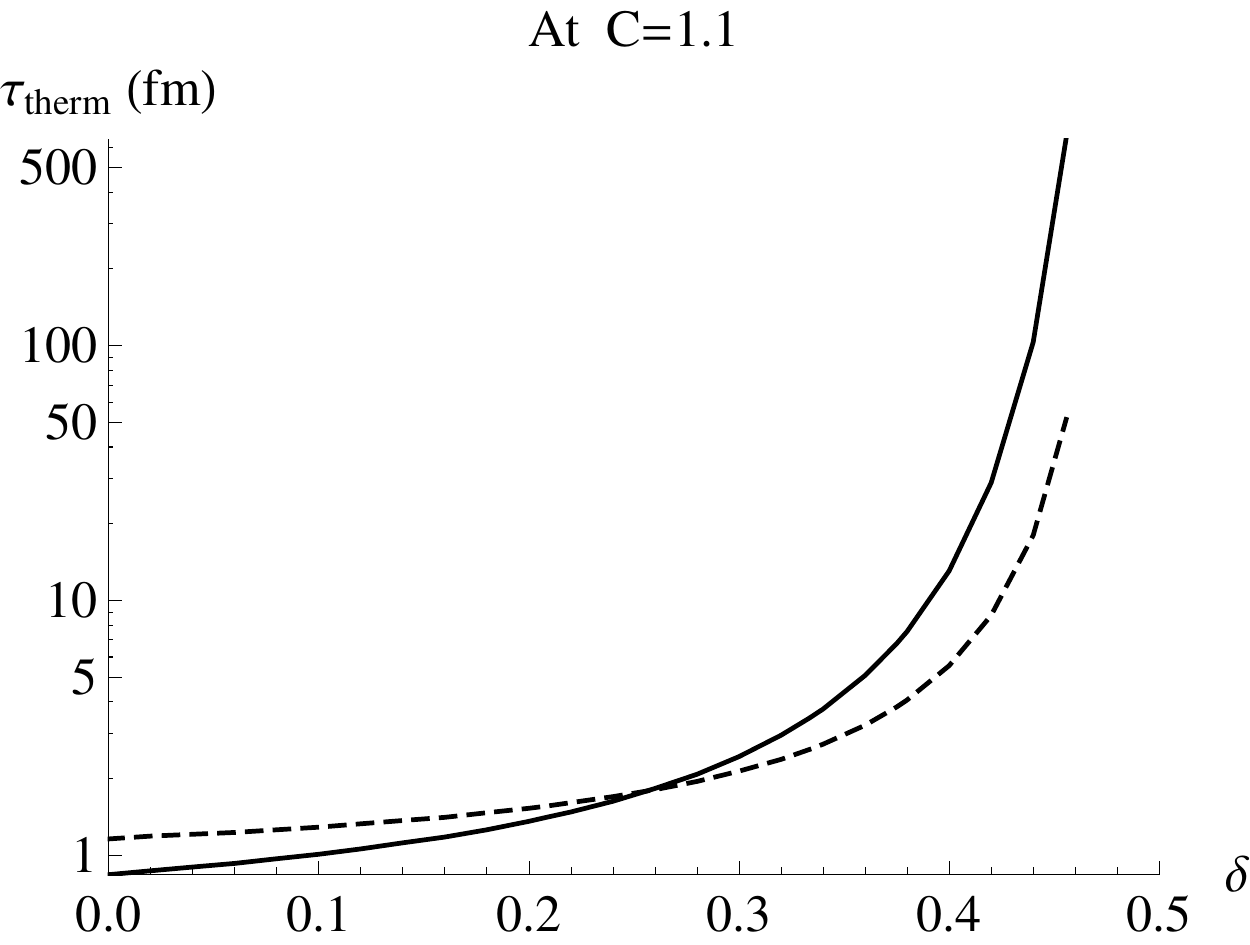}
\caption{The solid and dashed curves show the $\delta$-dependent $\tau_{therm}$ and $\tau_{1}$ at $C = 1.1$, respectively.}
\label{Omega_tau1}
\vspace{-0.2cm}
\end{center} 
\end{figure}

\begin{figure}[h!]
\begin{center}
\vspace{0.0cm}
\includegraphics[width=8.25cm]{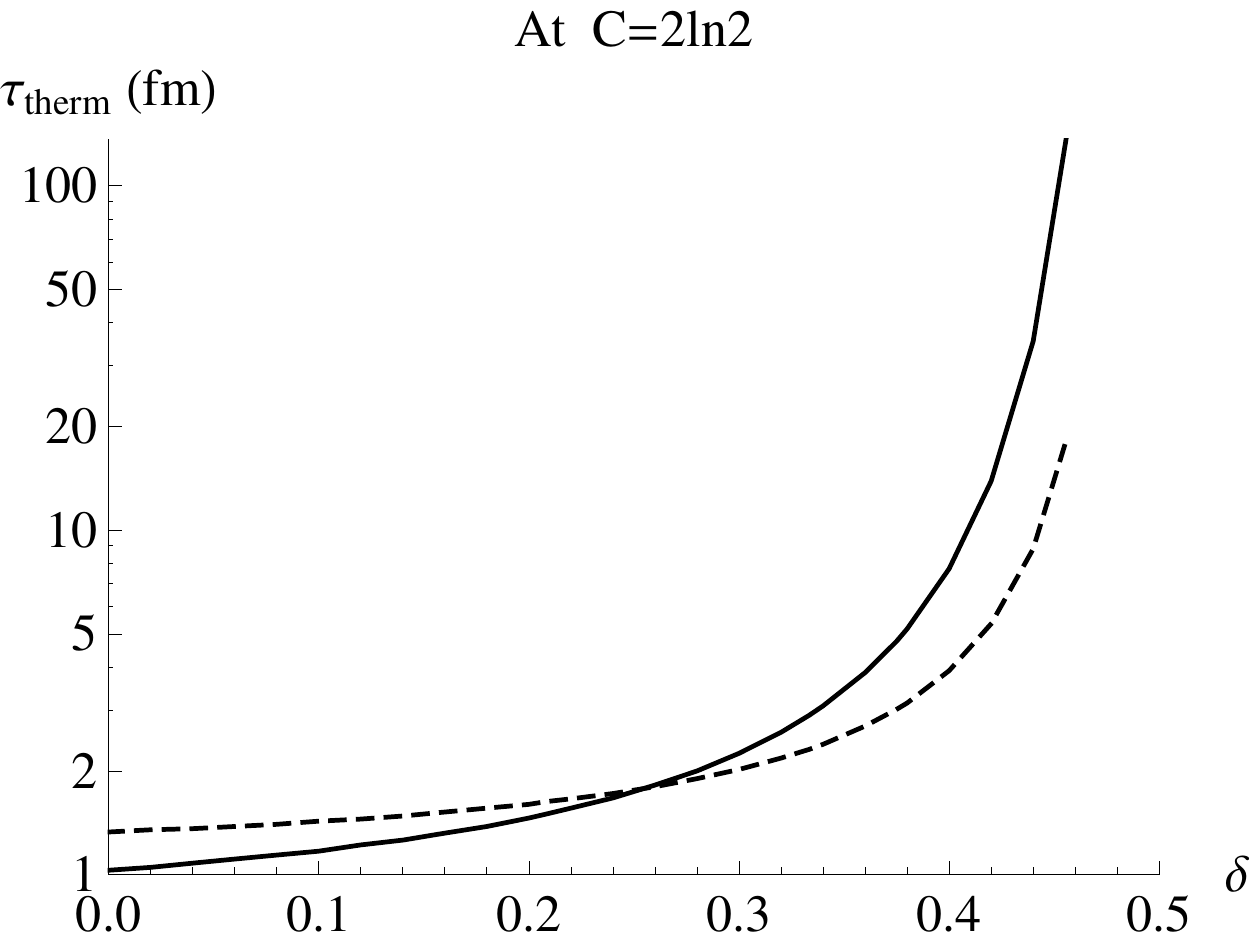}
\caption{The solid and dashed curves show the $\delta$-dependent $\tau_{therm}$ and $\tau_{1}$ at $C = 2\ln{\!2}$, respectively.}
\label{Omega_tau2}
\vspace{-0.2cm}
\end{center} 
\end{figure}

\begin{figure}[h!]
\begin{center}
\vspace{0.0cm}
\includegraphics[width=8.25cm]{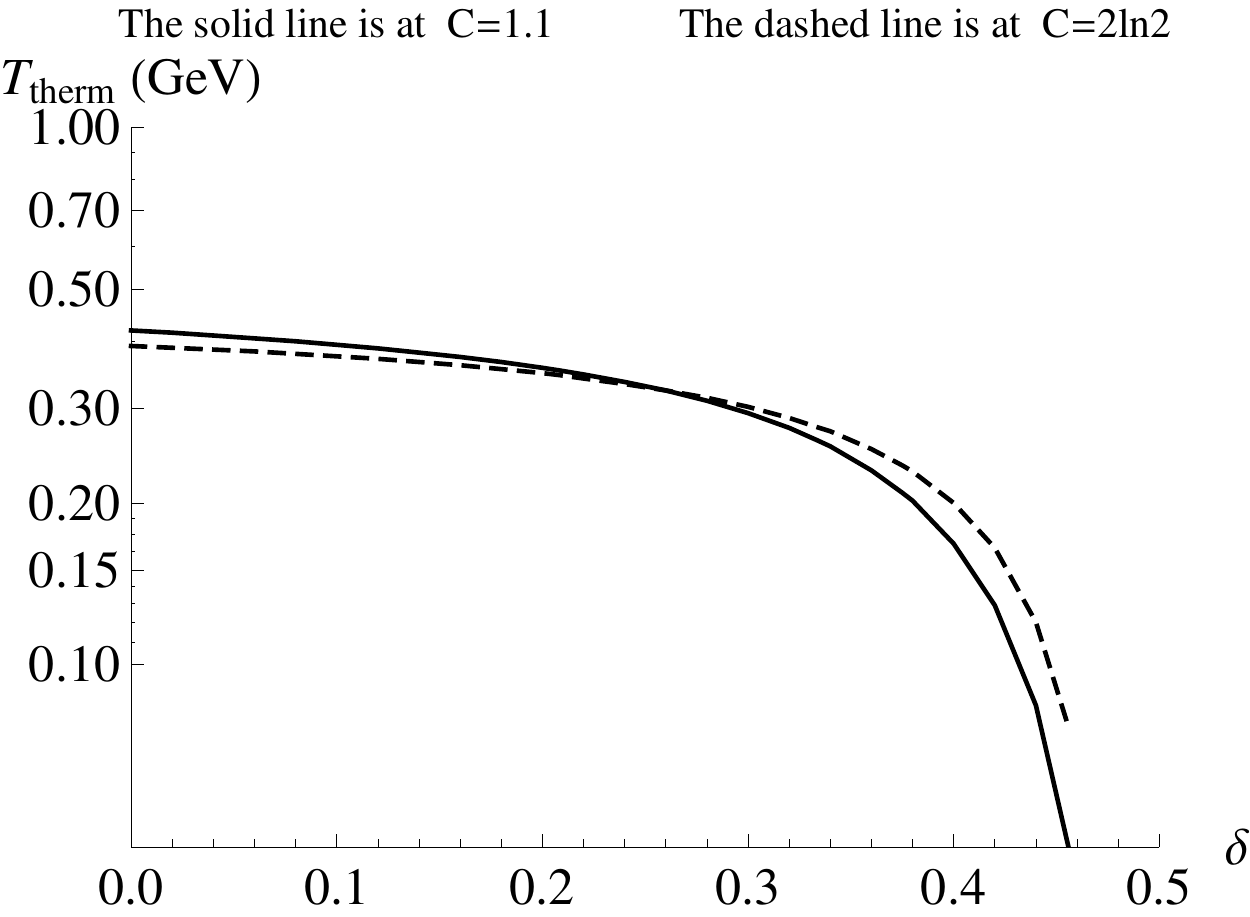}
\caption{The solid and dashed curves show the $\delta$-dependent $T_{therm}$ at $C = 1.1$ and $C = 2\ln{\!2}$, respectively.}
\label{Omega_T}
\vspace{-0.5cm}
\end{center} 
\end{figure}

Let us now find $\tau_{therm}$ and $T_{therm}$ as functions of the parameter $\delta$. We will try two values for the ``gluon liberation" coefficient: the first one calculated in \cite{Kovchegov:2001} with $C = 2\ln{\!2} \simeq 1.386$, and the second one in \cite{Lappi:2001} with $C = 1.1$. For example, at $\delta = 1/3$, from Eq.\,(\ref{eqn_ratio}) and Eq.\,(\ref{eqn_multip2}) we have
\begin{displaymath}
C_{therm} \simeq {2.260 \over C^{3/4}}\,\,\,\,\,\mbox{and}\,\,\,\,\,C_{therm} \geq {1.978 \over C^{2/4}}\,,
\end{displaymath}
which result in
\begin{equation}
C \leq 1.704\,\,\,\,\,\mbox{and}\,\,\,\,\,C_{therm} \geq 1.515\,.
\label{eqn_delta1}
\end{equation}
Consequently, if we use the aforementioned values of $C$, then from Eq.\,(\ref{eqn_therm_time}) and Eq.\,(\ref{eqn_therm_temp2}) the following values of the thermalization time and temperature follow: 
\bea
\mbox{at}\,\,\,\,C = 1.1\,\,\,\,\,\Rightarrow\,\,\,\,\,\,\,\,\tau_{therm} \simeq 3.45\,\mbox{fm}\,,\,\,T_{therm} \simeq 262\,\mbox{MeV},
\nonumber \\
\mbox{at}\,\,\,\,C = 2\ln{\!2}\,\,\,\,\,\Rightarrow\,\,\,\tau_{therm} \simeq 2.90\,\mbox{fm}\,,\,\,T_{therm} \simeq 277\,\mbox{MeV}.
\label{eqn_delta_tau_T}
\eea
Also, 
\begin{equation}
\mbox{at}\,\,\,\,C = 1.1 \,\,\,\,\,\Rightarrow\,\,\,\,\,R \simeq 3.1,\,\,\,\,\,\,\,\,\,\,\,\,\,\,\,\,\,\,\,\,\mbox{at}\,\,\,\,C = 2\ln{\!2}\,\,\,\,\,\Rightarrow\,\,\,\,\,R \simeq 2.46. 
\end{equation}
In Fig.\,\ref{Omega_tau1} and Fig.\,\ref{Omega_tau2} the solid lines show the $\delta$-dependent thermalization time at both values of $C$. The $\delta$-dependent thermalization temperature at both values of $C$ is shown in Fig.\,\ref{Omega_T}.

One can put some constraints on possible values of the parameter $\delta$. It can be accomplished if we find the time $\tau_{1}$ at which $n_{h} = n_{s}$. Namely, by using Eq.\,(\ref{eqn_nsoft}), Eq.\,(\ref{eqn_nhard}) and Eq.\,(\ref{eqn_therm_temp2}), we can derive
\begin{equation}
Q_{s}\tau_{1} = \Lb \frac{1}{0.1306\,C^{2}} \Rb^{(39\delta - 10)/(35 - 69\delta)} \alpha_{s}^{-(95 - 195\delta)/(35 - 69\delta)}\,.
\label{eqn_numerical2}
\end{equation}
It is obvious that at $\delta = 1/3$ the time $\tau_{1}$ becomes parametrically the same as that of the original bottom-up picture, i.e., 
 $n_{h} = n_{s}$\, at \,$Q_{s}\tau_{1} \sim \alpha_{s}^{-5/2}$. Then if we calculate the time $\tau_{1}$ at $C = 1.1$ and at $C = 2\ln{\!2}$\, for \,$\delta = 1/3$, we obtain the following result:
\bea
& & \mbox{at}\,\,\,\,C = 1.1\,\,\,\,\,\Rightarrow\,\,\,\,\,\,\,\,\tau_{1} \simeq 2.60\,fm,
\nonumber \\
& & \mbox{at}\,\,\,\,C = 2\ln{\!2}\,\,\,\,\,\Rightarrow\,\,\,\tau_{1} \simeq 2.31\,fm.
\label{eqn_delta2}
\eea
In Fig.\,\ref{Omega_tau1} and Fig.\,\ref{Omega_tau2} the dashed lines show the $\delta$-dependent $\tau_{1}$ at both values of $C$. So since the soft gluons start to overwhelm, in terms of number, the primary hard gluons at $\tau_{1} \sim \alpha_{s}^{-(95 - 195\delta)/(35 - 69\delta)}Q_{s}^{-1}$, it means that the inequality $\tau_{therm} > \tau_{1}$ must always hold.  By comparing the results shown in Fig.\,\ref{Omega_tau1} and Fig.\,\ref{Omega_tau2}, we notice that the condition at $\delta \lesssim 0.26$, where $\tau_{therm} < \tau_{1}$, should be excluded. On the other hand, the evolutionary picture looks highly unlikely at larger values of $\delta$ close to $\delta = 10/21$, which shows that the system never gets thermalized. 

However, one should note that the results in Figs.\,\ref{Omega_tau1}, \ref{Omega_tau2} and \ref{Omega_T} are approximate because in our estimations we make use of the coefficient $C_{T}$ from Eq.\,(\ref{eqn_numerical}) calculated in the original bottom-up thermalization. It is possible that in reality the numerical constant $C_{T}$ can be $\delta$-dependent as well, which perhaps will make the curves less steep at large values of $\delta$, and/or make the lower limit $\delta \simeq 0.26$ even smaller. 

\section*{Appendix III: An estimate of di-electron (and associated photon) production in the mBU-Glasma scenario}
\label{sec:Appendix III}

In this appendix we attempt to give a qualitative estimate of $e^{+}e^{-}$ pair (di-electron) production as well as its associated photon production in the mBU-Glasma scenario. Although there are other approximations used here leading to derivations of Eq.\,(\ref{eqn_lowpT1}) and Eq.\,(\ref{eqn_lowpT2}), and we do not construct plots as we did, for example, in Sec.\,\ref{sec:Fitting2}, in any case it is worth to have some additional discussion related to Fig.\,\ref{Photons_pt} and Fig.\,\ref{Dileptons_pt} in view of  the mBU-Glasma.

\begin{figure}[ht]
\vspace{0.0cm}
\begin{center}
\scalebox{1.0}[1.0] {
\hspace{0.2cm}
  \includegraphics[scale=0.10]{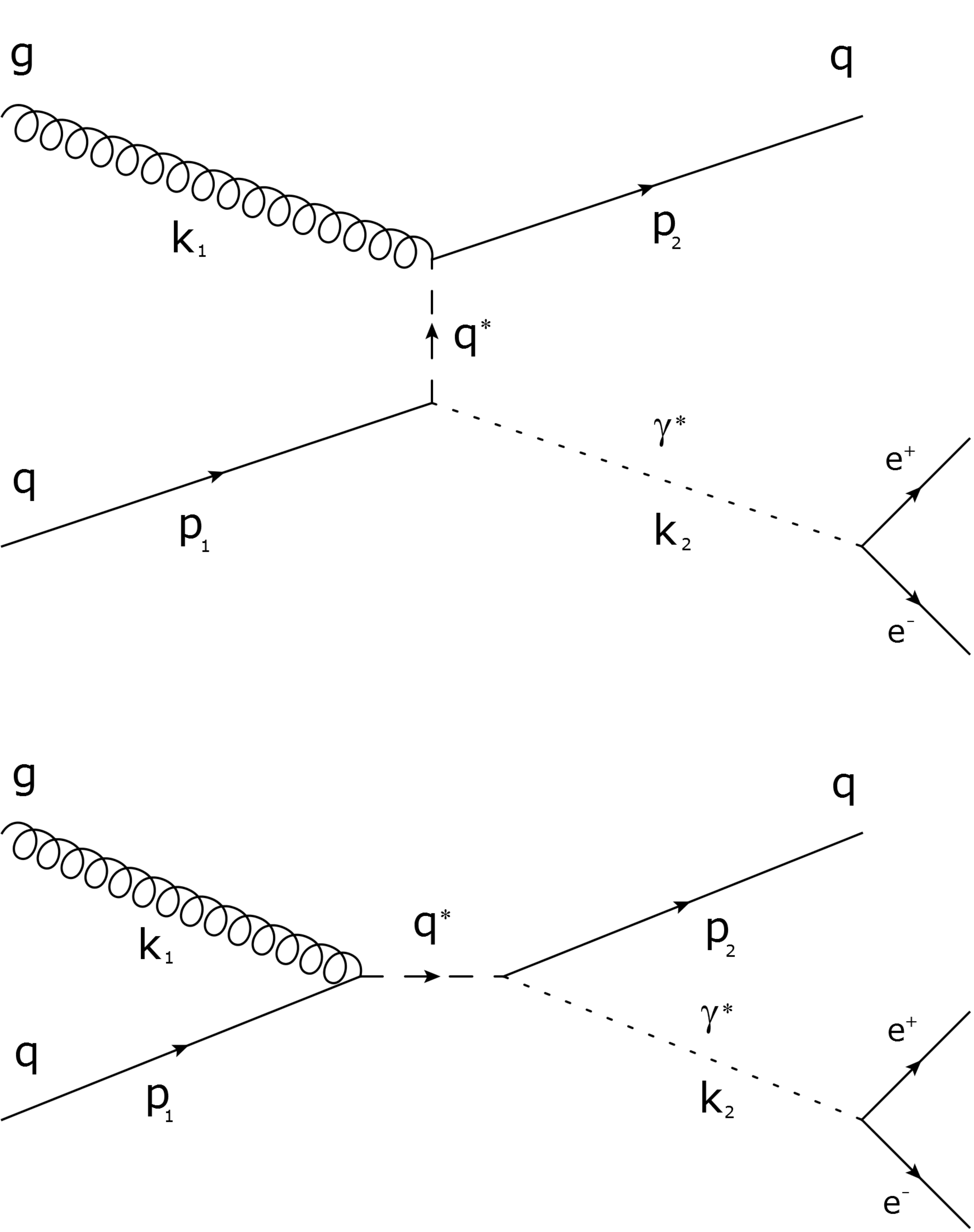} }
\end{center}
\caption{The leading order Feynman diagrams for the virtual photon and $e^{+}e^{-}$ pair production from the quark-gluon Compton scattering process in vacuum.}
\label{Feynman1}
\vspace{0.0cm}
\end{figure}

We start the appendix with a diagrammatic description of the direct photon production. Direct photons are produced by inelastic scattering processes between incoming partons. The lowest (leading) order processes are the quark-gluon QCD Compton scattering producing a (anti)quark and a photon ($q(\bar{q})g \rightarrow q(\bar{q})\gamma$) as well as the quark-antiquark annihilation into a gluon and a photon ($q\bar{q} \rightarrow g\gamma$)\footnote{The next to leading order process is dominated by bremsstrahlung and fragmentation.}. The direct photons are not only produced as massless real photons but also as virtual photons with nonzero invariant mass, which internally convert into $e^{+}e^{-}$ pairs. In general, any source of real photons, e.g., based on the Compton process, can also produce a virtual photon that is emitted as an $e^{+}e^{-}$ pair, such as shown in Fig.\,\ref{Feynman1}. Then one just has to measure such ``quasi-real" virtual photon yield. In contrast to the massless real photons, the virtual photons have an additional observable that is the aforementioned invariant mass. 


A spectrum of the invariant yield of direct photons in $0$-$20\%$, $20$-$40\%$ and $0$-$92\%$ (minimum bias) centralities in Au+Au collisions at $\sqrt{s_{NN}} = 200$\,GeV is shown in Fig.\,\ref{Photons_pt}, together with the direct photon cross section measured in p+p collisions at $\sqrt{s_{NN}} = 200$\,GeV. The measured invariant yield in Au+Au collisions is above the p+p fit. A significant photon excess is seen in the transverse momentum region, $1 < p_{T} < 3$\,GeV/c, for the three centrality classes. In turn, the p+p fit is well described by the power-law function (Eq.\,(\ref{eqn_B})):
\begin{equation}
E{d^{3}N \over d p ^{3}} =  T_{AA} \times \frac{A_{pp}}{(1 + p_{T}^{2}/p_{0}^{2})^n}\,,
\label{eqn_pow}
\end{equation}
as shown by the dashed lines in Fig.\,\ref{Photons_pt}, which is scaled by the corresponding Glauber nuclear overlap function, $T_{AA}$, for the three centralities under consideration. The values and units of the parameters in Eq.\,(\ref{eqn_pow}) are given in Sec.\,\ref{sec:Fitting1}, and also note that $T_{AA} = N_{coll}/\sigma_{in}$. Then the thermal photon excess in Fig.\,\ref{Photons_pt} is quantified by an exponential plus the  power-law fit to the Au+Au data\footnote{One can find the values of $T_{AA}$ at various centralities in \cite{Reygers:2005}. For the values of $A$ and $T$, see TABLE XII of Ref.\,\cite{Adare:2009qk}.}:

\begin{SCfigure}
\vspace{0.0cm}
\scalebox{1.0}[1.0] {
\hspace{0.0cm}
  \includegraphics[width=3.5in,height=3.5in]{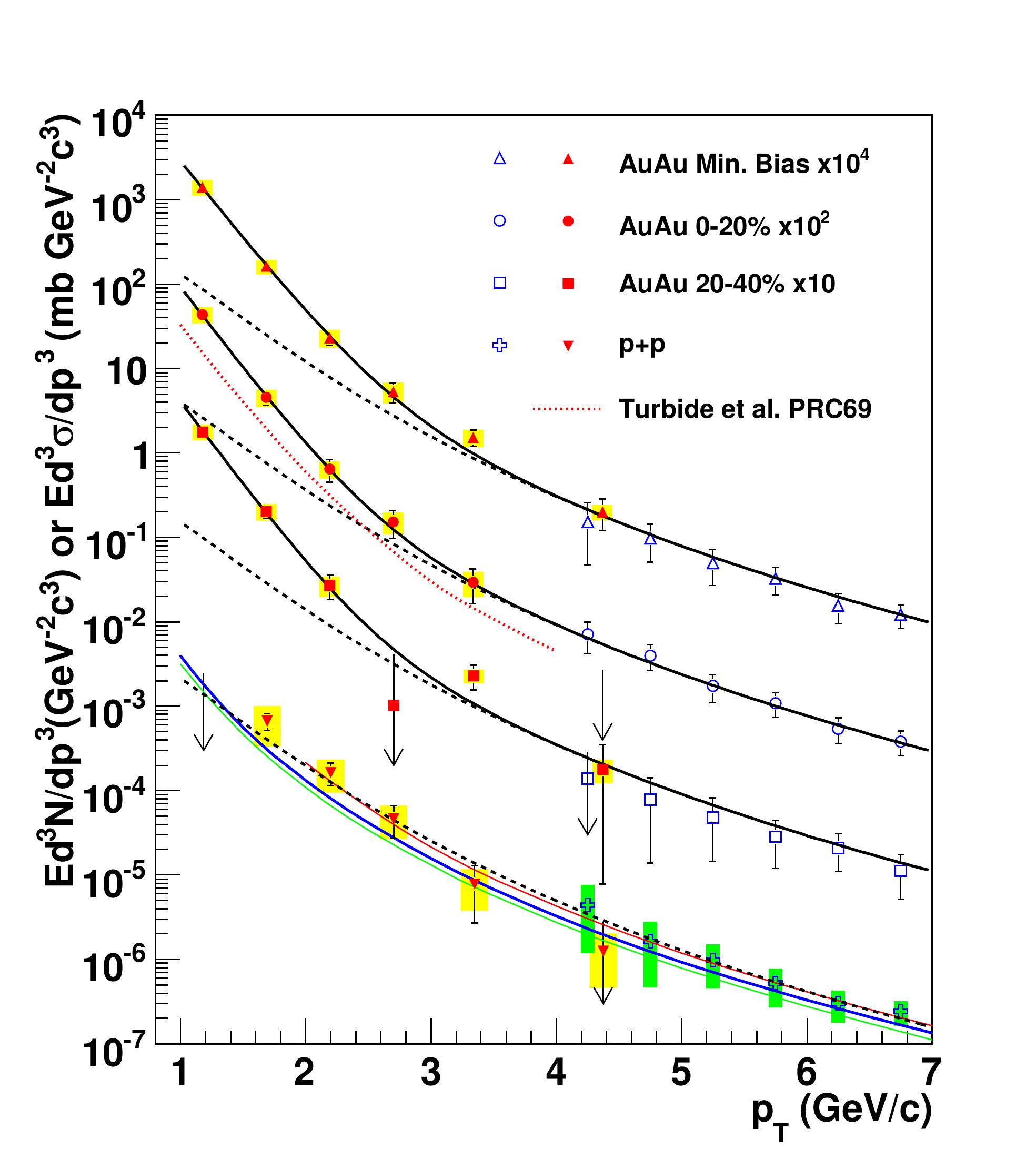} }
\caption{(Color online) The invariant cross section (p+p) and the invariant yield (Au+Au) of direct photons as a function of $p_{T}$ at $\sqrt{s_{NN}} = 200$\,GeV. The three curves on the p+p data represent NLO pQCD calculations, and the dashed curves show the power-law fit to the p+p data scaled by the Glauber nuclear overlap function, $T_{AA}$, for Au+Au. The black solid curves are an exponential plus the $T_{AA}$-scaled p+p fit. The dotted (red) curve near the $0$-$20\,\%$ centrality data is a theory calculation \cite{Turbide:2004}. This figure is from Refs.\,\cite{Adare:2008qk,Adare:2009qk}.}
\label{Photons_pt}
\end{SCfigure}

\begin{equation}
{1 \over 2\pi p_{T}}{d^{2}N_{\gamma} \over dp_{T}dy} = A\,e^{-p_{T}/T} + T_{AA}\,A_{pp} \Lb 1 + {p_{T}^{2} \over p_{0}^{2}}  \Rb^{-n}\,,
\label{eqn_prod1}
\end{equation}
where the exponential describes the thermal photon yield.

The relation between the photon production and the associated $e^{+}e^{-}$ pair production can be represented as (\cite{Adare:2008qk}, \cite{Adare:2009qk}, \cite{Dahms:2008})
\begin{equation}
{d^{2}N_{ee} \over dm_{ee}dp_{T}} = {2\alpha \over 3\pi}{1 \over m_{ee}} L(m_{ee}) S(m_{ee},p_{T}){dN_{\gamma} \over dp_{T}}\,,
\label{eqn_prod2}
\end{equation}
with
\begin{equation}
L(m_{ee}) = \sqrt{1 - {4m_{e}^{2} \over m_{ee}^{2}}} \Lb 1 + {2m_{e}^{2} \over m_{ee}^{2}} \Rb\,,
\label{eqn_L}
\end{equation}
where $\alpha$ is the electromagnetic coupling, $m_{ee}$ is the invariant mass of the $e^{+}e^{-}$ pair, and $m_{e} = 511$\,keV/c$^{2}$ is the mass of the electron. The function $S(m_{ee},p_{T})$ is process-dependent, accounting for differences between the real and virtual photon production, such as the phase space, cross section and form factors. For the quark-gluon Compton scattering, the factor \,$S(m_{ee},p_{T})$\, is determined to be
\begin{eqnarray}
S(m_{ee},p_{T}) \equiv S_{qg}(u, t, s) & = & 1+ {2u \over t^{2} + s^{2}}\,m_{ee}^{2} =
\nonumber \\
& = & 1- {2x \over \Lb x + \sqrt{1+x^{2}} \Rb \Lb 3x^{2} + 1 + 2x\sqrt{1 + x^{2}} \Rb}\,,
\label{eqn_factorS}
\end{eqnarray}
where $x = p_{T}/m_{ee}$, and $u$, $t$, $s$ are the Mandelstam variables defined as $u =(p_{1} - p_{2})^{2}$, $t =(p_{2} - k_{2})^{2}$, $s =(p_{1} + k_{1})^{2}$: with $p_{1}$, $k_{1}$, $p_{2}$ and $k_{2}$ being the $4$-momenta of the incoming quark, incoming gluon, outgoing quark and outgoing (virtual or real) photon, respectively. From Eq.\,(\ref{eqn_factorS}) it follows that $S \approx 1$ at $p_{T} \gg m_{ee}$.

The yield of the photons can be converted into a yield of $e^{+}e^{-}$ pairs in a mass bin $m_{ee,min} < m_{ee} < m_{ee,max}$ according to Eq.\,(\ref{eqn_prod2}) and Eq.\,(\ref{eqn_prod1}) \cite{Dahms:2008}:
\begin{eqnarray}
{1 \over 2\pi p_{T}}{d^{2}N_{ee} \over dp_{T}dy} & = & {1 \over 2\pi p_{T}}{d^{2}N_{\gamma} \over dp_{T}dy}
\int_{m_{ee,min}}^{m_{ee,max}}{2\alpha \over 3\pi}{1 \over m_{ee}} L(m_{ee}) S(m_{ee},p_{T})\,dm_{ee} =
\nonumber \\
& = & \left(A\,e^{-p_{T}/T} + T_{AA}\,A_{pp} \Lb 1 + {p_{T}^{2} \over p_{0}^{2}}  \Rb^{-n}\right) \times
\nonumber \\
& \times & \int_{m_{ee,min}}^{m_{ee,max}}{2\alpha \over 3\pi}{1 \over m_{ee}} L(m_{ee}) S(m_{ee},p_{T})\,dm_{ee}\,.
\label{eqn_prod3}
\end{eqnarray}
By making use of this conversion formula, one can use it for comparisons with the di-electron data, taking also into account that the photon-to-pair converted yield must be added to the di-electron contribution from hadronic decay cocktail and charmed mesons. The result is depicted in Fig.\,\ref{Dileptons_pt} as colored dashed curves for six mass bins.

\begin{figure}[hbt]
\vspace{0.2cm}
\begin{center}
\scalebox{1.0}[1.0] {
\hspace{-0.5cm}
  \includegraphics[scale=0.60]{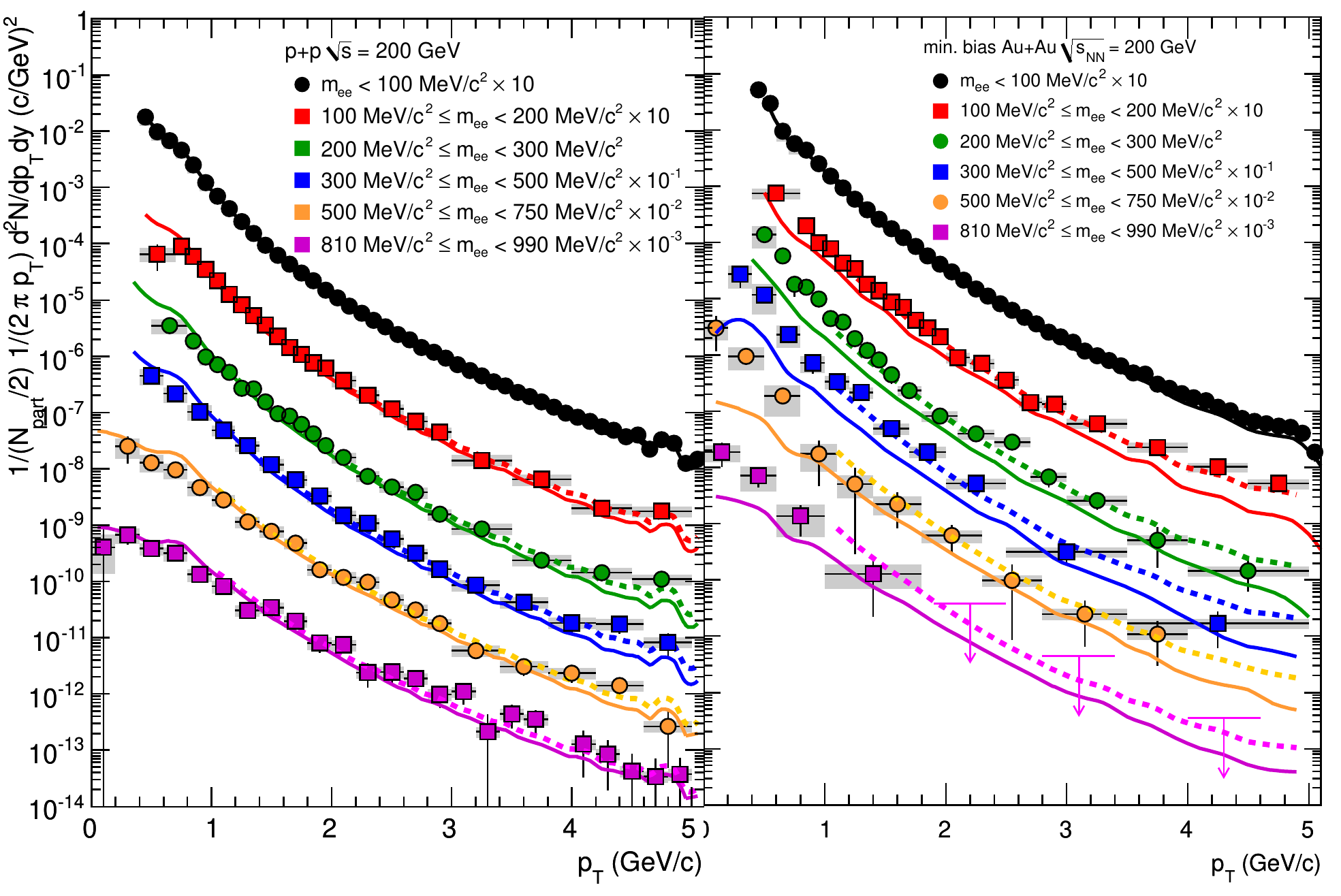} }
\end{center}
\caption{(Color online) The $p_{T}$ spectra of the $e^{+}e^{-}$ pairs in p+p (left) and Au+Au (right) collisions for six mass bins. The Au+Au spectra are divided by $N_{part}/2$. The solid curves show the expectations from the sum of contributions from the hadronic decay cocktail and charmed mesons. The dashed curves show the sum of the cocktail and charmed meson contributions plus the contribution from the direct photons calculated by converting the photon yield from Fig.\,\ref{Photons_pt} to a $e^{+}e^{-}$ pair yield using Eq.\,(\ref{eqn_prod3}). This figure is from Ref.\,\cite{Adare:2009qk}.}
\label{Dileptons_pt}
\end{figure}

In Eq.\,(\ref{eqn_prod3}) we can split apart the contribution of the thermal enhancement from the prompt contribution described by the power-law fit to the p+p data scaled by $T_{AA}$. Namely,
\begin{eqnarray}
\!\!\!\!\!\!\!\!\!\!\!\!
{1 \over 2\pi p_{T}}{d^{2}N_{ee} \over dp_{T}dy}|_{Th} & + & {1 \over 2\pi p_{T}}{d^{2}N_{ee} \over dp_{T}dy}|_{\gamma} = 
\nonumber \\
& = & A\,e^{-p_{T}/T} \int_{m_{ee,min}}^{m_{ee,max}}{2\alpha \over 3\pi}{1 \over m_{ee}} L(m_{ee}) S(m_{ee},p_{T})\,dm_{ee} +
\nonumber \\
& + & T_{AA}\,A_{pp}\Lb 1 + {p_{T}^{2} \over p_{0}^{2}}  \Rb^{-n} \int_{m_{ee,min}}^{m_{ee,max}}{2\alpha \over 3\pi}{1 \over m_{ee}} L(m_{ee}) S(m_{ee},p_{T})\,dm_{ee}\,.
\label{eqn_prod33}
\end{eqnarray}
Meanwhile, in the low $p_{T}$ region, as it can be seen from Fig.\,\ref{Photons_pt}, one can set up a condition between the exponential describing the thermal enhancement and the prompt contribution:
\begin{equation}
A\,e^{-p_{T}/T} \gg T_{AA}\,A_{pp}\Lb 1 + {p_{T}^{2} \over p_{0}}  \Rb^{-n}\,.
\label{eqn_cond}
\end{equation}
Therefore, instead of Eq.\,(\ref{eqn_prod33}) we have
\begin{eqnarray}
{1 \over 2\pi p_{T}}{d^{2}N_{ee} \over dp_{T}dy}|_{Th} = A\,e^{-p_{T}/T} \int_{m_{ee,min}}^{m_{ee,max}}{2\alpha \over 3\pi}{1 \over m_{ee}} L(m_{ee}) S(m_{ee},p_{T})\,dm_{ee}\,,
\label{eqn_photon_yield}
\end{eqnarray}
in the low $p_{T}$ region.

Now we can derive the di-electron rate based on the assumption, which led to Eq.\,(\ref{eqn_Photon_rate1}):
\begin{equation}
{d^{7}N \over {d^{4}x dy \,dm_{ee}^{2}}} \sim \alpha^{2}\Omega_{IR}\Omega_{UV}\,v(m_{ee}/\Omega_{UV})\,,
\label{eqn_Dielectron_rate1}
\end{equation}
where $v(m_{ee}/\Omega_{UV})$ is a function that cuts off when the invariant mass of an $e^{+}e^{-}$ pair is of the order of the ultraviolet cutoff scale $\Omega_{UV}$. Using the same kind of integration that resulted in Eq.\,(\ref{eqn_Photon_rate44}) we get
\begin{equation}
{{d^{3}N_{ee}} \over {dy\,dm_{ee}^{2}}} \sim \alpha^{2}R_{1}^{2}N_{part}^{2/3}{\Lb 3 + 2\Delta \over 3 \Rb} \left(Q_{s}[N_{part}, \sqrt{s}]  \over m_{ee} \right)^{\Delta}\,,
\label{eqn_Dielectron_rate2}
\end{equation}
where $Q_{s}$ is given by Eq.\,(\ref{eqn_sat}), and $R_{1} \sim R_{0}$. For obtaining the $p_{T}$-dependent production rate, one can derive it from Eq.\,(\ref{eqn_Dielectron_rate2})
\bea
{{d^{5}N_{ee}} \over {d^{2}p_{T} dy \,dm_{ee}^{2}}} & \sim & \alpha^{2}R_{1}^{2}N_{part}^{2/3}{\Lb 3 + 2\Delta \over 3 \Rb} \Lb \Lb \frac{\sqrt{s_{0}}}{\sqrt{s}} \Rb^{\lambda/(2 + \lambda)} \Rb^{(\lambda\Delta)/2} \times
\nonumber \\
& \times & 
\Lb Q_{s}[p_{T}, N_{part}, \sqrt{s}] \over m_{ee} \Rb^{\Delta}{e^{-p_{T}/\mu} \over \mu^{2 - (\lambda\Delta/2)}\!\cdot\!\Gamma[2 - (\lambda\Delta/2)]}\,,
\label{eqn_Dielectron_rate3}
\eea
such that the exponential function, $e^{-p_{T}/\mu}$, and the width parameter, $\mu$, is introduced to restore Eq.\,(\ref{eqn_Dielectron_rate2}) upon the full integration over $p_{T}$. In this case one can use $Q_{s}$ given by Eq.\,(\ref{eqn_momentum}). There can be some broadening in $p_{T}$ of $e^{+}e^{-}$ pairs due to both finite transverse size of the system and transverse collective expansion, which is accounted for by the inserted exponential function. Note that an analogous  formula is derived in Ref.\,\cite{Chiu:2013} for the di-electron production from the Bose-Einstein condensate of gluons that is theorized to exist in the overpopulated Glasma. 

Thereby, Eq.\,(\ref{eqn_Dielectron_rate3}) can be considered as a formula for obtaining the rate of the $e^{+}e^{-}$ pair production from the quark-gluon Compton scattering in the mBU-Glasma thermalization picture. Then it should be integrated over the invariant mass as well, in order to get the pair yield. By using also Eq.\,(\ref{eqn_momentum}) the integration gives the yield as a function of $p_{T}$:
\begin{eqnarray}
{1 \over 2\pi p_{T}}{{d^{2}N_{ee}} \over {dp_{T} dy}} & = & D_{\gamma} \Lb {2(3 + 2\Delta) \over 3(2 - \Delta) \cdot \Gamma[2 - (\lambda\Delta)/2]} \Rb N_{part}^{2/3} \times
\nonumber \\
& & \times \Lb m_{ee,max}^{2 - \Delta} - m_{ee,min}^{2 - \Delta}\Rb \Lb \Lb \frac{\sqrt{s_{0}}}{\sqrt{s}} \Rb^{\lambda/(2 + \lambda)} \Rb^{(\lambda\Delta)/2} \times
\nonumber \\
& & \times {e^{-p_{T}/\mu} \over \mu^{2 - (\lambda\Delta/2)}} \Lb Q_{0}^{2}[N_{part}]\Lb \frac{10^{-3}\!\cdot\!\sqrt{s}}{p_{T}} \Rb^{\lambda} \Rb^{\Delta/2}\,,
\label{eqn_lowpT1}
\end{eqnarray}
where \,$D_{\gamma} = {\rm const}\!\cdot\!\alpha^{2}R_{1}^{2}$\, describes the overall normalization to be found from a fit to data in one mass bin. 

Afterwards, the associated photon yield can be obtained by making use of the photon-to-pair conversion relation given by Eq.\,(\ref{eqn_photon_yield}). The extracted thermal photon yield as a function of $p_{T}$ is shown below:
\begin{eqnarray}
{d^{3}N_{\gamma} \over dy d^{2}p_{T}} & = & D_{\gamma} \Lb {2(3 + 2\Delta) \over 3(2 - \Delta) \cdot \Gamma[2 - (\lambda\Delta)/2]} \Rb N_{part}^{2/3} \times
\nonumber \\
& & \times \Lb m_{ee,max}^{2 - \Delta} - m_{ee,min}^{2 - \Delta}\Rb \Lb \Lb \frac{\sqrt{s_{0}}}{\sqrt{s}} \Rb^{\lambda/(2 + \lambda)} \Rb^{(\lambda\Delta)/2} \times
\nonumber \\
& & \times {e^{-p_{T}/\mu} \over \mu^{2 - (\lambda\Delta/2)}} \Lb Q_{0}^{2}[N_{part}]\Lb \frac{10^{-3}\!\cdot\!\sqrt{s}}{p_{T}} \Rb^{\lambda} \Rb^{\Delta/2} \times
\nonumber \\
& & \times \Lb \int_{m_{ee,min}}^{m_{ee,max}}{2\alpha \over 3\pi}{1 \over m_{ee}} L(m_{ee})\,S(m_{ee},p_{T})\,dm_{ee} \Rb^{-1}\,,
\label{eqn_lowpT2}
\end{eqnarray}
where $N_{part}$ dependence is the same as that in Eq.\,(\ref{eqn_E}), in spite of different $p_{T}$ shapes in both cases. 
One could try the mass range, e.g., from \,$m_{ee,min} = 300$\,MeV to $m_{ee,max} = 500$\,MeV of Fig.\,\ref{Dileptons_pt}, and carry out fitting with the data. The problem with Eq.\,(\ref{eqn_lowpT2}) is that there is an unknown parameter $\mu$, in addition to $\Delta$. The parameter $\mu$ might be a function of $\sqrt{s}$, which means that Eq.\,(\ref{eqn_lowpT2}) could be applicable for data comparison at various center-of-mass energies, but only if such a function of $\mu$ was determined. It can be determined in principle, however, its derivation is out of scope of our work in this paper, and can be calculated later in another work.

\end{document}